\documentclass[apj]{emulateapj}

\usepackage{amsmath}
\usepackage{xcolor}
\usepackage{longtable}
\PassOptionsToPackage{hyphens}{url}\usepackage[colorlinks,citecolor=blue]{hyperref}

\usepackage[hyphenbreaks]{breakurl}

\newcommand{\mum}{\ifmmode{\rm \mu m}\else{$\mu$m}\fi}

\newcommand{\sevenrm}{\rm\scriptsize}
\newcommand{\angstrom}{{\rm \AA}}

\newcommand{\hbeta}{H{$\beta$}}

\newcommand{\cosmology}{$\Omega _m = 0.3$,~$\Omega_{\Lambda}=0.7$,~and~$H_0=70$km~s$^{-1}$Mpc$^{-1}$}

\slugcomment{Published as Lyu et al. 2017, ApJ, 835, 257}

\begin{document}

\title{
    Dust-deficient Palomar-Green Quasars and the Diversity
    of AGN Intrinsic IR Emission
}

\author{Jianwei Lyu \altaffilmark{1}, G. H. Rieke \altaffilmark{1}, Yong Shi \altaffilmark{2}
}

\altaffiltext{1}{Steward Observatory, University of Arizona,
933 North Cherry Avenue, Tucson, AZ 85721, USA; \email{jianwei@email.arizona.edu}}
   \altaffiltext{2}{School of Astronomy and Space Science, Nanjing University, Nanjing 210093, China}

\shorttitle{IR SEDs of Quasars}
\shortauthors{Lyu, Rieke, \& Shi}

\begin{abstract}
    To elucidate the intrinsic broadband infrared (IR) emission properties of
    active galactic nuclei (AGNs), we analyze the spectral energy distributions
    (SEDs) of 87 $z\lesssim0.5$ Palomar-Green (PG) quasars.  While the Elvis AGN
    template with a moderate far-IR correction can reasonably match the SEDs of
    the AGN components in $\sim60\%$ of the sample (and is superior to
    alternatives such as that by Assef), it fails on two quasar populations:
    (1) hot-dust-deficient (HDD) quasars that show very weak emission
    thoroughly from the near-IR to the far-IR, and (2) warm-dust-deficient
    (WDD) quasars that have similar hot dust emission as normal quasars but are
    relatively faint in the mid- and far-IR. After building composite AGN
    templates for these dust-deficient quasars, we successfully fit the
    0.3-500~$\mum$ SEDs of the PG sample with the appropriate AGN template, an
    infrared template of a star-forming galaxy, and a host galaxy stellar
    template. 20 HDD and 12 WDD quasars are identified from the SED
    decomposition, including seven ambiguous cases.  Compared with normal
    quasars, the HDD quasars have AGNs with relatively low Eddington ratios and
    the fraction of WDD quasars increases with AGN luminosity.  Moreover, both
    the HDD and WDD quasar populations show relatively stronger mid-IR silicate
    emission. Virtually identical SED properties are also found in some quasars
    from $z=0.5$ to 6.  We propose a conceptual model to demonstrate that the
    observed dust deficiency of quasars can result from a change of structures
    of the circumnuclear tori that can occur at any cosmic epoch.
\end{abstract}


\section{Introduction}

The intrinsic UV to mid-IR spectral energy distributions (SEDs) of active
galactic nuclei (AGNs) seem universal. Albeit large SED variations are seen
among individual quasars, after removing the contamination from the host
galaxies and the extinction caused by dust, and averaging the measurements
of a sample of reasonable size, the mean SEDs of luminous AGNs are remarkably
similar \citep[e.g.,][]{Elvis1994, Richards2006, Shang2011}. Two broad and
prominent bumps dominate the UV-to-infrared SED -- one in the optical-UV (known
as the Big Blue Bump), contributed mainly by the thermal emission from the gas
in an accretion disk, and another in the near- to mid-IR, dominated by the
emission of dust heated by optical through soft X-ray photons
\citep[e.g.,][]{Rieke1978, Barvainis1987, Neugebauer1987, Sanders1989, Elvis1994}.  An
inflection around 1.25$~\mum$ separates these two bumps, as a result of dust
sublimation at a temperature of $\sim1800~$K
\citep[e.g.,][]{Barvainis1987}. The mean AGN SEDs of quasars show little
variation across a broad luminosity range as well as redshift
\citep[e.g.,][]{Wang2008a,Elvis2012,Hao2014,Lyu2016}, suggesting that these
massive black hole accreting systems and their nearby surroundings share
similar properties \citep[e.g.,][]{Scott2014}. Most importantly, the dusty
obscuration structure, typically traced by the near- to mid-IR emission and
termed as a `torus,' lay the foundation for AGN unification (e.g.,
\citealt{Antonucci1993}; see the recent review by \citealt{Netzer2015}), and its
formation is a vital part of the make-up of a quasar
\citep[e.g.,][]{Sanders1988a, King2003, Di_Matteo2005,Hopkins2005a,
Hopkins2012, Vollmer2008}.

Given the great importance of the AGN torus, it would be intriguing to look for
quasars with abnormal IR SED features that could be potentially linked with the
structure of the torus and its evolution. \cite{Jiang2010} reported the
non-detection of hot dust emission from AGNs for two $z\gtrsim5$ quasars,
claiming they are likely to be the first-generation quasars in a dust-free
environment.  \cite{Hao2010, Hao2011} reported a population of
``hot-dust-poor'' quasars at moderate redshift and argued that there is a
higher fraction at these redshifts compared to nearby quasars. Others have
explored the redshift evolution of this special population of quasars based on
large sky surveys but found contradictory results \citep{Mor2012,Jun2013}. Due
to the limited observations at high redshift, the identifications of these
quasars have been based on simple color selections, which is unlikely to be
completely accurate. Meanwhile, since the mid-IR to far-IR infrared SEDs of
these dust-deficient quasars are unknown, we lack a complete picture of the
dust distribution in these systems.

A simple characterization of the torus is provided by the dust-covering factor,
which is typically quantified by the relative luminosity between the torus and
the accretion disk emission \citep[e.g.,][]{Maiolino2007}.  In a number of
papers, the mid-IR to optical luminosity ratios of quasars are found to reduce
with increasing AGN luminosity (e.g., \citealt{Maiolino2007, Treister2008,
Mor2011, Calderone2012, Gu2013, Ma2013, Roseboom2013}), which is typically
viewed as support for the model that the torus recedes as the luminosity
increases \citep[e.g.,][]{Lawrence1991, Simpson2005, Assef2013}.  In contrast,
\cite{Richards2006} show that the mid-IR parts of the AGN SEDs are similar, but
the most luminous quasars are much brighter at mid-IR bands compared with the
least optically luminous quasars (see also, e.g., \citealt{Edelson1986,
Krawczyk2013}). Meanwhile, others have reported that the fraction of quasars
with weak hot dust emission was generally independent of the AGN luminosity
\citep{Hao2010, Hao2011, Mor2012}. These apparent discrepancies again
demonstrate our limited knowledge of quasars with weak dust emission as well as
of the connection between the weak dust emission and AGN properties.

This work reports on the dust-deficient quasars among 87 $z<0.5$ Palomar-Green
(PG) quasars from the Palomar Bright Quasar Survey \citep{Schmidt1983,
Boroson1992}. The PG sample is representative of bright optical-selected
quasars and has been a cornerstone for quasar studies in the past 40 years.
Thanks to the ample multiband observations made previously, we can calibrate
the host galaxy properties derived from SED analysis with other independent
methods and reveal the intrinsic AGN infrared emission.  We will use SED
decomposition to explore how well the classical AGN spectral template fits the
behavior of these low-redshift AGNs and to identify any with a deficiency of hot
or warm dust emission. The distributions of various AGN properties, such as AGN
luminosities, black hole masses, and Eddington ratios, will be compared between
the dust-deficient quasar population and the normal quasar population. Finally,
we will make comparative studies of the PG sample with the high-redshift
results \citep[e.g.,][]{Hao2010, Hao2011, Jiang2010, Leipski2014} and discuss
the nature of the dust-deficient quasars in general.

The paper is organized as follows. A description of the data collection and the
issue of AGN variability is provided in Section~\ref{sec:sample}. We present
the infrared SED templates for normal quasars as well as dust-deficient quasars
in Section~\ref{sec:templates}. Section~\ref{sec:pg_sed} introduces the SED
decomposition method.  Section~\ref{sec:diversity} contains the results from
the SED analysis. We discuss the diversity of AGN infrared intrinsic emission,
the characteristics of the dust-deficient quasars, as well as similar quasars
found at high-$z$ in Section~\ref{sec:discuss}.  A summary is given in
Section~\ref{sec:summary}.

Throughout this paper, we adopt the cosmology \cosmology.

\section{Data}\label{sec:sample}

\subsection{Data Compilation}\label{sec:pho-data}

To construct the IR SEDs of the 87 PG quasars, we compiled mid-IR to far-IR
photometry measured by {\it Spitzer}/MIPS at 24, 70, and 160$~\mum$
\citep{Shi2014}, and far-IR and sub-millimeter photometry observed by {\rm
Herschel} PACS and SPIRE at 70, 100, 160, 250, 350, 500$~\mum$
\citep{Petric2015}. We also gathered the near-IR photometry at $J$
(1.24$~\mum$), $H$ (1.66$~\mum$), and $K_s$ (2.16$~\mum$) from the 2 Micron All
Sky Survey (2MASS; \citealt{2MASS}) and the UKIRT Infrared Deep Sky Survey
(UKIDSS; \citealt{UKIDSS}), and mid-IR photometry at $W1$ (3.4$~\mum$), $W2$
(4.6$~\mum$), $W3$ (12$~\mum$), and $W4$ (22$~\mum$) from the AllWISE program
(\citealt{wise}).  Since we are going to compare the host galaxy stellar
emission retrieved from SED decomposition with that from image analysis,
particular attention has been paid to the selection of photometry data with
appropriate apertures to include the whole light (AGN plus the host galaxy) of
the quasar and reduce possible contamination. We used the Standard Photometry
with isophotal apertures based on the K$_s$ 20 mag/arcsec$^2$ elliptical
isophote from the 2MASS extended source catalog if the quasar light profile is
not identified to be a single point-spread-function by 2MASS. The {\it
Wide-field Infrared Survey Explorer} ({\it WISE}) $W1$ and $W2$ bands can also
be contaminated by the stellar emission. If a quasar has been identified as an
extended object by 2MASS and the {\it WISE} {\it W1}/{\it W2} aperture
photometry flux is larger than that based on the default profile-fit photometry
optimized for point sources, we chose the {\it WISE} {\it W1}/{\it W2} band
scaled-2MASS-aperture photometry with the largest aperture as long as no other
source was included. UKIDSS has observed 29 PG quasars from our sample up to
its Data Release 10. Compared with 2MASS, the UKIDSS data has a smaller time
gap with the {\it WISE} data, reducing the effect on the SED due to long-term
IR variability.  However, UKIDSS used a 2.0 arcsec diameter aperture to record
the quasar flux \citep{Dye2006}. We only use the UKIDSS data when the quasar is
a 2MASS point source.\footnote{We replaced the 2MASS data with the UKIDSS data
for the following quasars: PG 0003+158, PG 0026+129, PG 0043+039, PG 1001+054,
PG 1049$-$005, PG 1103$-$006, PG 1151+117, PG 1211+143, PG 1216+069, PG 1307+085,
PG 1309+355, PG 1552+085, PG 1612+261, PG 2251+113, and PG 2308+098. Although PG
1004+130 is also a 2MASS point source that was matched in the UKIDSS catalog,
we removed it from the list due to its lack of the UKIDSS $K$-band observation.}
For 2MASS point-source PG quasars without the complete UKIDSS near-IR data, we
used the profile-fit photometry in the 2MASS catalog. The {\it WISE} {\it W3}
and {\it W4} band profile-fit photometry was adopted for all the PG quasars.

For the UV-optical data, the Galaxy Evolution Explorer (GALEX; \citealt{GALEX})
has observed these PG quasars with far-UV (0.15$~\mum$) detections for 72 of
them and near-UV (0.23$~\mum$) detections for 78 in GALEX Release 7
\citep{GR7}. We have archival optical $u$ (0.35~$\mum$), $g$ (0.48~$\mum$), $r$
(0.62~$\mum$), $i$ (0.76~$\mum$), and $z$ (0.91~$\mum$) observations for 75 of our
quasars from the Sloan Digital Sky Survey (SDSS; \citealt{SDSS}) Data Release
12 \citep{SDSSDR12}. We adopted the SDSS Model Magnitude,
\href{http://www.sdss.org/dr12/algorithms/magnitudes/\#mag\_model}{\texttt{modelMag}},
to account for the optical emission of both resolved and unresolved objects.
For quasars without SDSS or GALEX observations, we complete their SEDs with
literature data\footnote{Broadband 1450~$\angstrom$ continuum flux for PG
0844+349, PG 0953+414, and PG 0804+761 \citep{Kaspi2005}; CIV continuum flux
(0.15~$\mum$) for PG 1535+547 and PG 2308+098 \citep{Baskin2005}; broadband
1350~$\angstrom$ and 5500~$\angstrom$ continuum flux for PG 1416$-$129
\citep{Labita2006, Hamilton2008}; {\it XMM}-{\it Newton} 0.21-0.29~$\mum$ UV
measurements for PG 0844+349, PG 0953+414, PG 1126$-$041, PG 1352+183, PG
1535+547, and PG 1626+554 \citep{Brocksopp2006, Ballo2008, Gallo2011,
Giustini2011, Page2012}; interpolated 0.25~$\mum$ flux from published data for
PG 0838+770 and PG 0804+761 \citep{Steffen2006}; integrated Johnson $B$
(0.44~$\mum$) and I (0.88~$\mum$) band flux for PG 0838+770, PG 1126$-$041, and PG
1613+658 \citep{Surace2001}; Broadband 5100~$\angstrom$ continuum flux for PG
1700+518 \citep{Kaspi2000}; Johnson R photometry for PG 1700+518
\citep{Carini2007}; Johnson $B$, $V$, $R$, $I$ photometry for PG 1302$-$102 and PG
0804+761 \citep{Ojha2009}; Johnson $V$ band from the original PG Catalog for PG
1011$-$040, PG 1310$-$108 \citep{Green1986}; $B$ ($\sim0.44~\mum$), $V$
($\sim0.55~\mum$), $g'$ ($\sim0.48~\mum$), or $r'$ ($\sim0.63~\mum$) photometry
for PG 1048$-$090, PG 1149$-$110, PG 1435$-$067 from the AAVSO Photometric All Sky
Survey \citep{Henden2016}; DENIS $I$-band (0.89~$\mum$) photometry for PG
1011$-$040, PG 1310$-$108 \citep{Paturel2003}.} with the aid of the {\it NASA/IPAC
Extragalactic Database} (NED) and the VizieR service \citep{VizieR}.

We compared the SED composed by us with the observed SED data of 27 PG quasars
presented in \cite{Elvis1994}\footnote{We only focus on the quasars in the UVSX
sample that have been analyzed in detail (Table 1A in \citealt{Elvis1994})}.
The optical-to-near-IR SED shapes of these quasars do not show substantial
changes over a timescale of $\sim$20 years. The mid- to far-IR SEDs of 16 of
these quasars show a clear drop in the most recent data, which is due to the
smaller beam sizes and higher sensitivities of {\it Spitzer}, {\it WISE}, and
{\it Herschel} compared with those of the {\it Infrared Astronomical Satellite}
({\it IRAS}) whose data product was used in \cite{Elvis1994}.

\subsection{Photometric Variability}\label{sec:var}

Since the photometric data used in this work were taken at various times
spanning $\sim$ 20 years, AGN variability could be a potential factor to
produce unphysical SED features between different data sets. Various programs
of quasar monitoring have demonstrated that the variability deceases at longer
wavelengths \citep[e.g.,][]{Cutri1985,Neugebauer1989,
di-Clemente1996,de_Vires2005}. Although the UV variability of some quasars can
be as large as $\sim$2-3 mag \cite[e.g.,][]{Paltani1994,Wheatley2008,
Welsh2011}, the optical variability amplitudes are typically at 0.2-1.5~mag
\citep[e.g.][]{Giveon1999,de_Vires2005}, and the near-IR bands only at
$\sim$0.2-0.3~mag \citep[e.g.,][]{Enya2002a,Enya2002b}. For example, in the
heterogeneous sample of $\sim$ 200 quasars observed by
\cite{Enya2002a,Enya2002b}, excluding blazars, only 23 exceeded 0.1 dex in
variation and only 5 exceeded 0.2 dex. This sample had disproportionate numbers
of AGNs expected to vary, such as radio-loud ones.  More relevant to our study,
\cite{Neugebauer1989} reported on a comprehensive study of near infrared
variability of PG quasars. Only 6 of 108 sources in their study varied by more
than 0.15 mag and only 3 sources varied by more than 0.25 mag (0.1 dex). Of the
32 quasars that will be classified as being deficient in warm or hot dust by us
(see section~\ref{sec:dd_ident}), 28 quasars were included in their study; only
one (PG 0049+171) varied by more than 0.15 mag (= 0.06 dex).  A longer baseline
can be studied by comparing the photometry in \cite{Neugebauer1989} with 2MASS
and UKIDSS photometry. However, differences in the measured fluxes could also
be due to instrumental changes (i.e., aperture photometer versus array camera),
so we can only identify {\it candidates} PG 1048$-$090, PG 1115+407, PG
1216+069, PG 1307+085, PG 1426+015, and PG 1617+175. For PG 1115+407 and PG
1426+015, the host galaxies are comparable in brightness to the AGNs
(\citealt{McLeod1994}, \citealt{Veilleux2009} respectively), so the evidence
for variations is ambiguous, since different measurement strategies could
include more or less of the host galaxy in the signal. PG 1226+023 (3C273) is
one of the few blazars in the PG sample \citep{Massaro2015}.  None of the
changes are larger than 0.25 dex, even over the 25 year baseline represented by
these measurements.

\begin{figure}[htp]
	\centering
	\includegraphics[width=1.0\hsize]{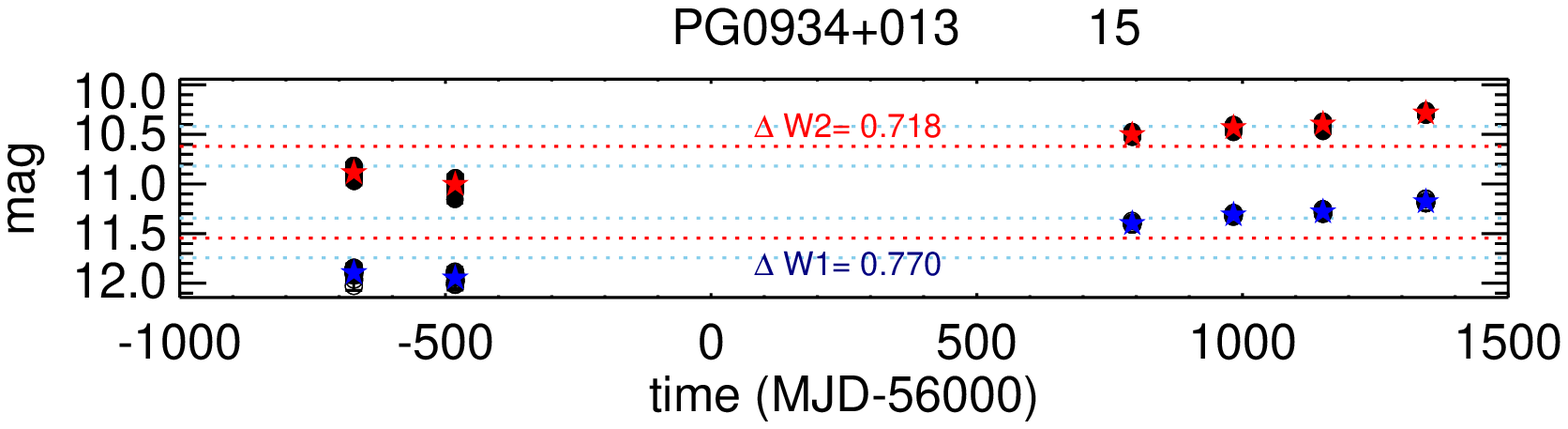} %0.77
	\includegraphics[width=1.0\hsize]{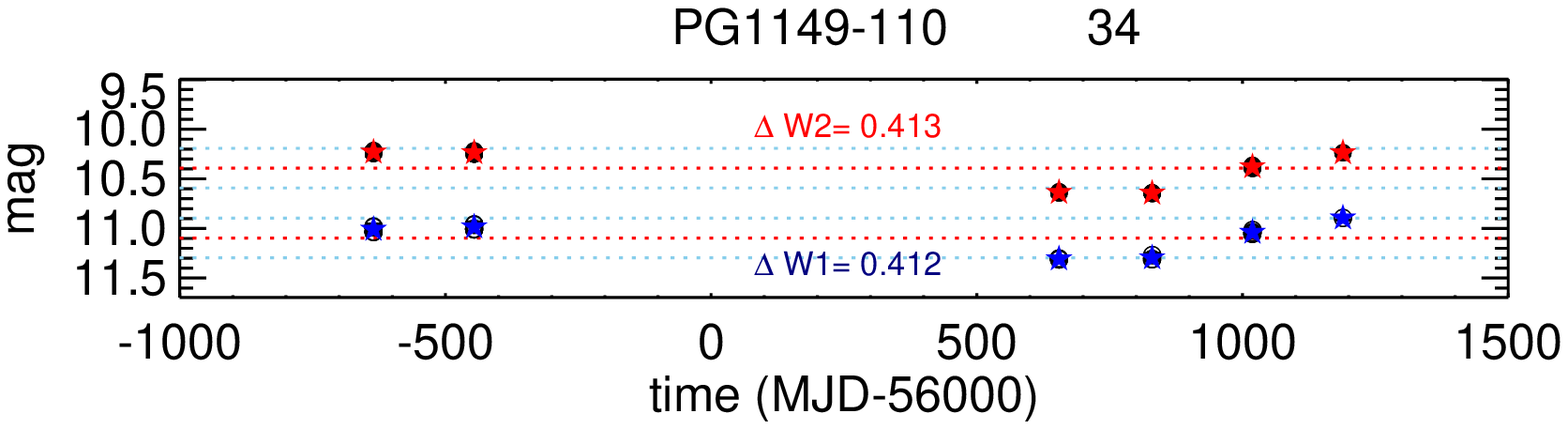} %0.41
	\includegraphics[width=1.0\hsize]{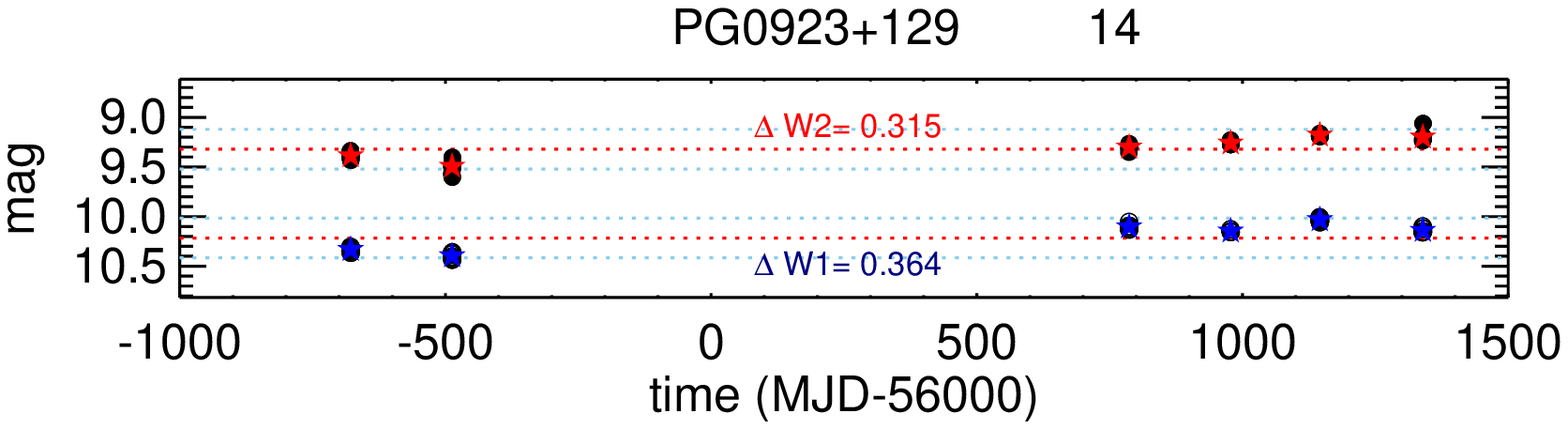} %0.36
	\includegraphics[width=1.0\hsize]{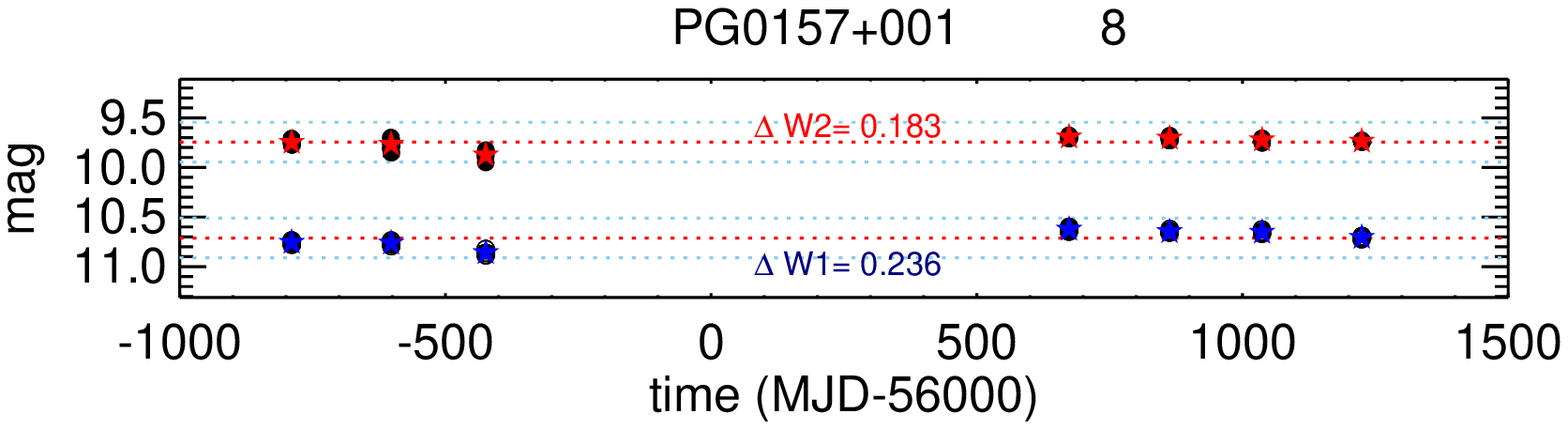} %0.24
	\includegraphics[width=1.0\hsize]{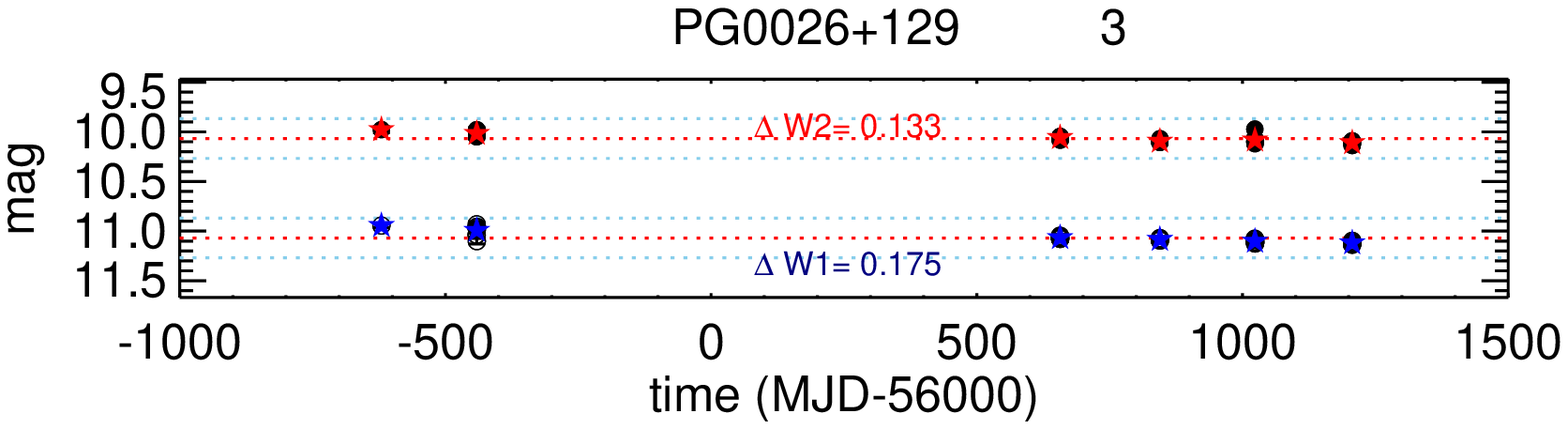} %0.175
	\includegraphics[width=1.0\hsize]{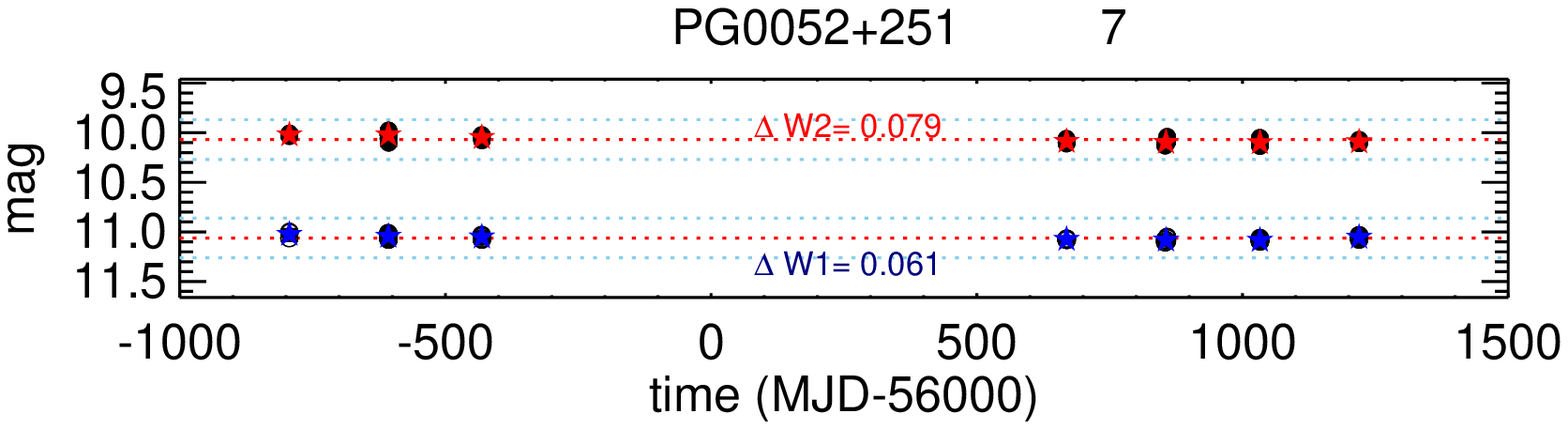}  % 0.06
	\caption{ 
	    Example {\it WISE} $W1$ ($3.4~\mum$, blue), $W2$ ($4.6~\mum$, red) band
	    light curves for a few PG quasars. The variability amplitudes are
	    indicated as $\Delta W1$ and $\Delta W2$ in each plot. We denote the
	    {\it ALLWISE} photometry value as red dotted lines with $\pm0.15$
	    mag deviations for reference as blue dotted lines. We suggest
	    $\Delta W1$ or $\Delta W2$ $>$ 0.3 mag as strong variable signals.
}
	\label{fig:pg_wise_var}
\end{figure}

To explore the longer-wavelength infrared variability of these quasars, we
collected the $W1$ and $W2$ band observations from the WISE \citep{wise} and newly released data from the
Near-Earth Object {\it WISE} Reactivation mission (\emph{NEOWISE-R};
\citealt{neowise}). During a time period of five years (2010-2016), each quasar
was observed for five to seven epochs with 10-20 individual exposures in a
single epoch. To reduce the systematic uncertainties, we average the
photometric values during an epoch, after removing exposures with poor
quality.\footnote{e.g., being influenced by the moon light or contaminated by
halos due to nearby bright stars. See:
\burl{http://wise2.ipac.caltech.edu/docs/release/allsky/expsup/sec2_4b.html} }
Figure~\ref{fig:pg_wise_var} presents example {\it WISE} light curves for a few
PG quasars with different variability amplitudes.  For the mid-IR
(3.4-4.6$~\mum$) light curves probed by {\it WISE}, $\gtrsim80$ of the PG
quasars have W1 or W2 variability amplitudes smaller than $\sim0.4~$mag.  For a
few extreme cases (e.g., PG 0003+199, PG 0934+013, PG 1534+580, PG 2304+042),
the variability can be as large as $\sim$0.6-0.7~mag.  \cite{Neugebauer1999}
studied the mid-IR (3.7 and 10~$\mum$) light curves of 25 PG
quasars in 1974-1998, and reported similar variability amplitudes.
\cite{Runnoe2012b} compared the {\it ALLWISE} $W3$ and $W4$ magnitudes with
synthetic values derived from {\it Spitzer}/IRS spectra for 22 PG quasars in
our sample and reported less than 0.2~mag variability amplitudes (i.e., less
than 0.1 dex) over a period of four to five years. For these reasons, there could be a
non-physical shift between the {\it WISE} and 2MASS data in a small number of cases.
However, the influence is still within $0.2$ dex. 

Since the separation of the host galaxy and the AGN is mostly dependent on the
near- and mid-IR for the great majority of cases, we conclude that the
differing SED behavior we find below is not significantly influenced
by variability over the time span for the various measurements.

\section{The IR SED Templates for AGNs}\label{sec:templates}

\subsection{Normal Quasars}\label{sec:templates-normal}

After subtracting the near-IR and far-IR emission from host galaxies, a number
of common dust features emerge in AGN SEDs. One such feature is hot dust
emission peaked at $\sim2~\mum$ \citep[e.g.,][]{Sanders1989, Elvis1994,
Richards2006}, which originates from the innermost region of the torus
\citep[e.g.,][]{Netzer2007}. Another major feature is a relatively flat
continuum that spans $\lambda \sim$3-20~$\mum$ with silicate emission at
$\sim10$ and $\sim18~\mum$, contributed by the warm dust emission from the
torus \citep[e.g.,][]{Fritz2006, Nenkova2008a}. The mid-IR emission of AGNs
is strongly correlated with the hard X-ray flux \citep[e.g.,][]{Lutz2004,
Horst2006, Asmus2015, Mateos2015}, further suggesting the similar SED properties
of most luminous AGNs. In the far-IR, although the contamination from the host
galaxy can become severe in many cases, the SED of AGNs is believed to drop
quickly \citep[e.g.,][]{Deo2009}.

\cite{Elvis1994} built an AGN SED template based on a sample of both
optically selected and radio-selected quasars that have strong X-ray emission
and are optically blue.  Although a number of these quasars were not detected
by {\it IRAS} due to the limited sensitivity, the far-IR SED of
\cite{Elvis1994} agrees remarkably well with later work
\citep[e.g.,][]{Richards2006, Shang2011}. In \cite{Elvis1994}, the host galaxy
IR contribution of the AGN SED was not corrected since the dispersion of the
mean template of spiral galaxies used then was too large to be useful. The lack
of understanding of the galaxy IR SED was mitigated by later work (e.g.,
\citealt{Rieke2009, Rujopakarn2011, Rujopakarn2013}): for the same IR
luminosity surface density, the IR SEDs of star-forming galaxies are similar.
Based on the correlation between the 11.3$~\mum$ aromatic feature strength and
the infrared colors, \cite{Xu2015a} removed the IR contribution from star
formation in the \cite{Elvis1994} template. The validity of this updated AGN
template to fit the UV-to-IR SED of high-redshift quasars has been demonstrated
in \cite{Xu2015a} and \cite{Lyu2016}.

Based on an iterative algorithm to derive the SED templates of $\sim10^4$ AGNs
at $z\sim$0-5.6, \cite{Assef2010} reported an AGN empirical template with a
much deeper 1~$\mum$ inflection (or stronger hot dust emission) than the
\cite{Richards2006} AGN template. \cite{Assef2010} suggested that the
\cite{Elvis1994}-like AGN templates intrinsically have considerable amounts of
host stellar contamination in the near-IR. We disagree with this argument. For
the most luminous quasars at $z\gtrsim5$, where the host galaxy contamination
can be ignored, the \cite{Elvis1994}-like AGN templates match the observed
UV-to-mid-IR SEDs well in most cases \citep[e.g.,][]{Jiang2007, Jiang2010,
Wang2008a, Lyu2016}. In Section~\ref{sec:host-stellar}, we will show that the
\cite{Assef2010} AGN template systematically overestimates the host stellar
emission in fitting normal PG quasars with HST image decomposition results as
an independent calibrator.

Recently, some authors have claimed that the intrinsic AGN IR SED should contain
much more far-IR emission, even compared with the far-IR uncorrected
\cite{Elvis1994} template \citep[e.g.,][]{Kirkpatrick2015, Symeonidis2016}.
The derivations of these templates are highly dependent on the detection of
mid-IR spectral features related to host galaxy star formation and on use of
appropriate SED templates for star forming galaxies to relate these features to
the far IR emission. In fact, if we make a mock quasar SED by combining the
\cite{Rieke2009} $\log (L_{\rm IR}/L_\odot) = 11.25$ star-formation galaxy
template and \cite{Elvis1994} AGN template, when the host galaxy contribution
is only $\sim10\%$ in the mid-IR, the galaxy still contributes $\sim50\%$ of
the total IR (8-1000~$\mum$) luminosity of the system. Given the limited
signal-to-noise of the {\it Spitzer}/IRS spectra, the uncertainties in
extrapolating to the far infrared make the derived quasar SEDs highly
uncertain.

We will use the modified \cite{Elvis1994} AGN template by \cite{Xu2015a} to
represent the intrinsic AGN SED for normal quasars in this paper.

\subsection{Dust-Deficient Quasars}\label{sec:templates-dd}

PG quasars with weak near- and mid-infrared emission were first noted in
the early 1990s \citep{Barvainis1990, McDowell1992} and later by
\cite{Hao2011}. In this paper, we use the classical \cite{Elvis1994} quasar
template as a standard ruler to look for abnormal quasars in the near-IR or the
mid-IR bands. \cite{Elvis1994} reported that the dispersion of quasar SEDs has a 68
percentile distribution within a factor of two to three of the mean throughout.  Thus,
we adopt a near- or mid-IR SED deviation from the normal quasar template by
$\gtrsim0.3$ dex to look for quasars with persuasive evidence of weak emission.

\begin{figure*}[htp]
	\centering
	\includegraphics[width=1.0\hsize]{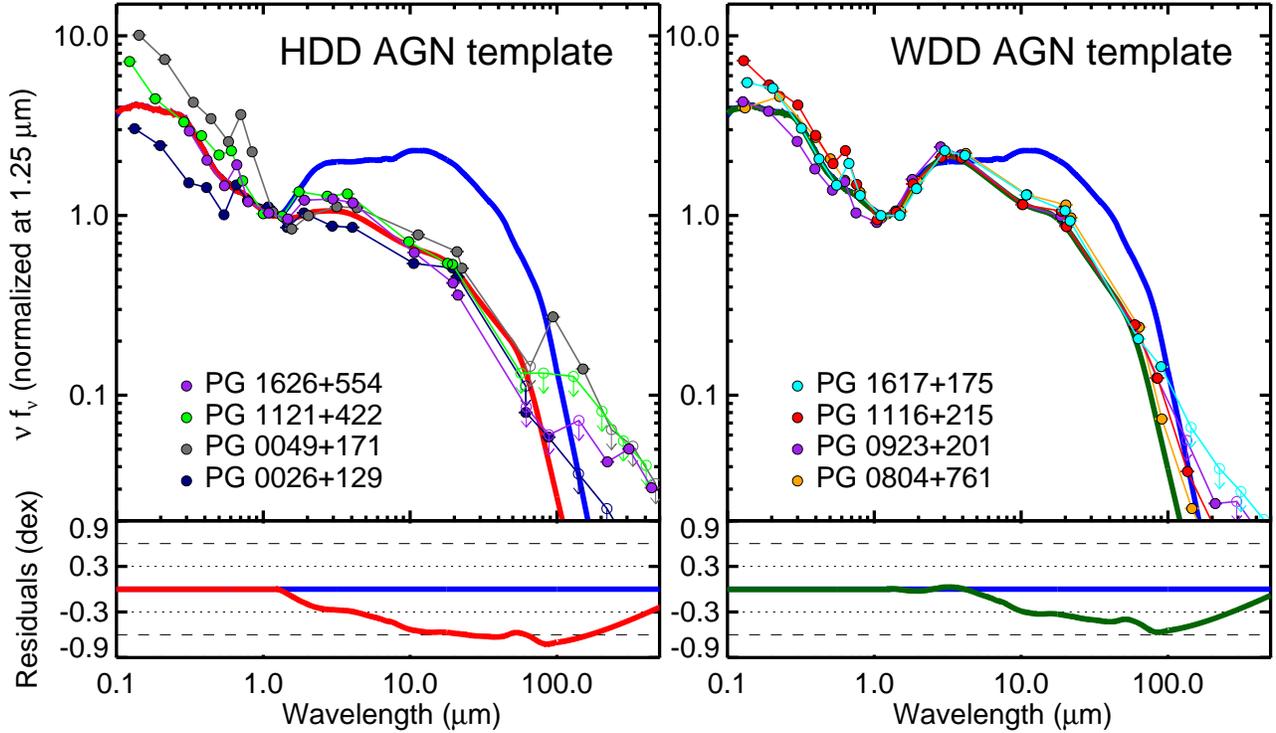}
	\caption{ 
	    Comparison of the hot-dust-deficient (HDD) AGN template (left
	    panel, red solid line) and the warm-dust-deficient (WDD) template
	    (right panel, green solid line) to the normal \cite{Elvis1994} AGN
	    template (far-IR corrected by \citealt{Xu2015a}; blue solid line).
	    We also show the SEDs of the four PG quasars that are used to
	    derive each template.
	}
	\label{fig:agn_template}
\end{figure*}

After visually inspecting the SEDs, we found a population of PG quasars
with a deficiency of hot dust emission and very weak far-IR emission. However,
host galaxy contamination in the near-IR (due to stellar emission) and the
far-IR (due to star formation in HII regions) can make the identification of
these quasars difficult. Therefore, we postulate that there is an intrinsic
hot-dust-deficient (hereafter HDD) SED and make a template by averaging the
SEDs of quasars that clearly present a deficiency of hot dust emission and do
not have strong host galaxy contamination in their SEDs. After a detailed
inspection of the observed SEDs of the whole sample, we ended up with four
purest examples of HDD quasars: PG 0026+129, PG 0049+171, PG 1121+422, and PG
1626+554. Normalized at 1.25$~\mum$, their IR SEDs look quite similar, as
shown in the left panel of Figure~\ref{fig:agn_template}. Three of the four are
not detected in the far-IR and one -- PG 0039+171 -- is not much above the
SPIRE confusion noise \citep{Nguyen2010}. Their infrared emission is not likely
to be strongly contaminated by radio synchrotron emission considering the small
value of radio loudness (see Table~\ref{tab:dd-quasar}).  Therefore, as found
in \cite{Xu2015a}, we scale a blackbody of 118 K and with a wavelength-dependent
emissivity proportional to $\lambda^{-1.5}$ to match the HDD template at
$\lambda<100~\mum$. The optical SEDs of these four quasars show some variation
but the SED obtained by averaging the individual quasar measurements is
still similar to that of normal quasars. Thus, we assume the same average SED
as that of normal quasars for the HDD template at $\lambda<1.0~\mum$.

\begin{deluxetable}{lccccc}
    %\tabletypesize{\scriptsize}
    \tablewidth{1.0\hsize}
    \tablecolumns{6}
    \tablecaption{PG quasars used to derive the templates\label{tab:dd-quasar}
    }
    \tablehead{
	\colhead{Source} & \colhead{$z$}  & \colhead{$f_{\rm nucleus,~H}$} & \colhead{Reference} & \colhead{Variability} & \colhead{$R$} \\
	\colhead{(1)} & \colhead{(2)} & \colhead{(3)} & \colhead{(4)} & \colhead{(5)} & \colhead{(6)}
    }
    \startdata
    \multicolumn{6}{c}{HDD template} \\
    PG 0026+129  &      0.14  &     0.80  &  1       & N   & 1.08 \\
    PG 0049+171  &      0.06  &     --    &  --      & N   & 0.32 \\
    PG 1121+422  &      0.23  &     0.93  &  2       & N   & 0.10 \\
    PG 1626+544  &      0.13  &     0.72  &  1       & N   & 0.11 \\
    \multicolumn{6}{c}{WDD template} \\ 
    PG 0804+761  &      0.11  &     0.90  &  3       & N   & 0.60 \\
    PG 0923+201  &      0.19  &     0.77  &  1       & N   & 1.74 \\
    PG 1116+215  &      0.18  &     0.95  &  1       & N   & 0.72 \\
    PG 1617+175  &      0.11  &     0.89  &  1       & Y   & 0.72
    \enddata
    \tablecomments{
	Column (1): object name; column (2): redshift; column (3): the contribution
	of the PSF component in the observed $H$ band from the HST image
	decomposition; column (4): references for HST image decomposition
	results: 1-\cite{Veilleux2009}; 2-\cite{McLeod2001};
	3-\cite{Guyon2006}.  column  (5): if the quasar shows strong near-IR
	variability; column (6): radio loudness, taken from \cite{Petric2015}.
    }
\end{deluxetable}

A standard procedure for making an AGN template should include the subtraction
of the host galaxy contribution in the near-IR \citep[e.g.,][]{Elvis1994,
Richards2006, Shang2011}. Nevertheless, the host galaxy contamination in these
four HDD quasars is small. For the three quasars with HST images, the quasar
light substantially outshines the stellar emission in the near-IR (see
Table~\ref{tab:dd-quasar}). The other quasar, PG 0049+171, is not resolved in
the 2MASS images despite its low redshift. It is barely resolved in $V$ band by
\cite{Smith1986}, who concluded that the host galaxy is two magnitudes fainter
than the quasar at this wavelength. Its optical to near-IR colors are bluer
than or identical to classical quasars, suggesting that the near-IR emission is
not significantly boosted by the host galaxy. As a result, we conclude that the
near-IR host galaxy stellar contamination for these four quasars is too small
to have any visible influence on the derived HDD AGN template.

In addition, we have found that there are a number of quasars whose SEDs cannot
be fitted adequately by either the normal or HDD templates, or by any
combination of them. Their near-IR SEDs present the typical bump peaked at
$\sim$ 2.0 $\mu$m, but they drop quickly at $\lambda >$ 5.0 $\mu$m.  In other
words, hot dust emission is present in these quasars, but their mid-IR emission
is relatively weak compared with normal quasars. Host galaxy contamination
would not produce this type of SED, and should be small in any case as shown in
Table~\ref{tab:dd-quasar}. We combine the IR SEDs of four such quasars, PG
0804+761, PG 0923+201, PG 0953+414, PG 1116+215, and make a composite SED
similarly to the HDD template (right panel of Figure~\ref{fig:agn_template}).
These four quasars all have small radio loudnesses and weak near-IR
variability. We will describe this SED as the warm-dust-deficient (WDD)
AGN template from now on.

\begin{figure}[htp]
    \centering
	\includegraphics[width=1.0\hsize]{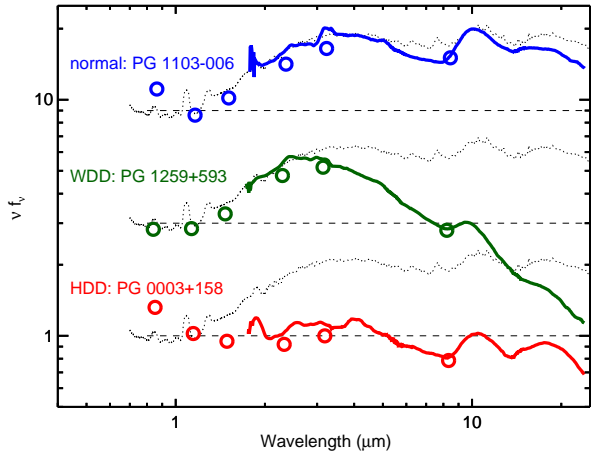} 
	\caption{ 
	    {\it Akari}+{\it Spitzer}/IRS combined infrared spectra for three
	    PG quasars: the normal quasar PG 1103$-$006 (blue), the WDD quasar
	    PG 1259+593 (green), and the HDD quasar PG 0003+158 (red). We also
	    show the 2MASS ($J$, $H$, $K_s$) and {\it WISE} ($W1$, $W2$, $W3$)
	    photometry for each quasar as open circles. We normalize each
	    quasar IR spectrum and the corresponding template at $1.25~\mum$
	    with respect to the dashed reference line. The dotted lines are the
	    spectral templates for normal quasars, in which we join the
	    5-30$~\mum$ quasar template by \cite{Hao2007} and 1-10$~\mum$
	    quasar template by \cite{Hernan-Caballero2016} at $6.7~\mum$.
}
	\label{fig:quasar_spec}
\end{figure}

We can confirm the distinct features of the three AGN SED templates for normal
quasars, HDD quasars and WDD quasars by the {\it Akari} and {\it Spitzer}/IRS
infrared spectra of example quasars, as shown in Figure~\ref{fig:quasar_spec}.
PG 0003+158, PG 1259+593, and PG 1103$-$006 were selected because their redshifts
($z>0.43$) allow {\it Akari} spectra to cover the near-IR band
($\sim1.7~\mum$). The differences among normal, WDD and HDD quasars are due to
the dust continuum, not any emission features like aromatic bands or silicate
features. With the same UV-optical luminosity, the HDD AGN template has only
$\sim$40\% of the emission of the normal AGN template at 1.25-1000~$\mum$.  For
the WDD AGN template, this value is $\sim$70\%.

\subsection{Are the HDD and WDD Templates Bona Fide?}

An artificial SED similar to that of quasars with weak infrared emission could
result if there is a strong contribution from the population of old stars in
the host galaxy, which would peak near 1 micron and fill in the typical minimum
near that wavelength in the normal (Elvis) quasar SED.  In this subsection, we
check if the HDD and WDD templates could be derived from the combination of
normal quasar SED and stellar template.

\begin{figure}[htp]
    \centering
	\includegraphics[width=1.0\hsize]{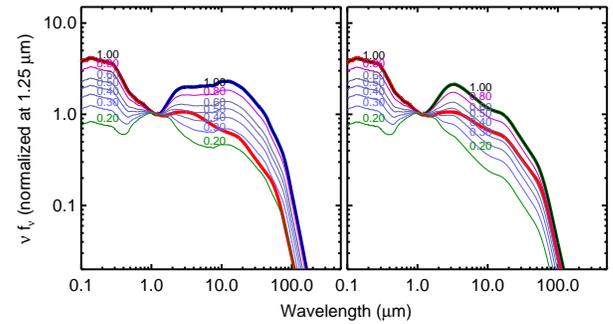} 
	\caption{ 
	    Mock SEDs of quasars. The left panel presents the composite SEDs of
	    the Elvis AGN template and an old stellar population template, with
	    the numbers indicating the fraction of the AGN contribution at
	    $1.25~\mum$. The normal AGN template is shown as the red line. The
	    right panel shows the composite SEDs of the WDD AGN template and an
	    old stellar population template. The WDD AGN template is denoted as
	    the green line. In both panels, the HDD template is shown as the
	    blue thick line.
}
	\label{fig:pg_hdd_mock}
\end{figure}

In the top panel of Figure~\ref{fig:pg_hdd_mock}, we combine the
\cite{Elvis1994} AGN template with a single stellar population template for 13
Gyr old stars from \cite{bc03} to explore the SEDs of normal quasars with
different stellar contamination in the near-IR (as indicated by the AGN light
fraction in the near-IR, $f_{\rm AGN,1.25~\mum}$). As the stellar emission in
the near-IR increases, the prominence of the near-IR dust spectral bump
gradually decreases. However, we also see a change of the optical slope away
from that of normal quasars. As shown in Section~\ref{sec:templates-dd}, the
HDD quasars share similar optical colors as other local quasars. Meanwhile,
compared with the composite SEDs, the HDD template has relatively weaker
emission in the mid-IR. For these reasons, we conclude that the HDD quasar
template cannot be produced by a normal quasar template plus stellar emission
in the near-IR. A similar argument can also be found in \cite{Hao2010,Hao2011}.

We also investigate if the HDD quasar template is a result of the combination
of the WDD quasar template and an old stellar population template. As shown in
the right panel of Figure~\ref{fig:pg_hdd_mock}, though the $f_{\rm
AGN,1.25~\mum}$=0.40-0.50 composite SEDs resemble the WDD template in the
infrared, their optical colors differ from the WDD quasars (see the right panel
of Figure~\ref{fig:agn_template}). Additionally, we expect the stellar
contributions to both HDD and WDD templates to be low (see the previous
Subsection). As a result, we conclude that the HDD and WDD templates represent
two different groups of quasars, rather than the results of different stellar
near-IR contaminations.

The WDD AGN template is distinguished from the normal AGN template by its very
weak mid-IR and far-IR emission. Normalized at 1.25$~\mum$, the former has only
$\sim45\%$ of the total infrared luminosity (8-1000$~\mum$) of the latter. As
stated in Appendix C of \cite{Xu2015a}, the IR modified \cite{Elvis1994} AGN
template provides a limiting case for the maximum plausible FIR contribution
from star formation in a quasar. Given the consistent star formation rates
(SFRs) based on the IR luminosities of the host galaxies and the mid-IR
11.3$~\mum$ aromatic feature strengths (see Section~\ref{sec:host-dust-cal}),
we have no reason to suggest that the intrinsic AGN emission of normal quasars could
be as low as the case of the WDD AGN template. In other words, although sharing
a similar hot dust emission feature, the normal AGN template and the WDD AGN
template reflect two populations of quasars with distinct mid- to far-IR
emission properties.

\section{SED Decomposition of PG quasars}\label{sec:pg_sed}

We now use the three AGN SED templates to fit the behavior of the full set
of 87 PG quasars. 

The wide wavelength range covered by the photometry of PG quasars enables
accurate SED decomposition of the emission from the AGN and the host galaxy.
SED fitting methods using multiple components have been developed and
demonstrated in a number of works
\citep[e.g.,][]{Bongiorno2007,Bongiorno2012,Pozzi2010, Lusso2011, Xu2015a}.
The decomposition depends on the large differences in the behavior of the
optical to near-IR SEDs of the stellar and AGN emission.  As summarized in the
Introduction, the AGN SED shows two broad maxima separated by a dip at
$\sim1.25~\mum$. Meanwhile, the galaxy emission is a result of multiple stellar
populations with different star formation histories. For the nearby massive
galaxies, the stellar SEDs generally peak at $\sim 1~\mum$, dominated by the
emission from stars in old stellar populations, and drop quickly as a
Rayleigh-Jeans tail toward the mid-IR.  Examples of using a galaxy plus AGN
component to decompose the optical-near-IR quasar SEDs can be seen in e.g.,
\cite{Bongiorno2007,Bongiorno2012,Pozzi2010}, and \cite{Lusso2011}. For the mid-IR and
far-IR emission of a quasar, both AGN and galaxy will contribute some emission
by dust. With proper consideration of templates used to represent the intrinsic
AGN emission and the host dust emission, separation of the contributions from
these two are possible \citep[e.g.,][]{Dale2014, Xu2015a, Lyu2016}

We model the 0.5-500~$\mum$ SED of each PG quasar with a combination of
three components.

\begin{enumerate}
    \item {\em AGN component with possible extinction}: As stated in
	Section~\ref{sec:templates-normal}, we select the AGN intrinsic
	template developed based on \cite{Elvis1994} by \cite{Xu2015a} (see
	their Appendix C) to represent the AGN emission for normal quasars. 
	This normal AGN template will be combined with the HDD or WDD AGN
	template to characterize the possible transitional infrared SEDs
	between the normal cases and dust-deficient cases, while keeping the
	SED of the UV-optical emission unchanged. To take into account the
	obscuration caused by dust in the AGN ambient regions as well as the
	host galaxy, we assume that the AGN component is reddened by an SMC-like
	($\lambda<1~\mum$) plus MW-like ($\lambda>1~\mum$) composite extinction
	curve, as in \cite{Xu2015a}.  Such a combination is proposed based on
	the statistical study of quasar UV-optical extinction based on SDSS
	\citep{Hopkins2004} and the lack of knowledge of the AGN near- to
	mid-IR extinction curves.

    \item {\em IR emission of a luminous infrared galaxy}: Many PG quasar host
	galaxies are found to have SFRs
	$\sim$10-100~$M_\odot {\rm yr}^{-1}$ \citep[based on the strength of
	$11.3~\mum$ aromatic feature, see ][]{Shi2014}, which roughly puts them
	in the LIRG ($L_{\rm IR}>10^{11}~L_\odot$) category. We adopt the
	templates in \cite{Rieke2009} with $\log_{10} L_{\rm IR}$=9.75-12.00.
	In \cite{Rieke2009}, the shape of the infrared SED of nearby galaxies
	is found to be dependent on their infrared luminosity. We will pick the
	one that best fits the far-IR SED of the quasar in combination with the
	other components, without imposing a luminosity constraint.

   \item {\em Stellar emission of an old stellar population}:  We use the
       stellar SEDs from \cite{bc03} with a Salpeter initial mass function,
	Padova evolutionary tracks, and solar metallicity. Considering the
	similarity between the SEDs for a range of ages, we pick one single
	stellar population template for each quasar with an age of 0.5 Gyr, 1.6 Gyr,
	7.2 Gyr, or 13 Gyr, to represent the overall SED properties of the
	stellar emission that contributes to the near-IR emission (in most
	cases, any differences in the optical and UV stellar SEDs are
	overwhelmed by the emission by the quasars).

\end{enumerate}

Our SED model can be summarized by the following equation
\begin{align*}
    f_{\rm quasar} = & \left( c_1 f_{\rm AGN, norm.}+ c_2 f_{\rm AGN, defi.} 
\right) e^{- c_3 (A_{\lambda}/A_{\rm V})} \\ 
    &  + c_4 f_{\rm gal., stars}  + c_5 f_{\rm gal.,dust}  ~~,
\end{align*}
where $(A_{\lambda}/A_{\rm V})$ is the normalized extinction curve, and $f_{\rm
AGN, norm.}$, $f_{\rm AGN, defi.}$,  $f_{\rm gal., stars}$, $f_{\rm gal., dust}$
are the SED templates for normal AGN, HDD or WDD AGN, stellar emission, and galaxy
infrared emission. There are five normalizing factors, denoted as $c_1,c_2,
c_3, c_4, c_5$. With the selection of the stellar templates and galaxy IR
templates, there are seven free parameters in total. The modeled photometry
will be computed by convolving the corresponding filter with the modeled SED.
Since the multiband photometry includes a large set of data with a diverse
level of relative uncertainty, to reflect the overall SED shape rather than
focusing on a few data points with the smallest uncertainties, we give the same
weight to all data points in $\nu$-$\nu f_\nu$ space during the fitting.  We will
use the classical definition of the $\chi^2$ statistic to evaluate the goodness
of the fitting at first, and then check the residuals of the points with large
$\chi^2$ values in detail.

\section{Results}\label{sec:diversity}

The SED fitting results for the 87 PG quasars are described in
Figure~\ref{fig:pg_sed_fitting} and Table~\ref{tab:PG-fit}.  As introduced
briefly in Section~\ref{sec:var}, the AGN variability potentially embedded in
different data sets could introduce some offset from the intrinsic SED ($\sim1$
dex in the UV, $\sim$0.1-0.2 dex in the near-IR). As a result, the UV-optical
SEDs are hard to interpret. We will mainly focus on the reproduction of the
overall shape of the SED, and the host galaxy contribution to the near-IR and
far-IR.

\onecolumngrid
\LongTables
\begin{deluxetable*}{clccccccccc}
    \tabletypesize{\scriptsize}
    \tablewidth{1.0\textwidth}
    \tablecolumns{11}
    \tablecaption{SED decomposition results for the 87 Palmor-Green quasars\label{tab:PG-fit}
    }
    \tablehead{
	\colhead{ID} &
	\colhead{Source} & 
	\colhead{$z$}  & 
	\colhead{Type} & 
	\colhead{$f_{\rm HDD/WDD}$} & 
	\colhead{$f_{\rm Stellar}$} &
	\colhead{$F_\text{star, H}/F_\text{quasar, H}$} 	& 
	\colhead{$L_{\rm IR}/10^{11}L_\odot$} & 
	\colhead{$f_{\rm host}$} &
	\colhead{$L_{\rm AGN}/10^{11}L_\odot$} &
	\colhead{SFR} \\
	\colhead{(1)} & 
	\colhead{(2)} & 
	\colhead{(3)} & 
	\colhead{(4)} & 
	\colhead{(5)} & 
	\colhead{(6)} &
	\colhead{(7)} &
	\colhead{(8)} &
	\colhead{(9)} &
	\colhead{(10)} &
	\colhead{(11)} 
    }
    \startdata
   0 &  PG 0003+158  &   0.45    &   HDD &  1.00   &   0.00   &  0.00  &   9.95     &    0.05  &  68.76 &   8.7     \\
   1 &  PG 0003+199  &   0.03    &   --  &  0.00   &   0.56   &  0.48  &   0.33     &    0.06  &   2.22 &   0.4     \\
   2 &  PG 0007+106  &   0.09    &   --  &  0.00   &   0.44   &  0.38  &   2.02     &    0.24  &  11.11 &   8.5     \\
   3 &  PG 0026+129  &   0.14    &   HDD &  0.81   &   0.11   &  0.10  &   1.62     &    0.00  &  11.81 &   0.0     \\
   4 &  PG 0043+039  &   0.38    &  HDD? &  0.84   &   0.17   &  0.18  &  10.40     &    0.35  &  49.50 &  62.2     \\
   5 &  PG 0049+171  &   0.06    &   HDD &  1.00   &   0.00   &  0.00  &   0.22     &    0.12  &   1.42 &   0.4     \\
   6 &  PG 0050+124  &   0.06    &   --  &  0.00   &   0.52   &  0.40  &   6.66     &    0.44  &  26.92 &  51.1     \\
   7 &  PG 0052+251  &   0.16    &   HDD &  0.57   &   0.10   &  0.09  &   3.37     &    0.21  &  19.49 &  12.0     \\
   8 &  PG 0157+001  &   0.16    &   --  &  0.00   &   0.12   &  0.10  &  25.35     &    0.69  &  56.82 & 303.4     \\
   9 &  PG 0804+761  &   0.10    &  WDD  &  1.00   &   0.00   &  0.00  &   3.19     &    0.04  &  22.34 &   2.1     \\
  10 &  PG 0838+770  &   0.13    &   --  &  0.00   &   0.52   &  0.47  &   2.66     &    0.49  &   9.91 &  22.5     \\
  11 &  PG 0844+349  &   0.06    &  HDD? &  0.70   &   0.40   &  0.36  &   0.62     &    0.23  &   3.49 &   2.4     \\
  12 &  PG 0921+525  &   0.04    &   --  &  0.00   &   0.56   &  0.47  &   0.22     &    0.11  &   1.45 &   0.4     \\
  13 &  PG 0923+201  &   0.19    &  WDD  &  1.00   &   0.00   &  0.00  &   3.39     &    0.09  &  22.59 &   5.0     \\
  14 &  PG 0923+129  &   0.03    &   --  &  0.00   &   0.82   &  0.77  &   0.40     &    0.54  &   1.33 &   3.7     \\
  15 &  PG 0934+013  &   0.05    &   --  &  0.00   &   0.64   &  0.58  &   0.40     &    0.62  &   1.10 &   4.3     \\
  16 &  PG 0947+396  &   0.21    &   --  &  0.00   &   0.02   &  0.02  &   6.71     &    0.29  &  34.75 &  33.5     \\
  17 &  PG 0953+414  &   0.24    &  WDD  &  1.00   &   0.00   &  0.00  &   7.31     &    0.00  &  53.24 &   0.0     \\
  18 &  PG 1001+054  &   0.16    &   --  &  0.00   &   0.31   &  0.29  &   2.15     &    0.04  &  14.99 &   1.6     \\
  19 &  PG 1004+130  &   0.24    &   --  &  0.00   &   0.37   &  0.36  &   8.87     &    0.16  &  54.24 &  24.6     \\
  20 &PG 1011$-$040  &   0.06    &   HDD &  0.75   &   0.42   &  0.39  &   0.60     &    0.47  &   2.31 &   4.8     \\
  21 &  PG 1012+008  &   0.19    &   --  &  0.00   &   0.50   &  0.48  &   3.76     &    0.24  &  20.82 &  15.6     \\
  22 &  PG 1022+519  &   0.05    &  HDD? &  0.63   &   0.61   &  0.58  &   0.30     &    0.65  &   0.77 &   3.3     \\
  23 &  PG 1048+342  &   0.17    &   --  &  0.00   &   0.56   &  0.54  &   2.05     &    0.43  &   8.53 &  15.2     \\
  24 &PG 1048$-$090  &   0.34    &  WDD  &  1.00   &   0.00   &  0.00  &   6.08     &    0.06  &  41.46 &   6.7     \\
  25 &PG 1049$-$005  &   0.36    &   --  &  0.00   &   0.00   &  0.00  &  30.49     &    0.31  & 152.93 & 164.0     \\
  26 &  PG 1100+772  &   0.31    &   HDD &  0.67   &   0.00   &  0.00  &  11.87     &    0.32  &  58.76 &  65.7     \\
  27 &PG 1103$-$006  &   0.43    &   --  &  0.00   &   0.10   &  0.11  &  13.13     &    0.00  &  95.56 &   0.0     \\ 
  28 &  PG 1114+445  &   0.14    &   --  &  0.00   &   0.05   &  0.04  &   4.04     &    0.03  &  28.58 &   2.0     \\ 
  29 &  PG 1115+407  &   0.15    &   HDD &  0.71   &   0.04   &  0.04  &   3.70     &    0.60  &  10.81 &  38.3     \\
  30 &  PG 1116+215  &   0.18    &  WDD  &  1.00   &   0.00   &  0.00  &   7.23     &    0.00  &  52.60 &   0.0     \\
  31 &  PG 1119+120  &   0.05    &   --  &  0.00   &   0.43   &  0.37  &   0.90     &    0.40  &   3.96 &   6.2     \\
  32 &  PG 1121+422  &   0.23    &   HDD &  1.00   &   0.00   &  0.00  &   1.73     &    0.00  &  12.62 &   0.0     \\
  33 &PG 1126$-$041  &   0.06    &   --  &  0.00   &   0.48   &  0.42  &   1.76     &    0.34  &   8.46 &  10.4     \\
  34 &PG 1149$-$110  &   0.05    &   --  &  0.00   &   0.82   &  0.78  &   0.53     &    0.56  &   1.69 &   5.2     \\
  35 &  PG 1151+117  &   0.18    &   --  &  0.00   &   0.27   &  0.25  &   1.77     &    0.00  &  12.85 &   0.0     \\
  36 &  PG 1202+281  &   0.17    &   --  &  0.00   &   0.01   &  0.01  &   3.36     &    0.26  &  18.10 &  15.2     \\
  37 &  PG 1211+143  &   0.09    &   --  &  0.00   &   0.22   &  0.18  &   2.57     &    0.00  &  18.72 &   0.0     \\
  38 &  PG 1216+069  &   0.33    &   HDD &  1.00   &   0.13   &  0.14  &   5.55     &    0.00  &  40.38 &   0.0     \\
  39 &  PG 1226+023  &   0.16    &   WDD &  1.00   &   0.00   &  0.00  &  30.11     &    0.21  & 173.72 & 108.0     \\
  40 &  PG 1229+204  &   0.06    &   --  &  0.00   &   0.61   &  0.56  &   0.90     &    0.29  &   4.66 &   4.5     \\
  41 &  PG 1244+026  &   0.05    &   --  &  0.00   &   0.31   &  0.26  &   0.37     &    0.42  &   1.53 &   2.7     \\
  42 &  PG 1259+593  &   0.47    &   WDD &  1.00   &   0.00   &  0.00  &  16.16     &    0.00  & 117.63 &   0.0     \\
  43 &PG 1302$-$102  &   0.29    &   HDD &  0.76   &   0.00   &  0.00  &  14.82     &    0.33  &  72.63 &  83.7     \\
  44 &  PG 1307+085  &   0.16    &   HDD &  0.00   &   0.26   &  0.23  &   2.79     &    0.01  &  20.18 &   0.3     \\
  45 &  PG 1309+355  &   0.18    &   --  &  0.00   &   0.19   &  0.18  &   5.47     &    0.12  &  34.95 &  11.5     \\
  46 &PG 1310$-$108  &   0.04    &   --  &  0.00   &   0.45   &  0.37  &   0.20     &    0.20  &   1.18 &   0.7     \\
  47 &  PG 1322+659  &   0.17    &   --  &  0.00   &   0.21   &  0.19  &   3.53     &    0.28  &  18.65 &  16.8     \\
  48 &  PG 1341+258  &   0.09    &  HDD? &  0.43   &   0.56   &  0.53  &   0.62     &    0.35  &   2.95 &   3.7     \\
  49 &  PG 1351+236  &   0.05    &   --  &  0.00   &   0.89   &  0.86  &   0.63     &    0.79  &   0.98 &   8.6     \\
  50 &  PG 1351+640  &   0.09    &   --  &  0.00   &   0.06   &  0.04  &   4.55     &    0.35  &  21.64 &  27.2     \\
  51 &  PG 1352+183  &   0.16    &   --  &  0.00   &   0.31   &  0.29  &   1.63     &    0.00  &  11.84 &   0.0     \\
  52 &  PG 1354+213  &   0.30    &   --  &  0.00   &   0.00   &  0.00  &   5.65     &    0.19  &  33.33 &  18.6     \\
  53 &  PG 1402+261  &   0.16    &   --  &  0.00   &   0.00   &  0.00  &   6.79     &    0.24  &  37.39 &  28.6     \\
  54 &  PG 1404+226  &   0.10    &  HDD? &  0.49   &   0.38   &  0.34  &   0.64     &    0.33  &   3.11 &   3.6     \\
  55 &  PG 1411+442  &   0.09    &   --  &  0.00   &   0.46   &  0.40  &   2.06     &    0.03  &  14.58 &   1.0     \\
  56 &  PG 1415+451  &   0.11    &   --  &  0.00   &   0.64   &  0.61  &   1.61     &    0.34  &   7.69 &   9.5     \\
  57 &PG 1416$-$129  &   0.13    &   --  &  0.00   &   0.36   &  0.32  &   0.89     &    0.16  &   5.46 &   2.4     \\
  58 &  PG 1425+267  &   0.37    &   --  &  0.00   &   0.14   &  0.14  &  13.00     &    0.21  &  75.05 &  46.5     \\
  59 &  PG 1426+015  &   0.09    &   --  &  0.00   &   0.46   &  0.40  &   2.86     &    0.18  &  17.18 &   8.7     \\
  60 &  PG 1427+480  &   0.22    &   --  &  0.00   &   0.19   &  0.18  &   3.94     &    0.32  &  19.40 &  22.0     \\
  61 &PG 1435$-$067  &   0.13    &   HDD &  0.67   &   0.03   &  0.03  &   1.14     &    0.02  &   8.07 &   0.5     \\
  62 &  PG 1440+356  &   0.08    &   --  &  0.00   &   0.58   &  0.53  &   2.92     &    0.50  &  10.64 &  25.2     \\
  63 &  PG 1444+407  &   0.27    &   --  &  0.00   &   0.01   &  0.01  &   9.36     &    0.07  &  63.43 &  11.2     \\
  64 &  PG 1448+273  &   0.06    &   --  &  0.00   &   0.63   &  0.57  &   0.74     &    0.31  &   3.75 &   3.9     \\
  65 &  PG 1501+106  &   0.04    &   --  &  0.00   &   0.44   &  0.37  &   0.67     &    0.30  &   3.41 &   3.5     \\
  66 &  PG 1512+370  &   0.37    &   --  &  0.00   &   0.05   &  0.05  &   9.43     &    0.03  &  66.95 &   4.1     \\
  67 &  PG 1519+226  &   0.14    &   --  &  0.00   &   0.07   &  0.06  &   2.37     &    0.15  &  14.67 &   6.2     \\
  68 &  PG 1534+580  &   0.03    &   --  &  0.00   &   0.55   &  0.48  &   0.22     &    0.24  &   1.24 &   0.9     \\
  69 &  PG 1535+547  &   0.04    &   --  &  0.00   &   0.37   &  0.29  &   0.18     &    0.16  &   1.10 &   0.5     \\
  70 &  PG 1543+489  &   0.40    &   --  &  0.00   &   0.00   &  0.00  &  41.26     &    0.45  & 165.92 & 319.3     \\
  71 &  PG 1545+210  &   0.27    &   WDD &  1.00   &   0.19   &  0.19  &   4.17     &    0.00  &  30.29 &   0.1     \\
  72 &  PG 1552+085  &   0.12    &  HDD? &  0.67   &   0.25   &  0.22  &   0.86     &    0.11  &   5.51 &   1.7     \\
  73 &  PG 1612+261  &   0.13    &   --  &  0.00   &   0.17   &  0.15  &   3.47     &    0.40  &  15.15 &  24.1     \\
  74 &  PG 1613+658  &   0.13    &   --  &  0.00   &   0.56   &  0.52  &   8.49     &    0.42  &  35.89 &  61.6     \\
  75 &  PG 1617+175  &   0.11    &   WDD &  1.00   &   0.00   &  0.00  &   1.19     &    0.05  &   8.28 &   1.0     \\
  76 &  PG 1626+554  &   0.13    &   HDD &  1.00   &   0.03   &  0.02  &   0.77     &    0.00  &   5.61 &   0.0     \\
  77 &  PG 1700+518  &   0.28    &   --  &  0.00   &   0.09   &  0.09  &  33.94     &    0.26  & 181.59 & 155.5     \\
  78 &  PG 1704+608  &   0.37    &   --  &  0.00   &   0.25   &  0.26  &  32.05     &    0.19  & 189.74 & 103.5     \\
  79 &  PG 2112+059  &   0.47    &   --  &  0.00   &   0.20   &  0.21  &  37.35     &    0.05  & 257.26 &  34.8     \\
  80 &  PG 2130+099  &   0.06    &   --  &  0.00   &   0.34   &  0.29  &   1.99     &    0.25  &  10.92 &   8.5     \\
  81 &  PG 2209+184  &   0.07    &  HDD? &  0.80   &   0.72   &  0.70  &   0.37     &    0.43  &   1.54 &   2.8     \\
  82 &  PG 2214+139  &   0.07    &  WDD  &  1.00   &   0.52   &  0.47  &   0.74     &    0.13  &   4.68 &   1.6     \\
  83 &  PG 2233+134  &   0.32    &   --  &  0.00   &   0.00   &  0.00  &  11.03     &    0.11  &  71.29 &  21.3     \\
  84 &  PG 2251+113  &   0.32    &  WDD  &  1.00   &   0.37   &  0.37  &   7.24     &    0.07  &  49.18 &   8.3     \\
  85 &  PG 2304+042  &   0.04    &   --  &  0.00   &   0.91   &  0.89  &   0.07     &    0.02  &   0.51 &   0.0     \\
  86 &  PG 2308+098  &   0.43    &  HDD &  0.88    &   0.00   &  0.00  &   9.22     &    0.00  &  67.13 &   0.0     %\\
\enddata
    \tablecomments{
	Column (2): object name; column (3): redshift; column (4): identification of
	HDD and WDD; column (5): the relative contribution of the HDD template
	(for HDD quasars) or the WDD template (for WDD quasars) to the AGN
	emission at rest-frame 1.25$~\mum$; column (6): the contribution fraction
	of the stellar template to the total quasar emission at rest-frame
	1.25$~\mum$; column (7): the host galaxy contribution to the total quasar
	light in observed $H$ band from SED decomposition; column (8): the total
	infrared luminosity (8-1000~$\mum$) of the object; column (9): the host
	SF template fraction for $L_{\rm IR}$; column (10): the bolometric
	luminosity of the AGN.  We adopted $L_{\rm AGN} = 5.29L_{\rm IR, AGN} =
	5.29L_{\rm IR}(1-f_{\rm host}) $ for all the quasars; column (11): the
	derived star formation rate based on $L_{\rm IR, host}=L_{\rm IR}
	f_{\rm host}$ following the \cite{Kennicutt1998} star-formation law.
    }
\end{deluxetable*}
\twocolumngrid

\begin{figure*}[htp]
    \centering
	\includegraphics[width=1.0\hsize]{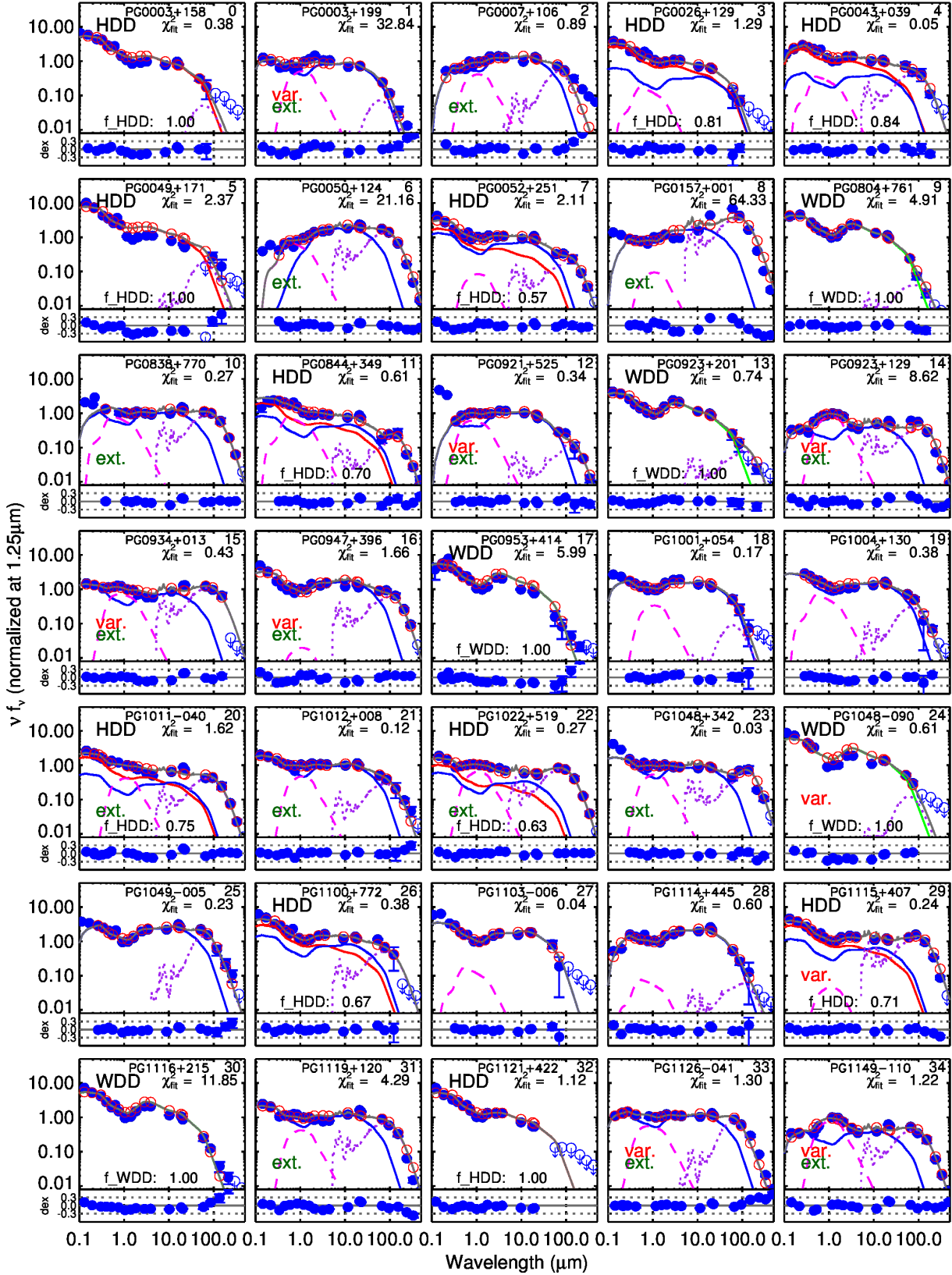} 
	\caption{ 
	    SED decomposition results (upper panels) and the residual plots
	    (lower panels) for the PG sample. Data points are shown as blue
	    dots (detection) and open circles with arrows (non-detection). The
	    AGN components of the SED model are shown as solid lines: blue --
	    the normal AGN template; green -- the WDD AGN template; red -- the
	    HDD AGN template.  The stellar template and the far-IR star
	    formation template of the host galaxy are shown as magenta dashed
	    lines and purple dotted lines, respectively.  The final composite
	    model SED is shown as the dark gray line, with the modeled points
	    (convolved with the corresponding photometry filters) as open red
	    circles. We also indicate the quasar type in the upper left corner
	    of each figure.  The fractional contribution of the HDD/WDD
	    template in the AGN component at $1.25~\mum$ is indicated as
	    $f_{\rm HDD}$, $f_{\rm WDD}$, respectively. We use var. (red) to
	    indicate quasars with infrared varability according to the {\it WISE}
	    light curves or literature near-IR data (see Section~\ref{sec:var})
	    and ext. (green) to indicate quasars that are not identified as
	    point sources in the 2MASS images (see Section~\ref{sec:pho-data}).
	}
	\label{fig:pg_sed_fitting}
\end{figure*}

\addtocounter{figure}{-1}
\begin{figure*}[htp]
    \centering
	\includegraphics[width=1.0\hsize]{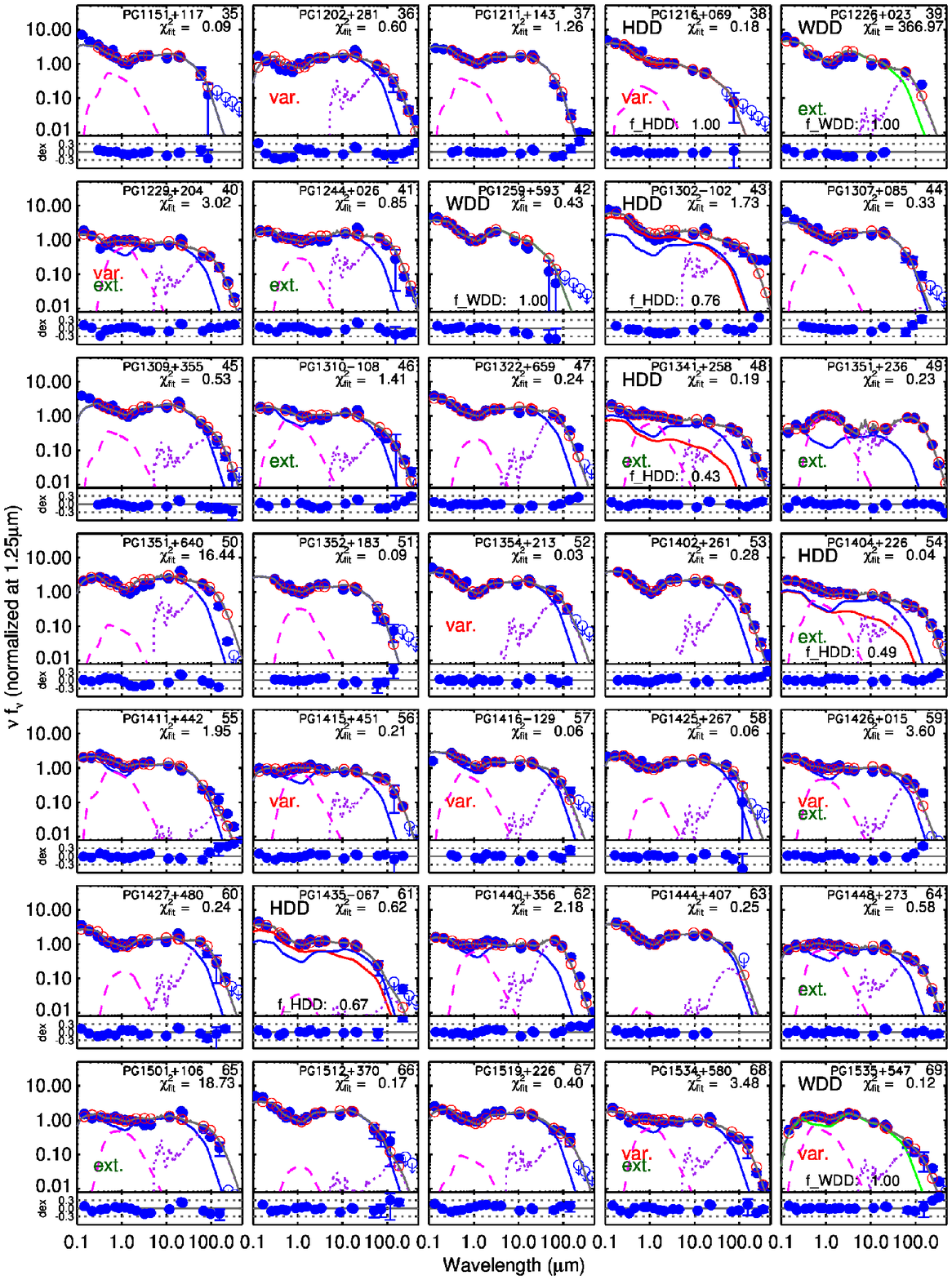} 
	\caption{ 
	    (continued.) SED decomposition results (upper panels) and the
	    residual plots (lower panels) for the PG sample.
	}
	\label{fig:pg_sed_fitting}
\end{figure*}

\addtocounter{figure}{-1}
\begin{figure*}[htp]
    \centering
	\includegraphics[width=1.0\hsize]{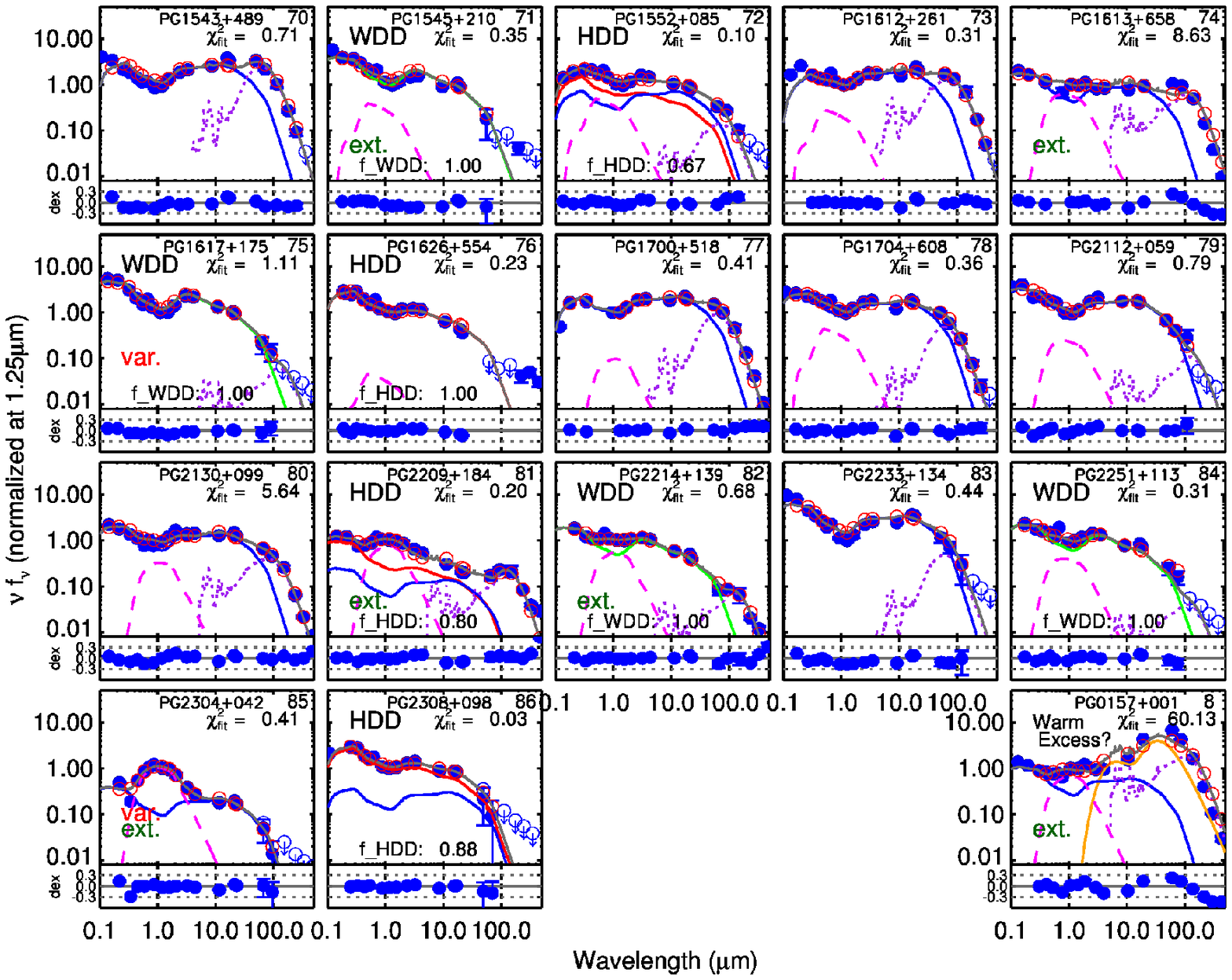} 
	\caption{ 
	    (continued.) SED decomposition results (upper panels) and the
	    residual plots (lower panels) for the PG sample. In the final SED
	    plot, we show the model fitting with a warm infrared component
	    (orange solid line) for PG 017+001.
	}
	\label{fig:pg_sed_fitting}
\end{figure*}

%\clearpage

\subsection{Quality of the SED Reproduction}\label{sec:quality}

The optical-to-far-IR SEDs of most PG quasars studied in this work have been
well reproduced by our model. The fitting residuals of the 0.5-100~$\mum$ SEDs
of all 87 quasars are less than 0.3 dex (see the residual panels in
Figure~\ref{fig:pg_sed_fitting}).  For nearly all cases where the far-IR
emission is strong, the fit selected a host galaxy template of luminosity
$10^{11}$-$10^{12}~L_\odot$ as expected. The fitted extinction levels are
small, usually zero. In 72 of the 87 cases the reduced $\chi^2$ is $<$ 3 and
only 11 have $\chi^2 > 5$.  The HDD template is required by 20 quasars
to best reproduce the SED.  Twelve quasars required the WDD template to be
included in the model. We comment on the fitted SEDs and discuss some notable
discrepancies between observations and model results below.

In the optical/UV bands, a few quasars show broadband excess emission if we
normalize the model SED to the observed near-IR data points (e.g., PG 0003+158
and PG 0049+171). This behavior can be explained by contamination from AGN
optical variability or the emission of young stellar populations in the AGN
host galaxies. As shown in \cite{Giveon1999}, the most variable quasars become
bluer when they are brighter. For the second possibility, even a relatively
small population of young stars can change the color of the host galaxy since
their emitting power is very strong. In the UV band (0.1-0.3~$\mum$), we see
some moderate flux excess above the AGN templates in a few cases (e.g., PG
0921+525), which might arise from UV emission by very young stars.

With the combination of the normal, WDD, and HDD AGN templates with a stellar
template, the near-IR to mid-IR SEDs of all quasars are reasonably reproduced.
The template selections for some quasars (e.g., PG 1001+054 and PG 2251+113)
are ambiguous, since some transition among these three populations of AGN are
possible.  Notably, PG 0157+001 has an exceptionally strong mid- to far-IR
excess (20-100$~\mum$).  PG 0923+129 and PG 1613+658  also show warm excess
with a moderate strength. We will discuss these quasars later.

The far-IR SEDs of most quasars are fitted with our model, suggesting that
their dust emissivities lie between normal galaxies ($\beta$=0.7-1,
\citealt{Rieke2009}) and typical AGNs ($\beta$=1.5 for large grains, as assumed
in \citealt{Xu2015a} and this work). However, we find 12 quasars that show
a much slower drop of the far-IR emission: PG 0003+199, PG 0007+106, PG
0947+396, PG 0953+414, PG 1116+215, PG 1211+143, PG 1302$-$102, PG 1341+258, PG
1411+442, PG 1426+015, PG 1535+547, and PG 2214+139. Only two of them -- PG
1211+143 and PG 1302$-$102 -- are radio-loud, in which case synchrotron emission
may have a substantial influence on the far-IR slope.  The reasons for the
different dust emissivity of the other quasars are unknown.

\subsection{Identification of the HDD and WDD Quasars}\label{sec:dd_ident}

Based on the SED decomposition with different AGN templates, we identify
13 confirmed HDD quasars and another 7 candidate HDD quasars. For the
10 confirmed HDD quasars with image decomposition results, the stellar
contribution in the observed $H$-band is less than 30\%.  The $\chi^2$ of the fit
with the HDD template is at least a factor of two better than that with the
normal AGN template only (see Table~\ref{tab:dd-quasar-list}). Additionally,
the HDD template model fits yield stellar near-IR contributions roughly
consistent with the image decomposition results. For another four HDD quasars
without image decomposition data, the host galaxy contamination is low.
According to our fits, the near-IR AGN luminosities of these quasars are mainly
contributed ($>85\%$) by the HDD AGN component, suggesting the dominant
HDD behavior. In addition, they have the same blue optical colors as the pure
AGN template. Besides PG 1011+772, all the confirmed HDD quasars have not been
resolved in 2MASS images, suggesting their weak host galaxy contamination in
the near-IR. For PG 1011+772, we confirmed its HDD character because of a
strong deficiency (with the peak discrepancy $>$ 0.3 dex) of the near- to
mid-IR emission if the normal AGN template is used.

\begin{figure*}[htp]
	\centering
	\includegraphics[width=0.55\hsize]{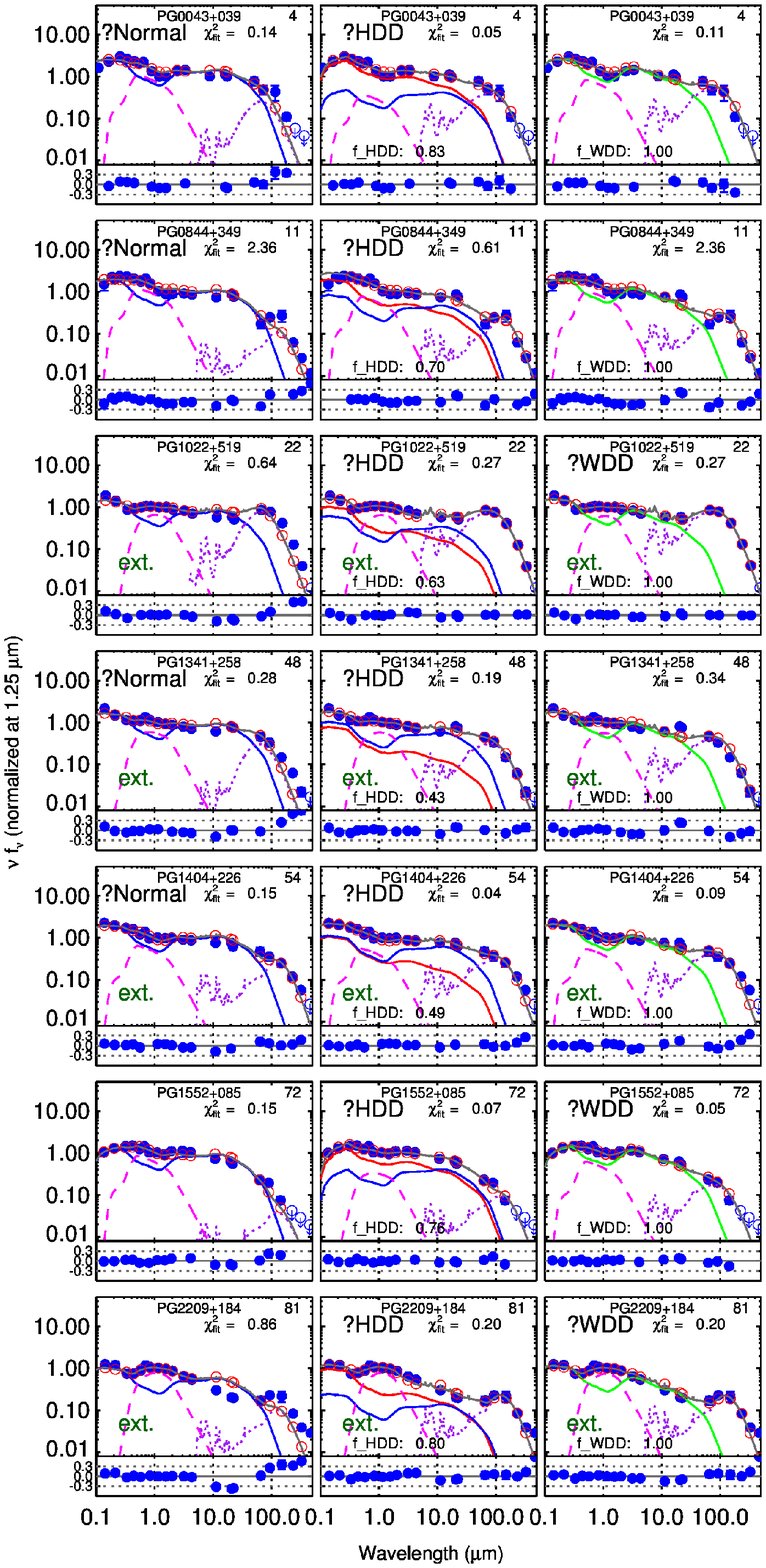} 
	\caption{ 
	    SED decomposition of quasars with ambiguous classifications with
	    normal AGN model (left), HDD AGN model (middle) and WDD model
	    (right). The meanings of the lines and the symbols are the same as
	    Figure~\ref{fig:pg_sed_fitting}.
}
	\label{fig:hdd_pick}
\end{figure*}

The identifications of hot dust deficiency are less secure for the HDD
candidate quasars. We show the comparison of fitting results with various AGN
templates in Figure~\ref{fig:hdd_pick}. For PG 0043+039, PG 0844+349, PG
1341+258, and PG 1404+226, the mid- to far-IR SEDs are better produced by the
introduction of the HDD AGN template with reduced $\chi^2$ values. However, the
host galaxy contribution in the near-IR seems to be strong ($\sim$30-50\%), and
these four quasars present relatively flat slopes in the optical. For PG
1022+519, PG 1552+085, and PG 2209+184, their dust deficiency is revealed by the
overestimated mid-IR emission from the normal AGN template, however, whether
they should be picked as HDD or WDD quasars is a question. Consequently, we
suggest the HDD quasar fraction in this PG sample is 15\%-23\%.

The number of identified WDD quasars in the PG sample is 15 with 12
confirmed cases and three candidates.  All confirmed WDD quasars show weak
mid-infrared emission and a clear hot dust emission peak at $\sim$3~$\mum$.  As
stated below, there are three HDD candidates that may be WDD candidates as
well. Given these numbers, we estimate the WDD quasar fraction in the PG sample
$\sim$14-17\%.

\onecolumngrid
\LongTables
\begin{deluxetable*}{clccccc}
    \tabletypesize{\scriptsize}
    \tablewidth{1.0\hsize}
    \tablecolumns{7}
    \tablecaption{Dust-deficient Quasars in $z<0.5$ PG sample\label{tab:dd-quasar-list}
    }
    \tablehead{
	\colhead{ID}  &\colhead{Source} & \colhead{$z$}  & \colhead{$f_{\rm nucleus,~H}$} & \colhead{Reference} & \colhead{Extended?} & \colhead{$\chi^2_{\tiny \text{0.5-30}\mum}$} \\
	\colhead{(1)} & \colhead{(2)} & \colhead{(3)} & \colhead{(4)} & \colhead{(5)}  & \colhead{(6)}  & \colhead{(7)} 
    }
    \startdata
    \multicolumn{6}{c}{HDD Quasars} \\
   0 & 0003+158         &      0.45  &     --  (1.00)   &  0    & N   & 1.8 (13.2)   \\
   3 & 0026+129*        &      0.14  &     0.80  (0.64) &  1    & N   & 1.5 (22.2)   \\
   5 &  0049+171*       &      0.06  &     --  (1.00)   &  --   & N   & 9.4 (53.1)   \\
   7 &  0052+251        &      0.16  &     --  (0.91)   &  --   & N   & 2.7 (11.3)   \\
   20 & 1011$-$040      &      0.06  &     --  (0.61)   &  --   & Y   & 10.7 (34.1)   \\
   26 & 1100+772        &      0.31  &     0.78  (1.00) &  4    & N   & 0.3 (4.5)   \\
   29 & 1115+407        &      0.15  &     --  (0.96)   &  --   & N   & 3.1 (12.2)   \\
   32 & 1121+422*       &      0.23  &     0.93  (1.00) &  2    & N   & 0.4 (11.1)   \\
   38 & 1216+069        &      0.33  &     0.91  (0.95) &  4    & N   & 1.2 (15.7)   \\
   43 & 1302$-$102      &      0.29  &     0.81  (1.00) &  1    & N   & 10.9 (20.3)   \\
   61 & 1435$-$067      &      0.13  &     0.73  (0.97) &  1    & N   & 0.4 (6.6)   \\
   76 & 1626+544*       &      0.13  &     0.72  (0.98) &  1    & N   & 0.5 (10.8)   \\
   86 & 2308+098        &      0.43  &     --  (0.87)   &  --   & N   & 0.5 (5.2)   \\
   4  & 0043+039?       &      0.38  &     --  (0.92)   &  --   & N   & 0.32 (0.62)        \\
   11 & 0844+349?       &      0.06  &     0.40 (0.40)  &  1    & N   & 7.00 (12.6)       \\
   22 & 1022+519?       &      0.05  &     --  (0.32)   &  --   & Y   & 0.7 (2.8)   \\
   48 & 1341+258?       &      0.09  &     --  (0.47)   &  --   & Y   & 0.6 (1.2)   \\
   54 & 1404+226?       &      0.10  &     --  (0.66)   &  --   & Y   & 0.4 (1.2)   \\
   72 & 1552+085?       &      0.12  &     --  (0.29)   & --    & N   & 0.3 (1.4) \\
   81 & 2209+184?       &      0.07  &     --  (0.75)   &  --   & Y   & 1.5 (7.0)   \\
    \multicolumn{6}{c}{WDD Quasars} \\ 
   9  & 0804+761*       &      0.11  &     0.90  (1.00) &  3    & N   & 18.7 (152.8)   \\
   13 & 0923+201*       &      0.19  &     0.77  (1.00) &  1    & N   & 2.5 (16.9)   \\ 
   17 & 0953+414        &      0.24  &     --  (1.00)   &  --   & N   & 16.3 (18.5)   \\
   24 & 1048$-$090      &      0.34  &     --  (1.00)   &  --   & N   & 2.7 (5.2)   \\
   30 & 1116+215*       &      0.18  &     0.95  (1.00) &  1    & N   & 46.0 (171.7)   \\
   39 & 1226+023        &      0.16  &     --  (1.00)   &  --   & Y   & 947.4 (1740.6)   \\
   42 & 1259+593        &      0.47  &     0.94  (1.00) &  4    & N   & 1.4 (8.4)   \\
   69 & 1535+547        &      0.04  &     --  (0.37)   &  --   & Y   & 2.1 (5.2)\\
   71 & 1545+210        &      0.27  &     --  (0.81)   &  --   & N   & 3.5 (9.6)   \\
   75 & 1617+175*       &      0.11  &     0.89  (1.00) &  1    & N   & 2.6 (63.4)   \\
   82 & 2214+139        &      0.07  &     0.47  (0.63) &  1    & Y   & 2.1 (37.6   \\
   84 & 2251+113        &      0.32  &     0.94  (0.66) &  1    & N   & 1.0 (3.6)  \\ 
   22 & 1022+519?       &      0.05  &     --  (0.61)   &  --   & Y   & 0.6 (2.8)\\
   72 & 1552+085?       &      0.12  &     --  (0.41)   &  --   & N   & 0.4 (1.4)\\
   81 & 2209+184?       &      0.07  &     --  (0.72)   &  --   & Y   & 1.2 (7.0) 
    \enddata
    \tablecomments{
	Column (1): object id; column (2): object name (``PG" is omitted). We
	denote the ambiguous cases with ``?'', and the quasars used to derive the
	template with `*'. ; column (3): redshift; column (4): the contribution of
	PSF component in the observed $H$ band from the HST image decomposition
	with the same quantity based on the SED decomposition in the brackets;
	column (5): references for HST image decomposition results:
	1-\cite{Veilleux2009}; 2-\cite{McLeod2001}; 3-\cite{Guyon2006};
	4-\cite{Shang2011}; column (6): whether the object is picked out as an
	extended source by 2MASS; column  (7): the $\chi^2$ values for data
	points in the rest-frame 0.5-30~$\mum$ range, the numbers outside and inside
	the brackets corresponding to the fitting with the dust-deficient
	template and the fitting with the classical AGN template only.
    }
\end{deluxetable*}
\twocolumngrid

\subsection{Host Galaxy Contamination}

To test if the three SED templates introduced in Section~\ref{sec:templates}
are convincing representatives for the infrared emission coming from the AGN
component in these PG quasars, we compare the strength of the AGN host galaxy
emission deduced from the SED decomposition with that from other independent
methods.

\subsubsection{Near-IR Stellar Emission}\label{sec:host-stellar}

Detailed morphology decomposition of quasars can be applied on deep and
high-resolution image observations in the near-IR \citep[e.g.,][]{McLeod2001,
Guyon2006, Veilleux2009}. By comparing the magnitudes of the AGN and the host
components, we can calculate the host galaxy fraction at given bands. A similar
host galaxy fraction can also be derived from the SED decomposition model. We
can compare the host galaxy fractions from these two methods to check the
validity of our model. 

\begin{figure}[htp]
	\centering
	\includegraphics[width=1.0\hsize]{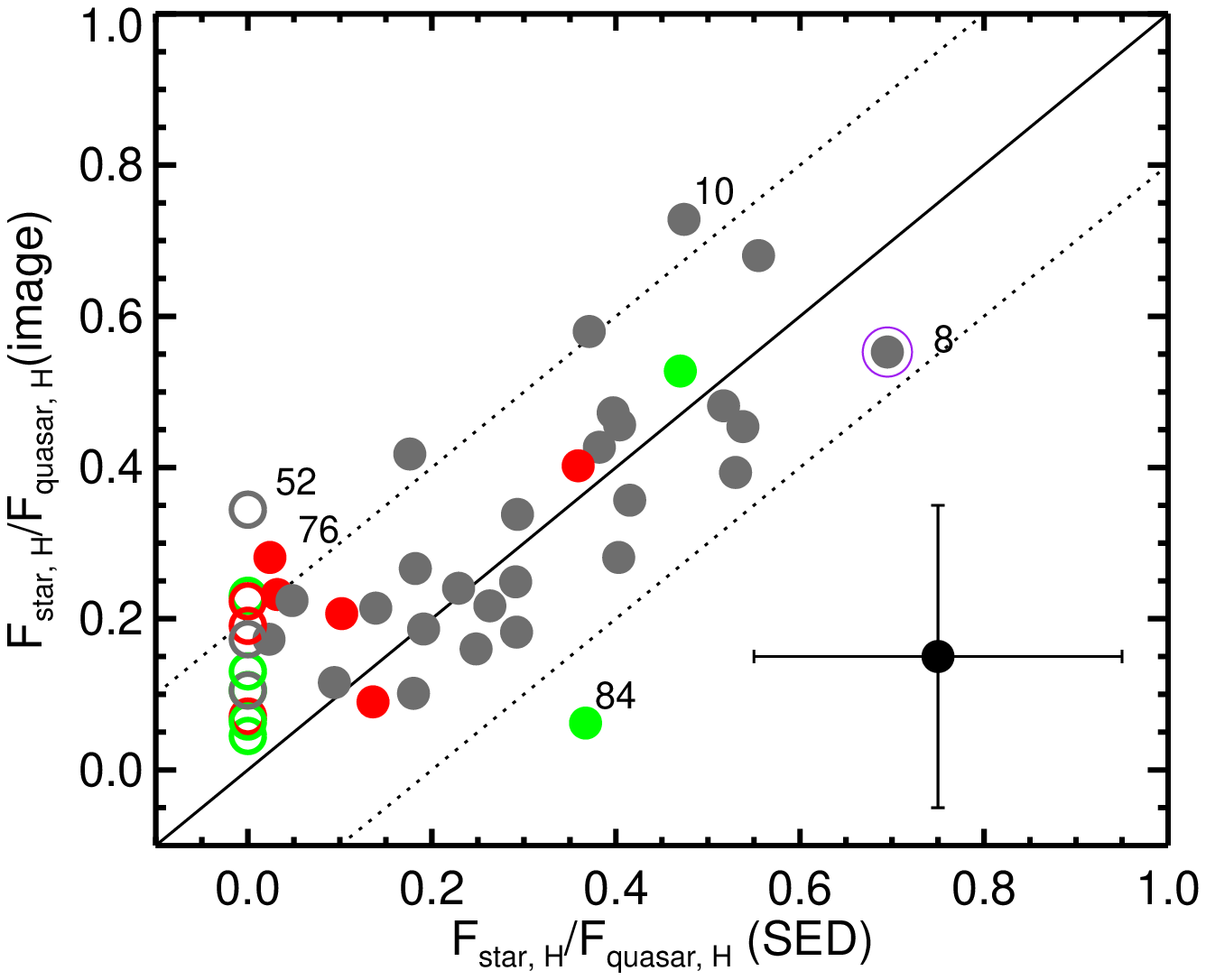} 
	\caption{ 
	    Comparison of the host galaxy stellar emission to the total quasar
	    light in observed $H$ band, $F_\text{star, H}/F_\text{quasar, H}$,
	    based on image decomposition and that from SED decomposition. We
	    show the 1:1 relation and $\pm0.2$ value deviations as solid and
	    dotted black lines.  Normal, WDD, and HDD quasars are indicated by
	    gray, green, and red colors. Empty circles indicate that the SED
	    decomposition yielded a near zero host contribution in the
	    corresponding bands. PG 0157+001 (\#8, as indicated with a purple
	    circle) has a consistent host galaxy contribution from the SED
	    decomposition and HST results if a warm excess component is
	    introduced (see Section~\ref{sec:dd_color}).
}
	\label{fig:nir_star}
\end{figure}

Based on the SED decomposition, we computed the host galaxy contribution to the
total quasar emission in the observed frame $H$-band, $F_\text{star,
H}/F_\text{quasar, H}$.  Figure~\ref{fig:nir_star} shows the comparison of the
literature results on the host light fraction of 44 PG quasars retrieved from
HST/ground-AO image decomposition \citep{McLeod2001, Hamilton2002,Marble2003,
Guyon2006,Hamilton2008, Veilleux2009}\footnote{See \cite{Zhang2016} for a
summary of the image decomposition results of PG quasars in the literature.}
with our results based on the SED model.  For the majority of these quasars,
the host galaxy contributions derived from the SED decomposition and image
decomposition are consistent, with an offset of less than 20\%. 

However, it seems the correlation disappears in the bottom-left corner of
Figure~\ref{fig:nir_star}: when the relative contribution of the near-IR
stellar light is small (as indicated by the low $F_\text{star,
H}/F_\text{quasar, H}$ values from image decomposition), our SED model
underestimated -- or even failed to identify -- the host galaxy emission
compared with the HST image decomposition.  This is a known systematic bias of
such SED models (see Section 5.3.1 in \citealt{Xu2015a}). Meanwhile, we note
that the image decomposition technique suffers a number of systematics. For
example, \cite{Kim2008} found that the flux of the host galaxy can be easily
overestimated from image decompositions when $F_\text{star, H}/F_\text{quasar,
H}\lesssim0.5$ due to realistic PSF mismatches. It is likely that the
systematics and uncertainties present in both AGN-host decomposition methods
contribute to the discrepancies. However, the cases with weak host stellar
emission are also the ones where the identification of HDD behavior is least
likely affected.

Given the fact that our SED decomposition gives consistent results on the
near-IR host stellar contamination with that based on the 2D image
decompositions, the stellar emission contribution is not likely to lead to any
incorrect identifications of HDD or WDD quasars.

\begin{figure}[htp]
	\centering
	\includegraphics[width=1.0\hsize]{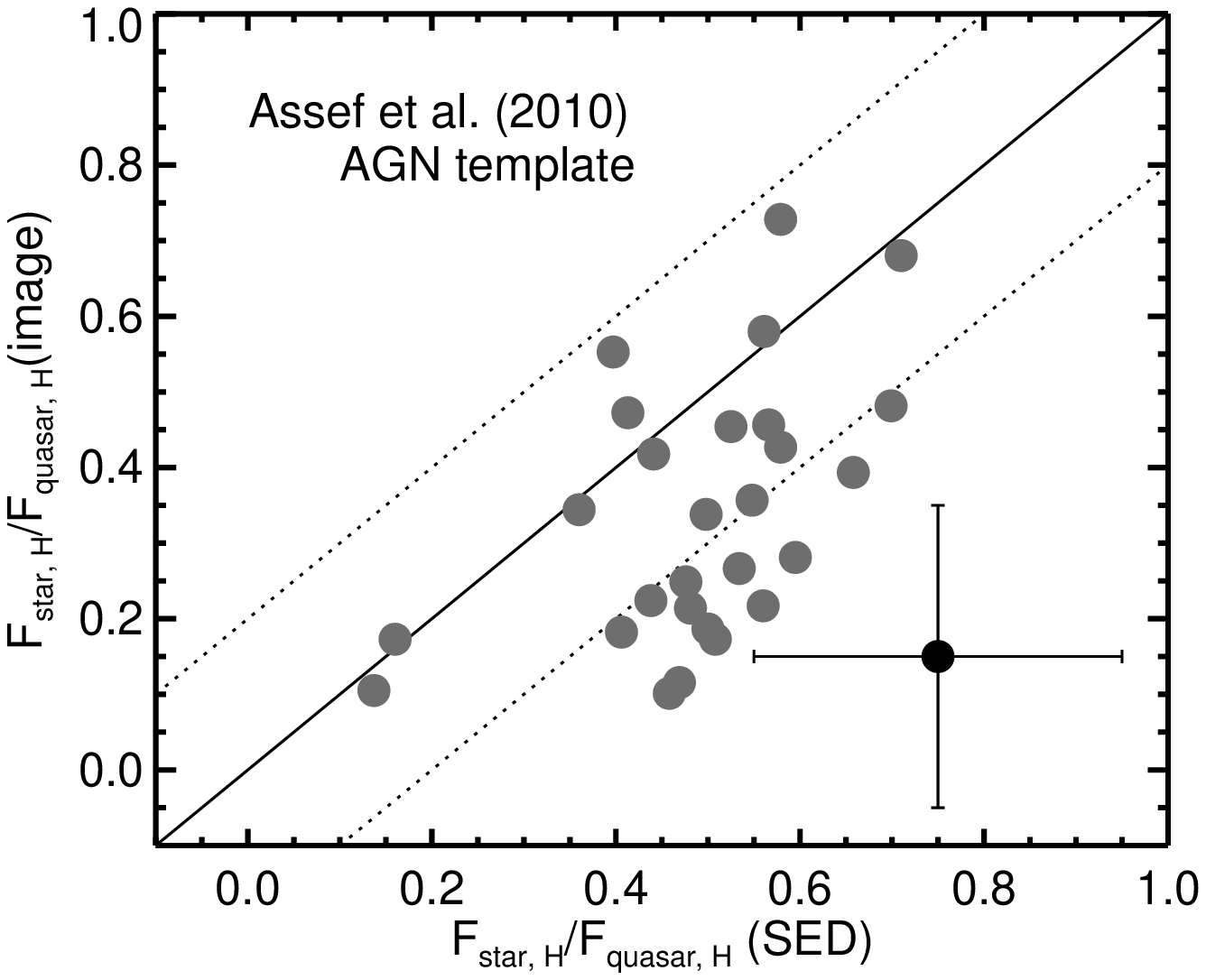} 
	\caption{ 
	    Comparison of imaging and SED deconvolution results for normal
	    quasars and using the Assef et al. (2010) AGN template. The symbols
	    and axes have the same meanings as in Figure~\ref{fig:nir_star}. 
	}
	\label{fig:assef_star}
\end{figure}

With the $H$-band image decomposition results for 28 normal PG quasars, we can
also test the validity of the \cite{Assef2010} AGN template to represent the
AGN emission. In Figure~\ref{fig:assef_star}, we compare the relative
observed-frame $H$-band stellar emission strength derived from the SED model with
the \cite{Assef2010} AGN template with the image decomposition results. It is
clear that the \cite{Assef2010} template model overestimates the stellar
contamination for these quasars, suggesting that the much stronger 1~$\mum$ dip of
the \cite{Assef2010} is unphysical. In other words, the \cite{Elvis1994}-like
templates are preferred over the \cite{Assef2010} template to represent the AGN
intrinsic near-IR emission of normal quasars.

\subsubsection{Mid- to Far-IR Dust Emission}\label{sec:host-dust-cal}

\begin{figure}[htp]
	\centering
	\includegraphics[width=1.0\hsize]{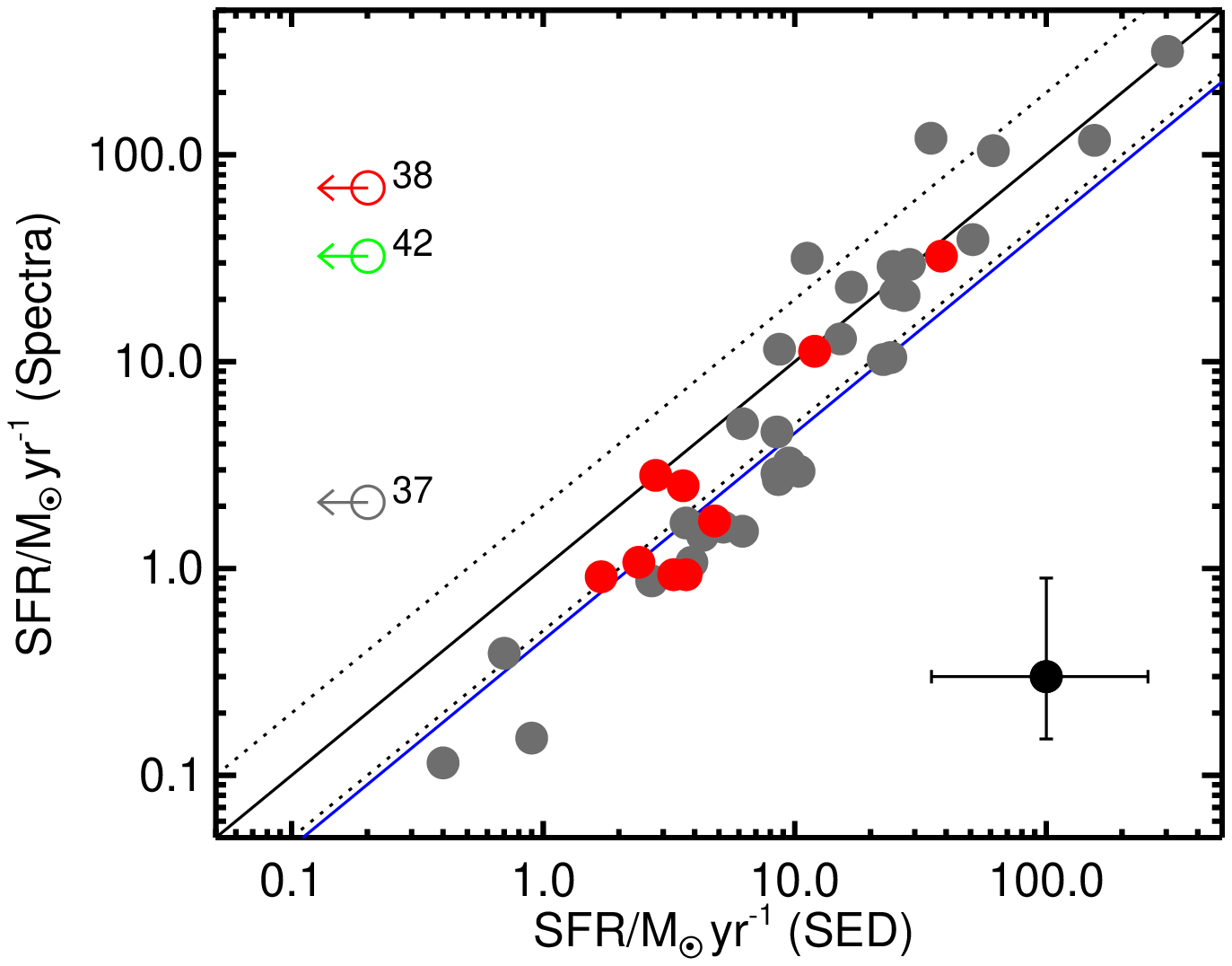} 
	\caption{ 
	    Comparison of the star formation rates derived through our fitting
	    of the FIR SEDs with those in \cite{Shi2014} from the 11.3$~\mum$
	    aromatic feature strength. Normal, WDD, HDD quasars are shown as
	    gray, green, and red dots, respectively. Open circles with a leftside
	    arrow indicate that the quasars do not have sufficiently strong host
	    galaxy contributions in the far-IR to derive a meaningful SFR (we
	    arbitrarily set their SFRs = 0.2 $M_\odot {\rm yr}^{-1}$ to plot these
	    quasars in the figure). We show the 1:1 relation and its 0.3 dex
	    deviations as solid and dotted black lines. The blue line denotes
	    that the SFRs from SED decomposition for normal quasars should be
	    increased by 2.2 times if the WDD template is adopted to represent
	    the AGN intrinsic emission, resulting in a 0.34 dex offset from the
	    1:1 relation.
	}
	\label{fig:sfr}
\end{figure}

The bolometric infrared luminosity of a galaxy is believed to be dominated by
the dust thermal emission heated by hot, young stars, providing a measure of
the SFR \citep[e.g.,][]{Kennicutt1998}. To calibrate the
infrared emission of the quasar host galaxies, we can compare the SFRs based on
the IR luminosities of the host galaxies from the SED decompositions with the
SFRs based on another independent method.  For quasars, the $11.3~\mum$
aromatic feature is perhaps the only plausible spectroscopic star formation
indicator where the AGN influence is negligible \citep{Diamond-Stanic2010,
Esquej2014,Alonso-Herrero2014}.  Assuming the \cite{Kennicutt1998} law, we
compare the IR-derived SFRs with the SFR measurements based on the $11.3~\mum$
aromatic feature strength by \cite{Shi2014} in Figure~\ref{fig:sfr}. The SFR
results for the majority of PG quasars are consistent\footnote{Among the three
outlier quasars, PG 1216+069 (\#38) and PG 2308+098 (\#86) have low
signal-to-noise spectra, which make the spectral measurements uncertain. PG
1259+593 (\#42) shows a very broad and prominent silicate feature. The two
Gaussian functions used to fit the silicate profile are widely separated,
producing a local dip at $\sim11~\mum$ (see the online Figure 1 of
\citealt{Shi2014}) that is unphysical. As a result, for PG 1258+593, the
equivalent width of the 11.3$~\mum$ aromatic feature above such a silicate
feature continuum could be overestimated.}. There is an indication that the
SFRs based on the aromatic feature are lower than the IR-derived SFRs when the
SFRs are low (e.g., SFR$~\lesssim10~M_\odot {\rm yr}^{-1}$).  This could be due to
(1) the difficulty of measuring the aromatic features when the AGN dominates
the {\it Spitzer}/IRS infrared spectra; (2) that the $11.3~\mum$ aromatic
feature strength is reduced by the prominent AGN emission in luminous quasars;
(3) the possible dust far-IR emission heated by old stars
\citep[e.g.,][]{Devereux1989,Popescu2002} in quasar host galaxies.
Nevertheless, for the science goals of this paper, such deviations are only a
secondary effect. At high SFRs, where the fits are best-constrained, there is
no evidence for such a shift in the calibration. Thus, we suggest the infrared
emission of the host galaxies is properly retrieved from our model.

The consistent SFR results with both of the methods in Figure~\ref{fig:sfr}
show that both the WDD and normal AGN templates are being applied
appropriately. For the same near-IR luminosities, the WDD AGN template produces
$\sim45\%$ of the total infrared luminosity (8-1000$~\mum$) of the normal AGN
template. As there is no necessity to shift the positions of the normal quasars
to match the 1:1 line in Figure~\ref{fig:sfr}, we can conclude that the WDD
template is not the best choice to represent the intrinsic IR emission of
normal quasars.

\section{Discussion}\label{sec:discuss}

\subsection{The Diversity of Infrared Colors of PG Quasars}\label{sec:dd_color}

\begin{figure*}[htp]
	\centering
	\includegraphics[width=1.0\hsize]{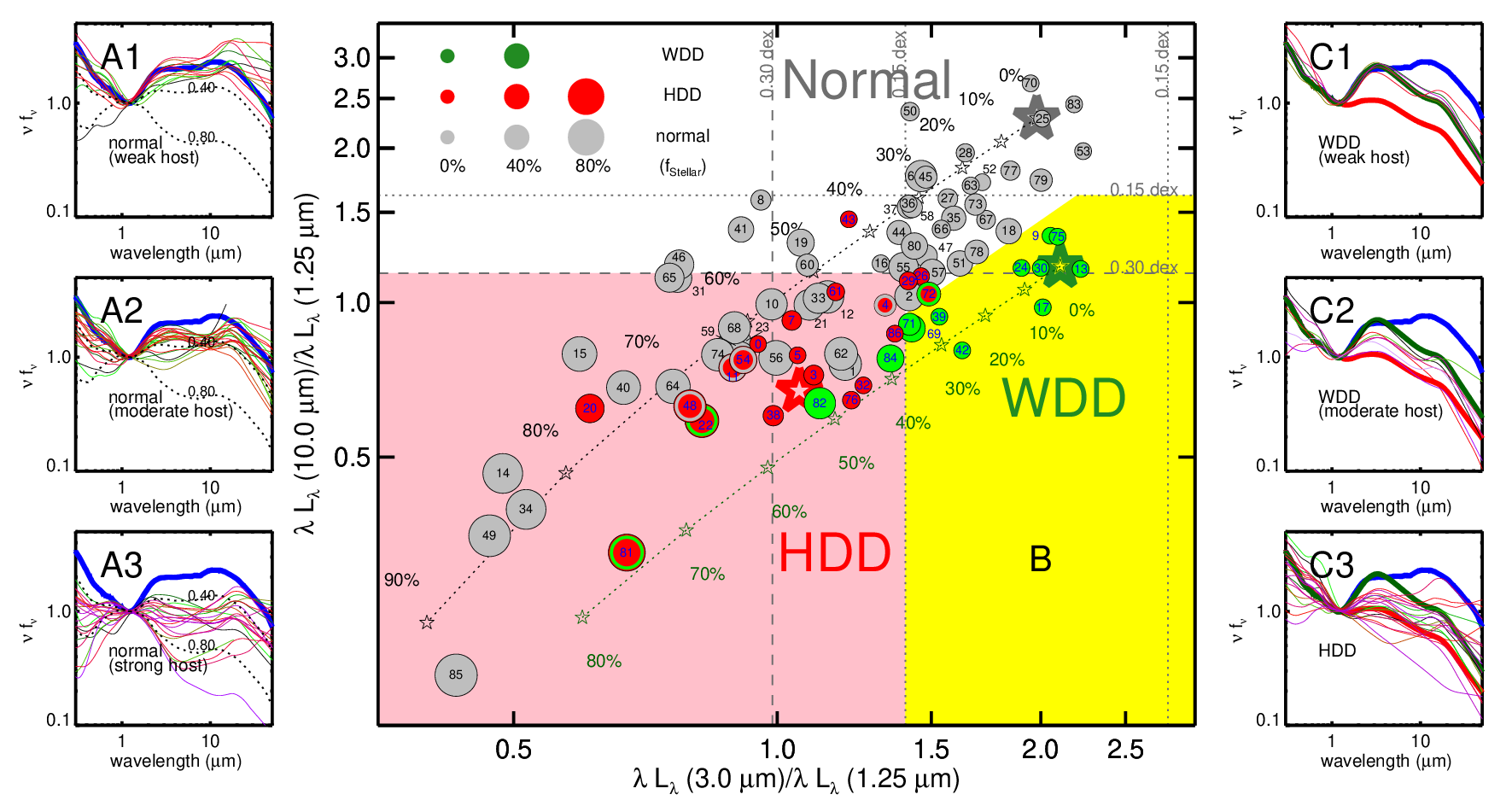} 
	\caption{ 
	    Infrared color distribution of the 87 PG quasars (Panel B) and
	    various SEDs of normal quasars (Panels A1, A2, A3) and
	    dust-deficient quasars (Panels C1, C2, C3). In panel B, the shaded
	    regions denote quasars with different infrared properties. The
	    colors of the templates for normal, WDD, and HDD quasars are
	    plotted as black, green, and red big five-pointed stars. We also
	    show the locations of normal and WDD quasars with different levels
	    of host galaxy contributions as small five-pointed stars connected
	    with dotted lines.  Individual quasars are shown as dots with the
	    sizes determined by the stellar near-IR contribution from the SED
	    fittings. We denote the ambiguous cases with thick circles inside
	    the dots. Panels A1, A2, A3 show the SEDs of individual normal
	    quasars with $10.0/1.25$ colors deviating from the Elvis template
	    (red thick line) $<0.15$ dex, 0.15-0.30 dex, $>0.30$ dex,
	    respectively. Panels C1 and C2 present the SEDs of all WDD quasars
	    (green dots in panel B). Panel C3 shows the SEDs of all HDD quasars
	    (red dots in panel B).  
	    }
	\label{fig:pg_color_dis}
\end{figure*}

With the three AGN templates (Section~\ref{sec:templates}) and our SED model
(Section~\ref{sec:pg_sed}), we can explain the diversity of the near- to mid-IR
colors of the PG quasars. For a normal AGN, the infrared SED is characterized
by a quick upturn from the $1.25~\mum$ inflection and a broad mid-infrared
plateau in $\lambda$-$\nu f_\nu$ space from $\sim$3-20~$\mum$. The relative
strengths of these SED features can be found by normalizing the corresponding
peak flux by the flux at the $1.25~\mum$ inflection. As shown in
Figure~\ref{fig:agn_template}, a $\gtrsim$ 0.3 dex deviation from the classical AGN
template starts at $\sim3.0~\mum$ for the HDD templates and at $\sim10.0~\mum$
for the WDD template. As a result, we propose to use luminosities at 3.0
and $10~\mum$ to reflect the relative strengths of the hot and warm dust.
For the Elvis template, we have
\begin{itemize}
    \item intrinsic hot dust peak:    
	\begin{equation*}
	    \lambda f_{\rm normal, \lambda} [3.0\mum/1.25\mum]= 1.98~~;
	\end{equation*}
    \item intrinsic warm dust peak: 
	\begin{equation*}
	    \lambda f_{\rm normal, \lambda} [10.0\mum/1.25\mum]= 2.29~~.
	\end{equation*}
\end{itemize}
We derive the observed quasar SED continuum by logarithmic interpolation on the
UV-to-IR photometry and calculate the corresponding color $\lambda f_\lambda
[3.0\mum/1.25\mum]$, and $\lambda f_\lambda [10.0\mum/1.25\mum]$. In
Figure~\ref{fig:pg_color_dis}, we present the color distribution of all 87 PG
quasars as well as their individual continuum SEDs.

The most obvious feature of panel B in Figure~\ref{fig:pg_color_dis} is that
most quasars are distributed along the diagonal direction with changes of
both apparent hot and warm dust emission. Such a diagonal distribution is mainly
caused by the increasing near-IR contamination from the host stellar emission.
For a normal quasar, as the host galaxy contribution increases, the position in
Figure~\ref{fig:pg_color_dis} is shifted to the lower left corner. To
demonstrate this, we combine the normal (Elvis) AGN template with a single
old stellar population template, and calculate the colors of the composite SED
as a function of the galaxy contribution in the near-IR (shown as small
five-pointed stars connected with the dotted gray line in
Figure~\ref{fig:pg_color_dis}). As expected, all normal quasars are randomly
distributed along this normal quasar color line, and their galaxy contributions
in the near-IR from the individual SED decompositions are roughly consistent
with the model line. From panels A1, A2, to A3, as the apparent warm dust
emission decreases, we see a systematic decrease of the optical slope for the
normal quasars.  The observed SEDs in these panels are also roughly consistent
with the mock quasar SEDs using a combination of the normal AGN template and
old stellar template with different relative strengths.

As Figure~\ref{fig:pg_color_dis} shows, the colors of the HDD PG quasars
typically deviate from the Elvis template by more than $-0.15$ dex for $\lambda
f_{\rm normal, \lambda} [3.0\mum/1.25\mum]$ and $-0.3$ dex for $\lambda f_{\rm
normal, \lambda} [10.0\mum/1.25\mum]$ (see Figure~\ref{fig:agn_template}). Thus,
we denote an HDD region in Figure~\ref{fig:pg_color_dis} as 
\begin{itemize}
    \item  apparent hot dust strength:
	\begin{equation*}
	    \lambda f_\lambda [3.0\mum/1.25\mum]< 1.40~\text{($-$0.15 dex)}~,
	\end{equation*}
    \item  apparent warm dust strength:
	\begin{equation*}
	    \lambda f_\lambda [10.0\mum/1.25\mum]< 1.14~\text{($-$0.30 dex)}~.
	\end{equation*}
\end{itemize}
A large number of normal quasars are also located in the HDD region.
The cause of this contamination can be identified from Panels C3 and A3.  
HDD quasars have a blue UV-optical continuum similar to the AGN template, 
while normal quasars in the same region show strong galaxy contamination in the
optical and near-IR.

In the lower-right region of Figure~\ref{fig:pg_color_dis} are the WDD quasars,
with the $\lambda f_{\rm normal, \lambda} [3.0\mum/1.25\mum]$ color within
$-0.15$ dex of the classical AGN, but $\lambda f_{\rm normal, \lambda}
[10.0\mum/1.25\mum]$ deviation greater than 0.15 dex. We can define a WDD
region, where
\begin{itemize}
    \item  apparent hot dust strength:
	\begin{equation*}
	    \lambda f_\lambda [3.0\mum/1.25\mum]> 1.40~\text{($-$0.15 dex)} ~,
	\end{equation*}
    \item  apparent warm dust strength:
	\begin{equation*}
	    \lambda f_\lambda [10.0\mum/1.25\mum]< 1.62~\text{($-$0.15 dex)} ~,
    \end{equation*}
    \item   diagonal direction cut:
	\begin{align*}
	    \begin{split}
	    \log_{10}(\lambda f_\lambda & [10.0\mum/1.25\mum])> \\
      &1.98\log_{10}(\lambda f_\lambda [3.0\mum/1.25\mum]) -0.39~.
	    \end{split}
	\end{align*}
\end{itemize}
The third cut along the diagonal direction is based on the mock SED mixing
galaxy and AGN templates. The contamination of normal quasars in this WDD
region is quite low. We also see an increasing host galaxy contribution along
the diagonal direction in the central panel, consistent with the prediction
from the mock SEDs composed of the WDD template and a near-IR stellar template.
The increasing host contribution among WDD quasars can also be seen in panel C1
($\lambda f_\lambda [10.0\mum/1.25\mum]<0.30$ dex of the Elvis template) and
panel C2 ($\lambda f_\lambda [10.0\mum/1.25\mum]>0.30$ dex of the Elvis
template), for which the decrease of apparent warm dust emission is caused by
the stronger host galaxy contamination at $1.25~\mum$. This behavior emphasizes
the risk of host galaxy contamination in any purely photometric means to
identify HDD quasars. However, it appears that such simple methods may work
reasonably well for WDD objects, though there is still a mixture of normal
quasars in their color space in the figure.

Broad IR spectral features may also influence the infrared colors of quasars.
On average, type 1 quasars have moderate silicate emission at $\sim10~\mum$
with strength\footnote{
The silicate strength is defined as 
\begin{equation*}
S_{10} = \ln\left(\frac{I_{\lambda\ast, {\rm obs}}}{I_{\lambda\ast, {\rm cont}}}\right),
\end{equation*}
where $\lambda\ast$ is the wavelength of the 10$~\mum$ silicate feature peak,
$I_{\lambda\ast, {\rm obs}}$ and $I_{\lambda\ast, {\rm cont}}$ are the
corresponding observed and continuum intensities, respectively.  }$\sim$0.20
\citep[e.g.,][]{Hao2007}, which corresponds to 0.08 dex of the local continuum.
This is much smaller compared to the $0.3$ dex difference we picked to
separate the HDD quasars. However, for individual quasars, the 10~$\mum$
silicate feature strength can be very large, in which cases its influence
should be considered.

A number of normal quasars above the normal quasar color line present a smooth
SED gradually peaked at the mid-infrared, e.g., PG 0157+001, PG 0934+013, PG
1119+120, PG 1244+026, PG 1310$-$108, PG 1351+640, PG 1501+106. Our best fitted
model underpredicts the emission at 20-100$~\mum$, suggesting a warm excess.
Similar behavior has also been found for type-1 AGN at $z\sim$0.3-3
\citep{Xu2015a, Kirkpatrick2015} as well as the low-luminosity AGNs in
nearby Seyfert galaxies \citep[e.g.,][]{Ho1999, Ho2008, Prieto2010}. One
possibility to generate this component is a very compact ($\lesssim$1 pc)
starbursting disk in the nucleus of the galaxy \citep{Thompson2005,
Ballantyne2008}. In most cases, our test fitting with an additional warm
component following \cite{Xu2015a} hardly improved the $\chi^2$, suggesting
that even if the warm-excesses are present, their contribution should be
moderate.  Another possibility is the additional IR-processed AGN emission
by dust either in the galactic interstellar medium
\citep[e.g.,][]{Schneider2015, Roebuck2016} or in the polar region of the
nucleus \citep[e.g.,][]{Raban2009,Honig2012, Honig2013, Tristram2014,
Asmus2016,Lopez-Gonzaga2016}. For PG 0157+001, it could also be its strong
shocks \citep[e.g.,][]{Leipski2006}, which may break up the surrounding dust
into much smaller grains and boost the mid-IR emission.

\subsection{Characteristics of the Dust-deficient PG Quasars}\label{sec:char}

\begin{figure}[htp]
	\centering
	\includegraphics[width=1.0\hsize]{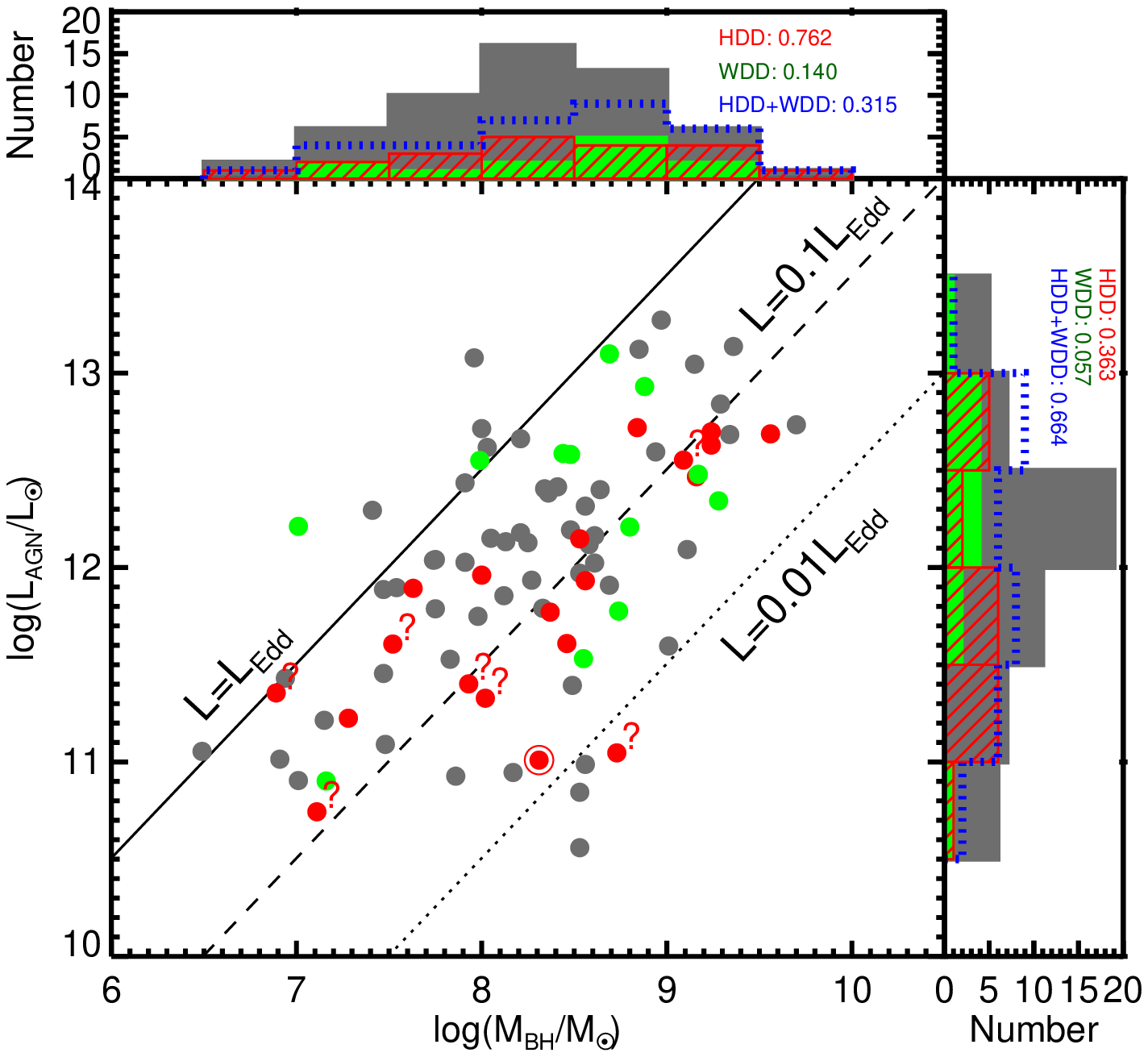} 
	\caption{ 
	    Distributions of AGN luminosities and black hole masses of 87
	    PG quasars color encoded by their infrared properties: gray for
	    normal quasars, green for WDD quasars, and red for HDD quasars. We
	    indicate the ambiguous HDD quasars with question marks and the most
	    HDD quasar, PG 0049+171, with a red circle. The histograms of the
	    properties of normal (gray shaded), WDD (green shaded), HDD (red
	    crossed), and WDD+HDD (blue dotted line) quasars are shown on the
	    corresponding sides. The K-S probabilities of the dust-deficient
	    quasar samples against the normal quasar sample for the
	    corresponding quantities are also presented.
	}
	\label{fig:pg_edd_lum}
\end{figure}

To judge if the dust-deficient population has special AGN properties, we
compare the distributions of the black hole masses, AGN luminosities, as well
as Eddington ratios between the dust-deficient quasars and normal quasars, as
shown in Figure~\ref{fig:pg_edd_lum}.  We collect the black hole mass
measurements from \cite{Peterson2004, Vestergaard2006}, and \cite{Denney2010} with a
virial factor from \cite{Woo2010}. The AGN luminosities are derived from the
AGN templates with the normalization from our SED decomposition results.
Although the HDD population has identical luminous AGNs and massive black holes
compared to the normal quasar population, its members have lower Eddington ratios
($f_{\rm Edd}=L_{\rm AGN}/L_{\rm Edd} \lesssim 0.1$) compared with normal
quasars ($f_{\rm Edd,~normal}$=0.1-1).  The Kolmogorov-Smirnov (K-S) test
yields a probability of only $\sim0.025$ that the Eddington ratios of the HDD
quasars are drawn from the same distribution as those of normal quasars.
Additionally, we note the most HDD PG quasar, PG 0049+171, has the lowest
Eddington ratio among the confirmed HDD quasars. All these observations suggest
the hot dust deficiency is possibly linked to the AGN Eddington ratio. 
Compared with normal quasars, the WDD quasars have higher AGN luminosities
with a K-S probability of $\sim$0.057 that the parameter is drawn from the same
distributions as for normal quasars. For both HDD and WDD quasar populations,
their black hole masses do not show strong differences compared with the normal
quasar population.  We also explore if the dust deficiency is related to the
quasar radio loudnesses, which was defined and measured by
\cite{Kellermann1989}. The fractions of radio-loud quasars among the HDD, WDD,
and normal sample are quite similar (12\%, 20\%, 16\%, respectively),
suggesting the radio properties may not influence the dust deficiency.

\begin{figure}[htp]
    \centering
	\includegraphics[width=1.0\hsize]{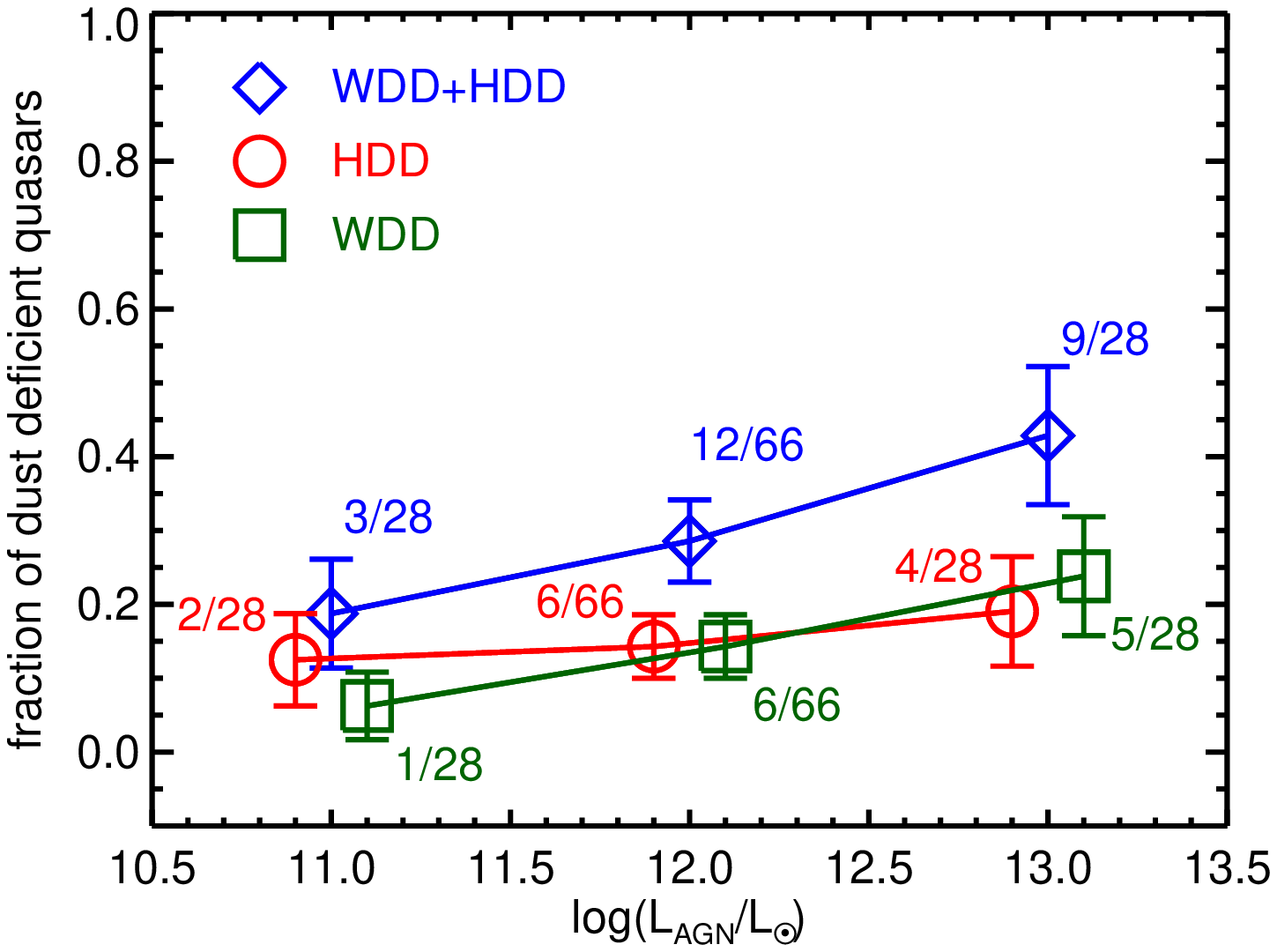} 
	\caption{ 
	    Fractions of dust-deficient quasars as a function of AGN
	    luminosity. We denote the numbers of the dust-deficient quasars and
	    the normal quasars in each luminosity bin. For clarity, data points
	    for HDD and WDD populations are arbitrarily shifted on the X-axis.
}
	\label{fig:covering_f}
\end{figure}

We now explore whether the fraction of dust-deficient quasars is dependent on
AGN luminosity. In Figure~\ref{fig:covering_f}, we show the number fraction of
dust-deficient quasars compared with normal quasars in three luminosity bins,
$\log (L_{\rm AGN}/L_\odot) = [10.5,11.5], [11.5,12.5], [12.5,13.5]$. The
ambiguous cases for the HDD and WDD quasars are removed in the analysis.  We
can see the HDD quasar fraction is not sensitive to AGN luminosity, confirming
similar conclusions reached by \cite{Hao2010,Hao2011} and \cite{Mor2011}. In
contrast, the fraction of WDD quasars shows a clear boost with increasing AGN
luminosity, which is generally consistent with the anti-correlation between the
IR-optical luminosity ratio and AGN luminosity, as found by many authors
\cite[e.g.,][]{Maiolino2007, Roseboom2013, Mateos2016}.

Results from previous studies on the relation between the hot dust deficiency
and AGN properties are contradictory. \cite{Jiang2010, Jun2013} suggested that
the quasars with weak hot dust emission tend to have relatively low black hole
masses ($M_{\rm BH}\sim 10^8~M_\odot$), and high Eddington ratios. On the
contrary, \cite{Hao2010, Mor2011} argued that the hot-dust-poor quasars are
identical to normal quasars in $M_{\rm BH}$ and $f_{\rm Edd}$. We firstly note
that all these studies focus on the very luminous quasars, with \cite{Hao2010,
Mor2011,Jun2013} at $L_{\rm AGN}\gtrsim 10^{12}~ L_\odot$ and \cite{Jiang2010}
at $L_{\rm AGN}\gtrsim 10^{13}~L_\odot$.  As argued in Section~\ref{sec:highz},
there is a bias toward more efficient dust-deficient quasar identification at
higher AGN luminosity (also at higher redshift) using an optical-to-NIR color
selection (e.g., \citealt{Jiang2010,Jun2013}). The AGN luminosities of the 87
PG quasars range from $10^{10.5}$ to $10^{13.5}~L_\odot$. Since the Eddington
ratios of AGN are positively correlated with their bolometric luminosities
\cite[e.g.,][]{Lusso2012}, we are probing the dust deficiency in the weak
accretion state of the black hole, different from previous studies.

\begin{figure}[htp]
	\centering
	\includegraphics[width=1.0\hsize]{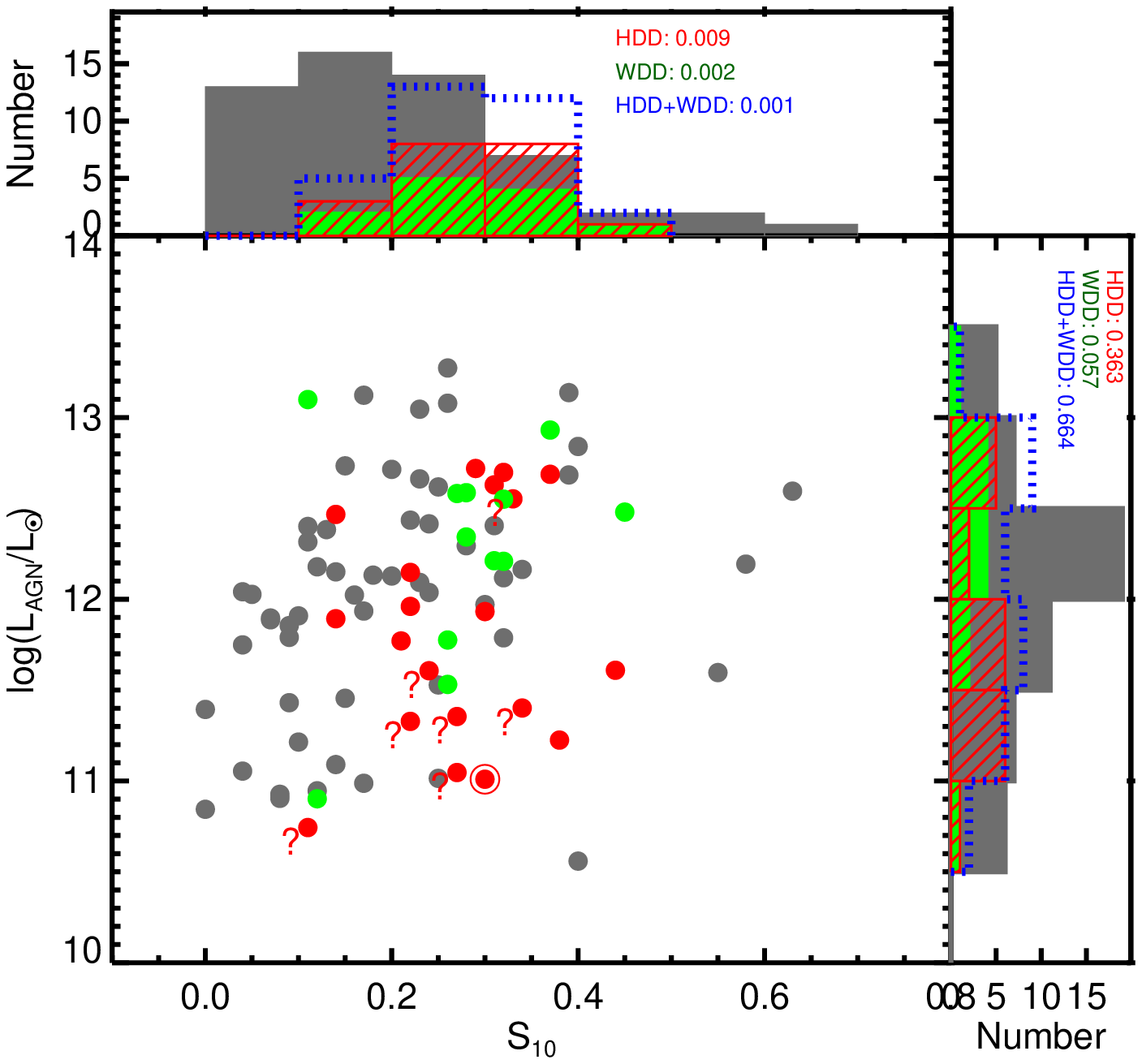} 
	\caption{ 
	    Distributions of AGN luminosities ($L_{\rm AGN}$) and 10$~\mum$
	    silicate strength ($S_{10}$) for normal, WDD and HDD quasars.
	    Symbols and styles are the same as Figure~\ref{fig:pg_edd_lum}.
}
	\label{fig:pg_silicate}
\end{figure}

The silicate feature is an important diagnostic of the dust structure around
the AGN. In Figure~\ref{fig:pg_silicate}, we investigate the strength of this
feature, $S_{10}$, for different types of quasars compared with the AGN
luminosity.  The silicate measurements are adopted from \cite{Shi2014}. We do
not find any strong dependence of the silicate emission strength on the AGN
luminosity, as seen by, e.g., \cite{Maiolino2007}, which is possibly due to the
limited dynamical range of the AGN luminosities of the PG sample. Meanwhile,
the two dust-deficient populations peak at stronger silicate emission
compared with the normal quasar population with K-S probabilities $<$ 0.01
of being drawn from the same distribution.

The K-S probabilities of the AGN properties discussed in this section are
summarized in Table~\ref{tab:K-S_prob}. We caution that the PG sample is
known to be incomplete \citep[e.g.,][]{Jester2005} and the sample size is
relatively small.  Statistical studies on a much larger sample with a
similar rich set of multiband observations are needed to solidify these
arguments.

\begin{deluxetable}{lccc}
    \tablewidth{1.0\hsize}
    \tablecolumns{4}
    \tablecaption{K-S probabilities of the HDD and WDD quasars \\ against normal quasars\label{tab:K-S_prob}
    }
    \tablehead{
	\colhead{Property} & \colhead{HDD}  & \colhead{WDD} & \colhead{HDD+WDD} 
    }
    \startdata
    $L_{\rm AGN}$               &      0.363               &      {\bf 0.057}\tablenotemark{*} &       0.664                \\
    $M_{\rm BH}$                &      0.762              &       0.140 &       0.315  \\
    $L_{\rm AGN}/L_{\rm Edd}$   &      {\bf 0.025} &       0.688              &       {\bf 0.074}  \\
    $S_{10}$                &      {\bf 0.009} &     {\bf 0.002} &       {\bf 0.001} 
    \enddata
    \tablenotetext{*}{We indicate significant differences in bold.}
\end{deluxetable}

\subsection{Are the High-$z$ Dust-deficient Quasars Abnormal \\ in Terms of Their SEDs?}
\label{sec:highz}

\subsubsection{$z\gtrsim5$}

\begin{figure}[htp]
	\centering
	\includegraphics[width=1.0\hsize]{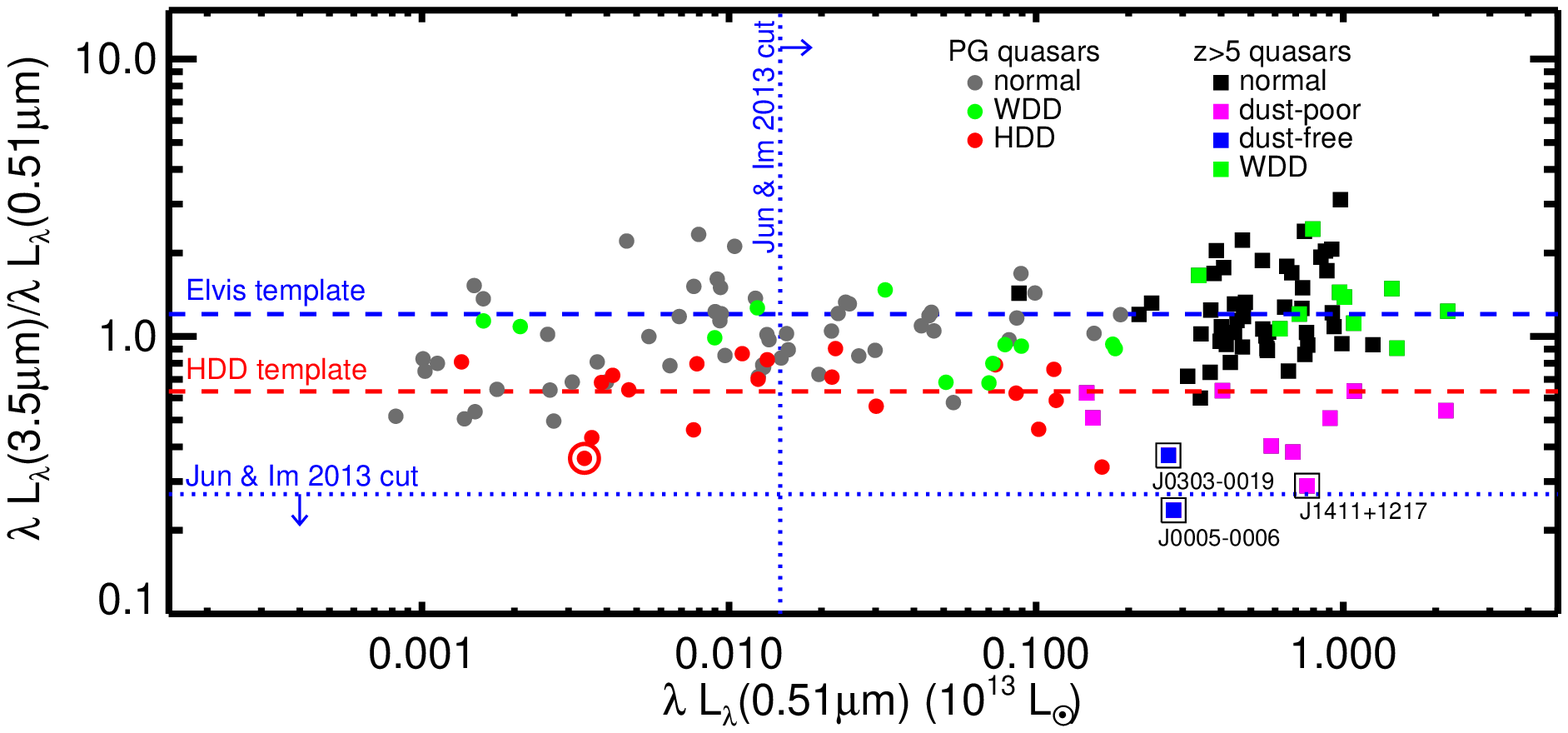} 
	\includegraphics[width=1.0\hsize]{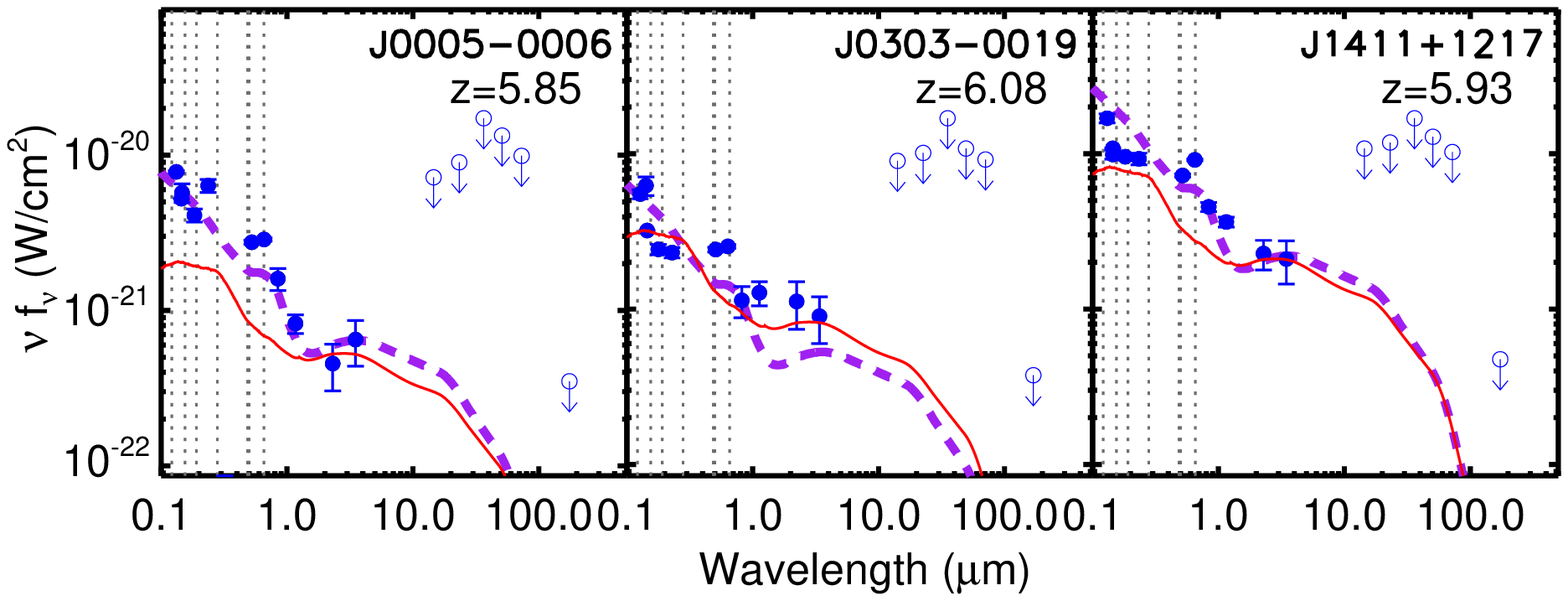} 
	\caption{ 
	    Upper panel: the $3.5~\mum$-to-$0.51~\mum$ (rest wavelengths)
	    luminosity ratio as a function of the optical luminosity at
	    $0.51~\mum$ for PG quasars (dots) and $z\gtrsim5$ quasars
	    (squares).  We use the PG quasar continuum SED based on this work
	    and updated photometry of $z\gtrsim5$ quasars \citep{Leipski2014}
	    to derive the corresponding quantities. $[\lambda
	    L_\lambda(3.5\mum)]/[\lambda L_\lambda(0.51\mum)]=0.5$ is denoted
	    as the dotted line. The $3.5~\mum$-to-$0.51~\mum$ luminosity ratio
	    for the most HDD quasar PG 0049+171 (marked with a red open circle)
	    is shown as the dashed red line. For PG quasars, the gray, green,
	    red colors represent normal, WDD, and HDD quasars. For $z\gtrsim5$
	    quasars, the blue color represents the two hot-dust-free quasars in
	    \cite{Jiang2010}, the magenta represents the other hot-dust-poor
	    quasars as suggested by \cite{Leipski2014}, and the green
	    represents the WDD quasars listed in Table~\ref{tab:dd-hiz}. The
	    three (possibly) hot-dust-free quasars, SDSS J005$-$0006, SDSS
	    J0303-0019 and J1411+1217, are marked with black open squares.
	    Lower panel: SEDs of the three (possibly) hot-dust-free
	    quasars at $z\sim6$.  Observational data points are shown as blue
	    dots (detection) and open circles with arrows (upper limit).  We
	    also scale the HDD template (red line) to the near-IR data points
	    ($\lambda\sim$2-3~$\mum$), and show the SED of the possible host
	    galaxy observed at that redshift (red line).  Comparisons with the
	    SED of PG 0049+171 (purple dashed line) that has an extreme
	    deficiency of hot dust emission are also included.  We also
	    indicate the locations of strong UV-optical lines, i.e.,
	    Ly$\alpha$, C$_{\rm IV}$, C$_{\rm III]}$, H$\beta$, H$\alpha$ (from
	    left to right), as vertical dotted lines.
	}
	\label{fig:jiang10_replot}
\end{figure}

\cite{Jiang2010} suggested that the two $z\sim6$ quasars, J0005$-$0006 and
J0303$-$0019, are dust-free due to their exceptionally low rest-frame $3.5~\mum$
to 5,100 \AA~ luminosity ratios. In the upper panel of
Figure~\ref{fig:jiang10_replot}, we plot the near-IR to optical ratios as a
function of luminosity for the $z>5$ sample with the photometry data in
\cite{Leipski2014} and the $z<0.5$ PG quasars. The 11 $z>5$ quasars with a
dearth of hot dust emission (see \citealt{Leipski2014}, \citealt{Lyu2016}) are
distributed around the value for the HDD AGN template. In the lower panel of
the same figure, we compare the SEDs of three extreme cases: SDSS J0005$-$0006,
SDSS J0303$-$0019, and SDSS J1411+1217, to the HDD AGN template as well as the SED of
the most extreme HDD PG quasar PG 0049+171.  We can see the SEDs of SDSS
J0005$-$0006 and SDSS J1411+1217 are quite similar to that of PG 0049+171,
suggesting that the latter could be a counterpart to the most dust-poor quasars
at $z\sim 6$.  Additionally, the {\it Spitzer}/IRAC $3.6~\mum$ band (rest-frame
$5100~\AA$) is possibly contaminated by the optical \hbeta~and
[N{\sevenrm\,II}] emission lines at $z\sim6$, thus a very low rest-frame
$3.5~\mum$ to $5,100~\AA$~ratio may result.  Therefore, the overall SEDs of the
$z\sim6$ hot-dust-free quasars resemble the HDD template, given the possible
variation of the UV/optical slopes and emission line contaminations.

\begin{figure*}[htp]
	\centering
	\includegraphics[width=1.0\hsize]{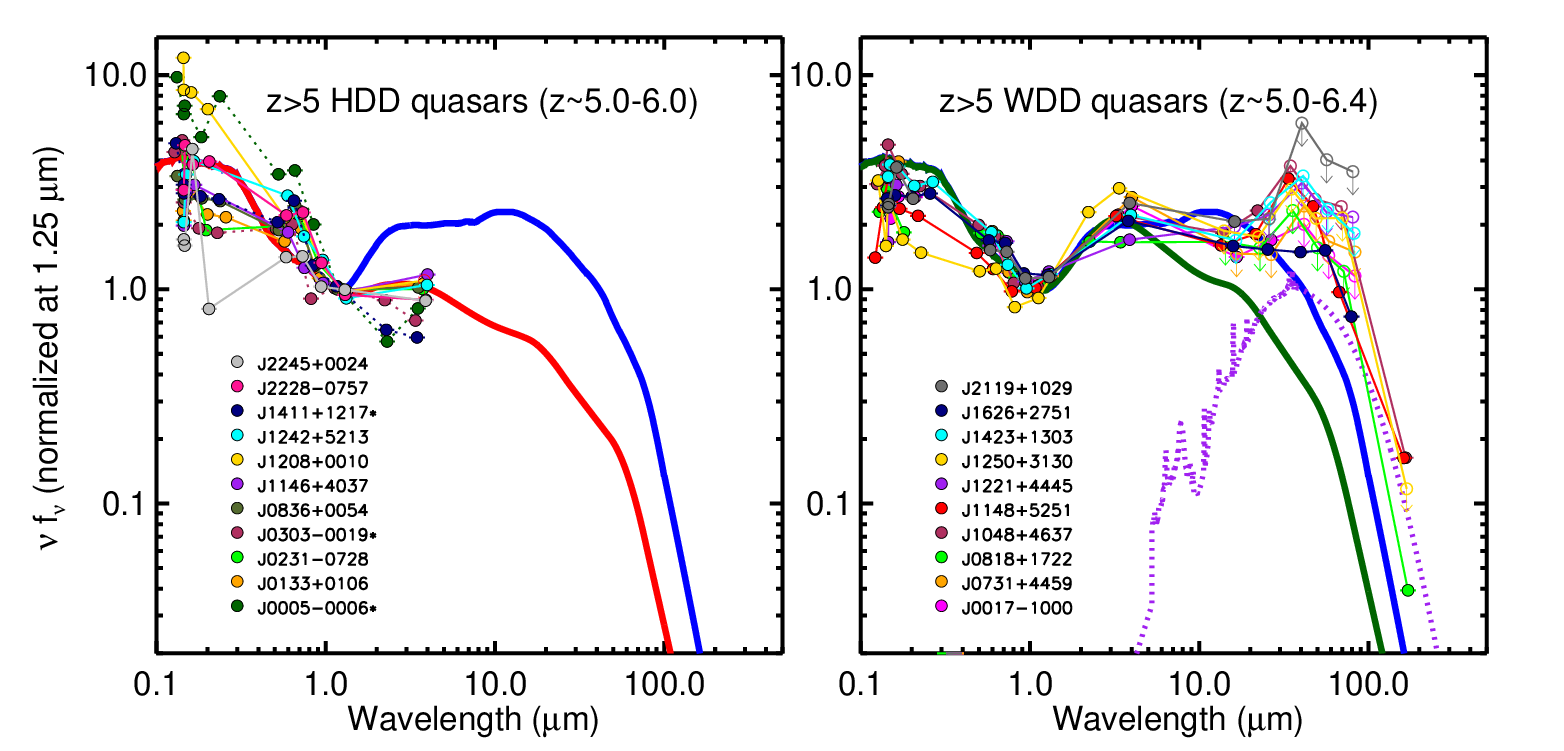} 
	\includegraphics[width=1.0\hsize]{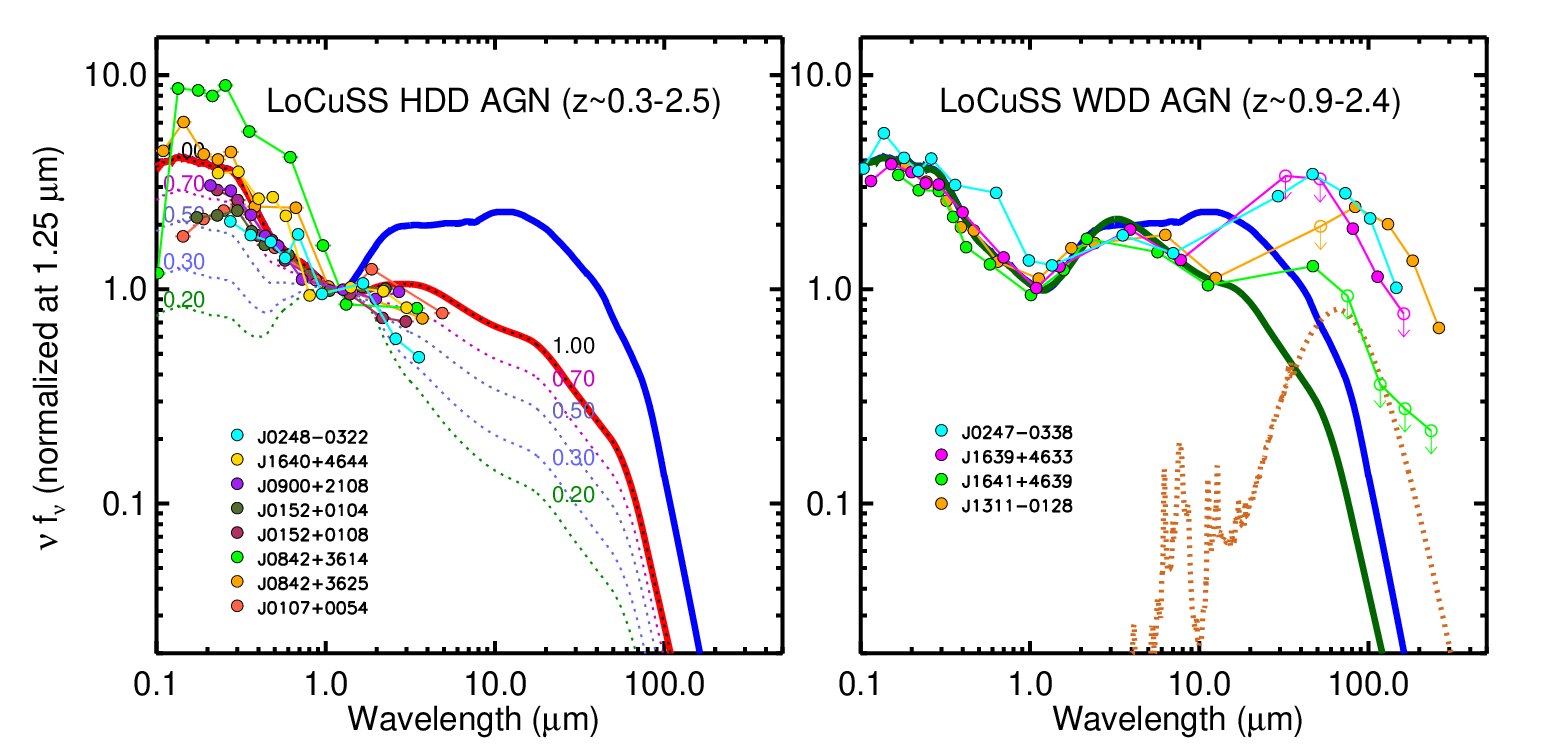} 
	\caption{
	    The SEDs of hot-dust-deficient and warm-dust-deficient quasars at high
	and intermediate redshifts.  We also show the normal \cite{Elvis1994}
	AGN template (far-IR corrected; blue solid line), the HDD template (red
	solid line) and the WDD template (green solid line) in corresponding
	panels. In the top-left panel, we denote the three most HDD quasars at
	$z\sim6$, J0005$-$0006, J0303$-$0019, and J1411+1217, with `*' and use
	dotted lines to connect the SED data points. Compared with the very
	luminous $z\gtrsim5$ quasars, the LoCuSS AGNs are not bright and near-IR
	stellar contamination in the SED is still possible. As a result, we
	also show composite quasar SEDs composed of the HDD template and an old
	stellar population template (dotted lines with numbers to denote the
	host galaxy contribution at $1.25~\mum$) in the bottom-left panel. In
	the right panels, we also plot the IR SEDs of most possible kinds of
	host galaxies: a low-metallicity, compact, starbursting galaxy as
	represented by Haro 11 for $z\gtrsim5$ quasars (\citealt{Lyu2016}; the
	purple dotted line in the top-right panel); a normal star-forming
	galaxy as represented by the \cite{Rieke2009} $\log(L_{\rm
	IR}/L_\odot)=11.5$ galaxy template for $z\sim2$ AGN (\citealt{Xu2015a};
	the orange dotted line in the bottom-right panel).  }
	\label{fig:hiz_dd}
\end{figure*}

Besides the three extreme HDD $z\sim6$ quasars above, \cite{Leipski2014}
also suggested another eight HDD candidates (see also \citealt{Lyu2016}), as
listed in Table~\ref{tab:dd-hiz} . We show their rest-frame SEDs in the
top-left panel of Figure~\ref{fig:hiz_dd}. These $z\gtrsim5$ quasar SEDs are
matched by our HDD AGN template reasonably well. At $z\gtrsim5$, we also
identified 10 WDD quasars from the \cite{Leipski2014} sample (as also listed in
Table~\ref{tab:dd-hiz}).  Their SEDs present a strong near-IR hot dust emission
bump but are weak in the mid-IR, as indicated by faint emission or even
non-detections in the {\it Herschel} PACS $70~\mum$ bands (the top-right panel
in Figure~\ref{fig:hiz_dd}). As argued in \cite{Lyu2016}, the host galaxies of
these quasars are likely to have a strong contribution to the mid-IR SEDs, due
to their low-metallicity and compact starbursting properties.  We expect that the
mid-IR emission contributed by the AGN is smaller than the rest-frame mid-IR
data points indicate. The number fraction of the $z\gtrsim5$ WDD quasars in
\cite{Leipski2014} is $\sim14\%$.

\begin{deluxetable}{lcc}
    \tabletypesize{\scriptsize}
    \tablewidth{1.0\hsize}
    \tablecolumns{3}
    \tablecaption{Dust-deficient Quasars at $z$=0.5-6\label{tab:dd-hiz}
    }
    \tablehead{
	\colhead{Source} & \colhead{$z$}  & \colhead{Type}  \\
	\colhead{(1)} & \colhead{(2)} & \colhead{(3)}   
    }
    \startdata
    \multicolumn{3}{c}{$z\gtrsim5$ quasars in \cite{Leipski2014}} \\
    SDSS J001714.67$-$100055.4  &     5.01  &  WDD\\ 
    SDSS J073103.12+445949.4  &       5.01  &  WDD\\ 
    SDSS J081827.40+172251.8  &       6.00  &  WDD\\ 
    SDSS J104845.05+463718.3  &       6.23  &  WDD\\ 
    SDSS J114816.64+525150.3  &       6.43  &  WDD\\ 
    SDSS J122146.42+444528.0  &       5.19  &  WDD\\ 
    SDSS J125051.93+313021.9  &       6.13  &  WDD\\ 
    SDSS J142325.92+130300.7  &       5.08  &  WDD\\ 
    SDSS J162626.50+275132.4  &       5.30  &  WDD\\ 
    SDSS J211928.32+102906.6  &       5.18  &  WDD\\ 
    SDSS J000552.34$-$000655.8*  &     5.85  &  HDD \\
    SDSS J013326.84+010637.7  &       5.30  &  HDD \\
    SDSS J023137.65$-$072854.5  &     5.41  &  HDD \\
    SDSS J030331.40$-$001912.9*  &     6.08  &  HDD \\
    SDSS J083643.85+005453.3  &       5.81  &  HDD \\
    SDSS J114657.79+403708.7  &       5.01  &  HDD \\
    SDSS J120823.82+001027.7  &       5.27  &  HDD \\
    SDSS J124247.91+521306.8  &       5.05  &  HDD \\
    SDSS J141111.29+121737.4*  &       5.93  &  HDD \\
    SDSS J222845.14$-$075755.2  &     5.14  &  HDD \\
    WFS J2245+0024           &        5.17  &  HDD \\
    \hline \\[-1.8ex]
    \multicolumn{3}{c}{AGN with MIPS $24~\mum$ Flux $>1~$mJy in \cite{Xu2015a}} \\
    LoCuSS J131107.34$-$012857.9&    0.92      & WDD\\
    LoCuSS J164116.66+463946.3&      1.13     & WDD\\
    LoCuSS J163950.35+463327.1&      2.09      & WDD\\
    LoCuSS J024725.09$-$033807.9&    2.42     & WDD\\
    LoCuSS J090021.93+210803.9&      0.70      & HDD\\
    LoCuSS J164025.01+464449.2&      0.54      & HDD\\
    LoCuSS J024851.43$-$032249.3&    0.30      & HDD\\
    \multicolumn{3}{c}{AGN with MIPS $24~\mum$ Flux $<1~$mJy ** } \\ 
    LoCuSS J010720.40+005435.2&      1.47      & HDD\\
    LoCuSS J084218.48+362504.1&      2.24      & HDD\\
    LoCuSS J084258.80+361444.2&      2.50      & HDD\\
    LoCuSS J015208.74+010823.6&      0.56      & HDD\\
    LoCuSS J015202.95+010445.3&      1.05      & HDD
    \enddata
    \tablenotetext{*}{The most extreme HDD quasars at $z\sim6$.}
    \tablenotetext{**}{The data of LoCuSS AGN with $f_{24~\mum}<1~$mJy is
    provided by L. Xu (2016, private communication).}
\end{deluxetable}

\begin{figure}[htp]
	\centering
	\includegraphics[width=1.0\hsize]{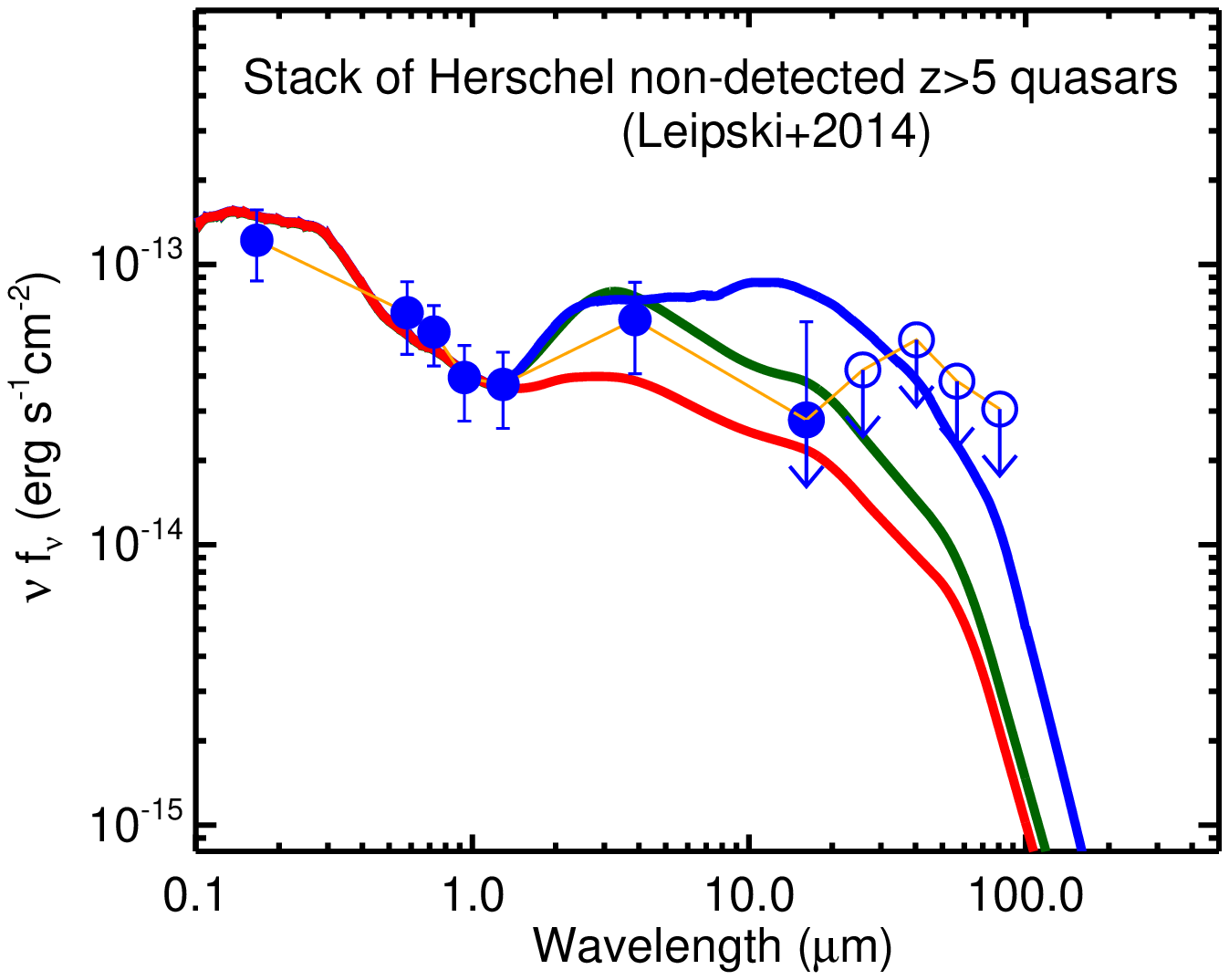} 
	\caption{ 
	    Comparison of the stacked SED of $z\gtrsim5$ {\it Herschel}
	    non-detected quasars in \cite{Leipski2014} with the normal
	    \cite{Elvis1994} AGN template (far-IR corrected; blue solid line),
	    the WDD template (green solid line), and the HDD template (red
	    solid line). 
	}
	\label{fig:leipski_undetected}
\end{figure}

As shown by \cite{Leipski2014}, the stacked SED of 33 {\it Herschel}
non-detected $z\gtrsim5$ quasars is not matched optimally with the classical
AGN template (also see \citealt{Lyu2016}). With the AGN templates derived in
this work, we find the average SED of the far-IR non-detected $z>5$ quasars
lies between the WDD and HDD AGN templates, as seen in
Figure~\ref{fig:leipski_undetected}. Combining our previous work
\citep{Lyu2016} with the discussion in this section, we can conclude that the
SEDs of the luminous quasars at $z\gtrsim5$ can be characterized by the normal
AGN template, WDD template, and HDD template derived from the $z<0.5$ PG
quasars. In other words, there is no indication of strong evolution of AGN
infrared SEDs at $z\gtrsim5$.

\subsubsection{$z\sim$0.5-2.5}

For the majority of AGNs with {\it Spitzer}/MIPS $24~\mum$ flux density $>1~$mJy
in the Local Cluster Substructure Survey
(LoCuSS\footnote{\url{http://www.sr.bham.ac.uk/locuss/}}), \cite{Xu2015a}
presented accurate decompositions with the same normal AGN template used in
this work. We also searched for HDD and WDD quasars in this survey. Eight
LoCuSS quasars with redshifts of 0.3-2.5 show indications of weak hot dust
emission (see Table~\ref{tab:dd-hiz} and the bottom-left panel of
Figure~\ref{fig:hiz_dd}). Besides J084218.48+362504.1, all the rest of these
LoCuSS HDD quasars are not detected by {\it Herschel}, indicating their weak
far-IR emission.  The variation of the SEDs of these quasars can be easily
explained by adding an old stellar component to the HDD template. Thus we again
confirm that similar HDD quasars are also seen at intermediate redshifts. 
We also find four WDD quasars at $z$=0.9-2.4 in the LoCuSS type-1 AGN sample
(see Table~\ref{tab:dd-hiz} and the bottom-right panel of
Figure~\ref{fig:hiz_dd}). Although the observed far-IR emission is high (due to
the host galaxy star-formation), the 0.1-10.0~$\mum$ SEDs of these AGNs show a
clear SED turnover at 3~$\mum$ and are matched well by the WDD AGN template.
Similarly to the situation at $z\gtrsim5$, no evidence among the LoCuSS sample
indicates the AGN SEDs at $z$=0.5-2.5 differ from those in the PG sample.

\begin{figure}[htp]
	\centering
	\includegraphics[width=1.0\hsize]{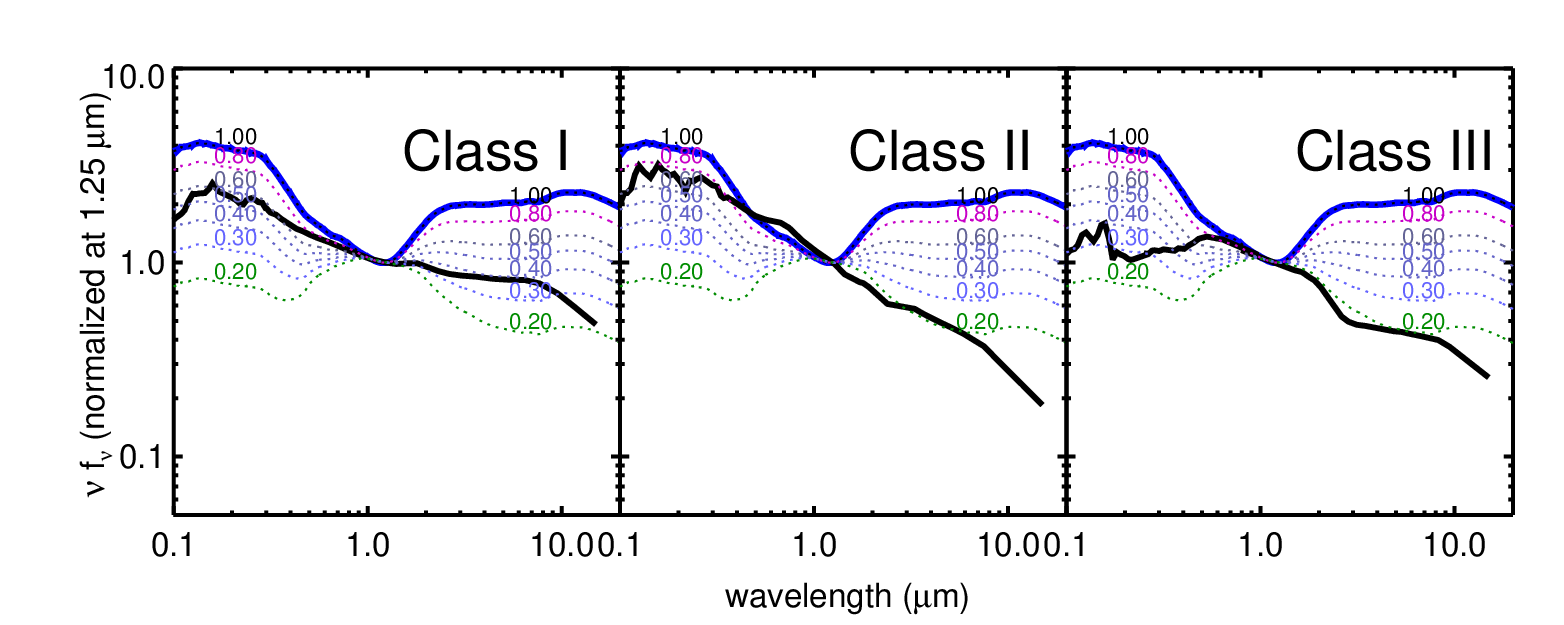} 
	\includegraphics[width=1.0\hsize]{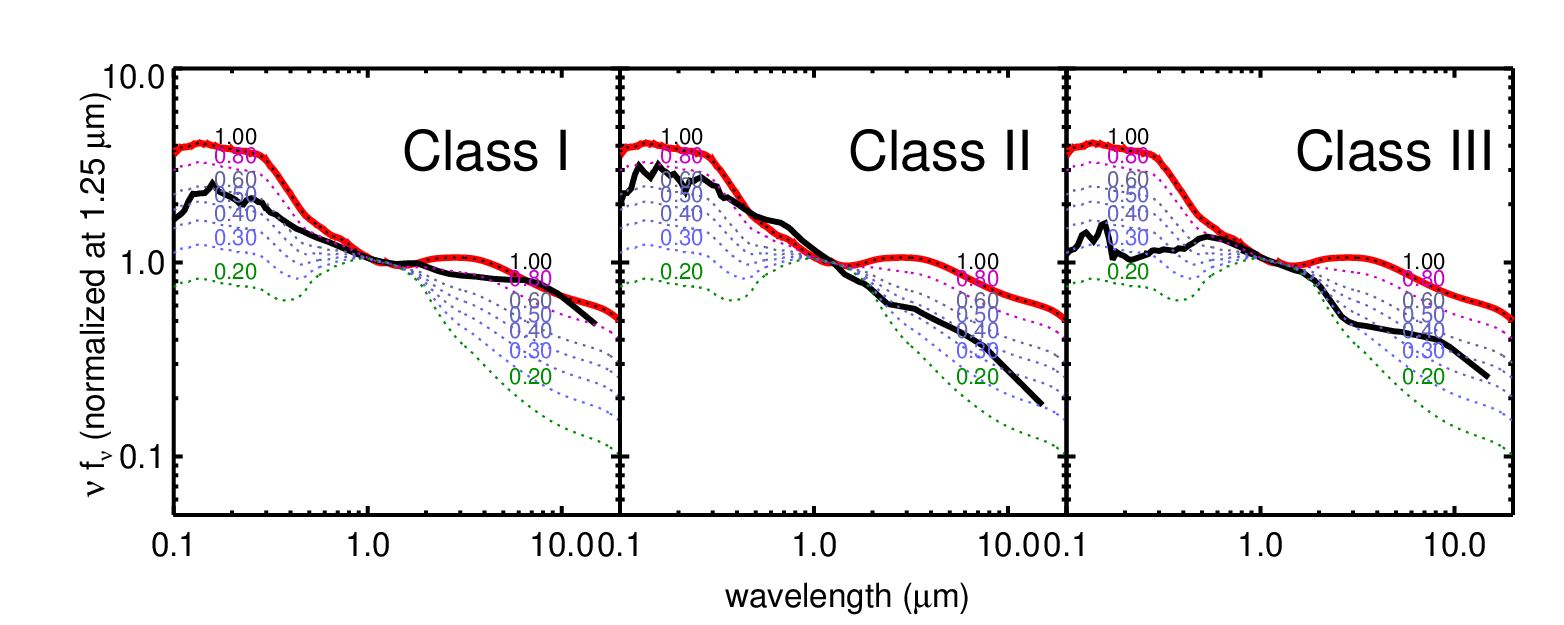} 
	\caption{ 
	    Comparison of the mean SEDs of three classes of hot-dust-poor
	    quasars in \citealt{Hao2010} (black lines) with the composite SEDs of
	    the normal AGN template (blue lines in the upper panels) or the HDD
	    AGN template (red lines in the lower panels). We also show the mock
	    SEDs of the AGN templates with an old stellar population template
	    as dotted lines with the numbers indicating the fraction of AGN
	    contribution at $1.25~\mum$.
	}
	\label{fig:hao11_sed}
\end{figure}

\cite{Hao2010,Hao2011} reported the discovery of hot-dust-poor (HDP) quasars at
$z$=0.1-3 in the XMM-{\it COSMOS} sample \citep{Elvis2012}, the {\it Spitzer}+SDSS
selected sample \citep{Richards2006}, and the \cite{Elvis1994} sample. These
quasars are identified by their special combination of optical (0.3-1~$\mum$)
and near-IR (1-3~$\mum$) slopes ($\alpha_{\rm opt}$ and $\alpha_{\rm NIR}$,
respectively), and are further grouped into three classes based on the
locations in a $\alpha_{\rm opt}$-$\alpha_{\rm NIR}$ plot (see details in
\citealt{Hao2010}). We compare our AGN templates to the mean SEDs of the three
classes of HDP quasars derived from the XMM-{\it COSMOS} sample
\citep{Hao2010} in Figure~\ref{fig:hao11_sed}.  Since the majority of HDP
quasars in \cite{Hao2010} are at $z$=1-3, their mid-IR SEDs (3-10$~\mum$) are
poorly constrained by the MIPS $24~\mum$ and IRAC $8~\mum$ photometry. Thus we
limit the comparison to $\lambda<3~\mum$. Firstly, we find that all three HDP
templates prefer the HDD AGN template to represent their AGN component. As
shown in the upper panels of Figure~\ref{fig:hao11_sed}, if the normal AGN
template is assumed, a strong contribution of the host galaxy is required to
match the IR SEDs of these quasars ($f_{\rm host, 1.25}\sim0.6$ for class I,
$f_{\rm host, 1.25}\sim0.8$ for class II and class III), leaving a strong
underestimation of the UV-optical observed SED. Adding an extremely strong
young stellar contribution to match such an SED deficiency is unlikely to be a
reasonable solution. In contrast, if the AGN components in class I HDP quasars
are represented by the HDD template, the host galaxy contamination in the
near-IR would be small ($f_{\rm host, 1.25}\sim0.20$). The difference between
the UV-optical SED of class I HDP quasars and the HDD AGN template can be
completely mitigated by introducing moderate extinction to the AGN component.
For the class II and class III HDP quasars, if a strong host galaxy
contribution ($f_{\rm host, 1.25}\sim$0.6-0.7) is added, the HDD AGN template
can also recover the HDP SEDs reasonably well. As a result, we do not find that the
HDP quasars presented by \cite{Hao2010, Hao2011} are atypical compared to the
PG sample.

In summary, quasars with weak hot dust emission at $z$=0.5-2.5 do not have
significantly different SEDs compared with the dust-deficient quasars in the PG
sample, if AGN extinction and the possible host galaxy near-IR contamination
are considered.

\subsection{Does the HDD Quasar Fraction Evolve with Redshift?}
\label{sec:hdd-z-evolution}

\subsubsection{Bias Due to the Parent Sample}

The HDD fractions are subject to the selection criteria of the parent sample.
For the MIPS $24~\mum>1~$mJy complete type-1 LoCuSS AGN sample \citep{Xu2015a},
we only identify 3 out of 107 quasars to be HDD, making the HDD fraction only
$\sim$ 3\%. For the 19 quasars identified by other means (e.g., SDSS) in the
same field but with MIPS $24~\mum<1~$mJy and relatively complete infrared SED
observations, 5 quasars are HDD, making an HDD fraction $\sim~26\%$.  Because
the LoCuSS type-1 quasars were selected on $24~\mum$ flux density, they are biased
against quasars with weak infrared emission. A similar effect for near-IR color
selection is also seen in the $z<0.3$ 2MASS sample in \cite{Shi2014}, where no
HDD quasars have been found.  The 2MASS sample is characterized by a red AGN
population with $J-K_s>2$ \citep{Cutri2001, Smith2002}, while the HDD quasars
tend to have $J-K_s\sim$1.0-1.5.

\begin{figure}[htp]
	\centering
	\includegraphics[width=1.0\hsize]{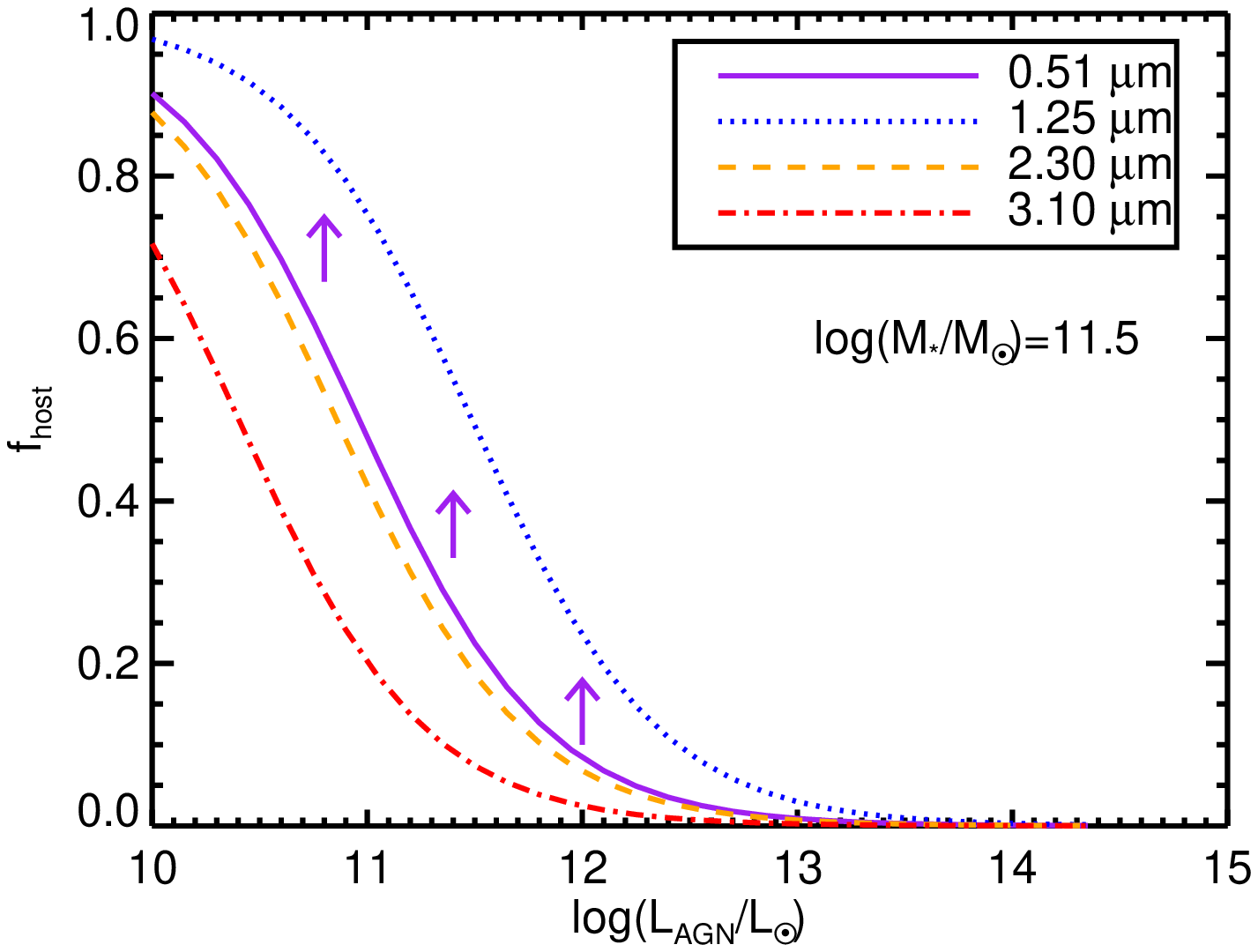} 
	\caption{ 
	    Host galaxy contribution vs. AGN luminosity from the SED model.
	    We assume $10^{11.5}~M_\odot$  as the maximum possible host galaxy
	    mass and adopt the mass to light ratio for local field galaxies
	    \citep{Bell2003}. The arrows indicate the possible shift of the
	    host galaxy fraction $f_{\rm host}$ if the galaxy contains young
	    stellar populations. See the text for details.
	}
	\label{fig:fhost_norm}
\end{figure}

A luminosity bias of the parent sample may also produce specious evolution of
the HDD fraction. As shown in Section~\ref{sec:dd_color}, strong near-IR SED
contamination by host galaxy stellar emission can mimic hot dust deficiency in
normal quasars. In Figure~\ref{fig:fhost_norm}, we combine the Elvis AGN
template and an elliptical galaxy template to explore the host galaxy fraction
$f_{\rm host}$ as a function of AGN luminosity. The host galaxy mass is assumed
to be $10^{11.5}~M_\odot$, which is likely to be the maximum value for most
quasars \citep[e.g.,][]{Reines2015,Bongiorno2016}. We convert the stellar mass
to the near-IR luminosity, adopting the mass to light ratio for local field
galaxies \citep{Bell2003}. We can see at $\log(L_{\rm AGN}/L_\odot)>13$, the
host galaxy contamination at the optical bands as well as the near-IR to mid-IR
bands is negligible (with $f_{\rm host}<0.05$), so the simple two color
identification adopted in \citealt{Jiang2010} should be good enough to pick out
dust-deficient quasars.  At $12<\log(L_{\rm AGN}/L_\odot)<13$, we see a gradual
increase of $f_{\rm host}$.  Below $\log(L_{\rm AGN}/L_\odot)=12$, the host
galaxy contamination increases rapidly from $f_{\rm host}\sim10\%$, in which
case the identification of HDD quasars becomes very difficult.

In the upper panel of Figure~\ref{fig:jiang10_replot}, we show the
$L_{0.51~\mum} > 10^{44.73}~{\rm erg~s^{-1}}$ cut in \citealt{Jun2013}. For
the same HDD quasar criteria ($\lambda
f_\lambda[3.5\mum/0.51\mum]<0.64$), no PG quasars are selected above the
luminosity cut. A large number of dust-deficient PG quasars are still hidden in
this simple selection. In contrast, similar HDD quasars at
$z\gtrsim5$ are easily identified. As we observe more luminous quasars at
higher redshift, this color selection becomes more productive. As a result, we
will get an increasing HDD quasar fraction even if the real value is
constant. Nevertheless, for the samples in \cite{Jiang2010} ($L_{\rm AGN}\sim
10^{13}~L_\odot$), \cite{Jun2013} ($L_{\rm AGN}> 10^{12}~L_\odot$), and
\cite{Mor2011} ($L_{\rm AGN} > 10^{12}~L_\odot$), the values of $f_{\rm host}$
at 0.51, 2.30, 3.10~$\mum$ are smaller than 10\%. In other words, the host
galaxy contribution has very limited impact on the dust-deficient quasar
selection in these papers.

\subsubsection{Bias Due to the Selection Methods}

\begin{figure}[htp]
	\centering
	\includegraphics[width=1.0\hsize]{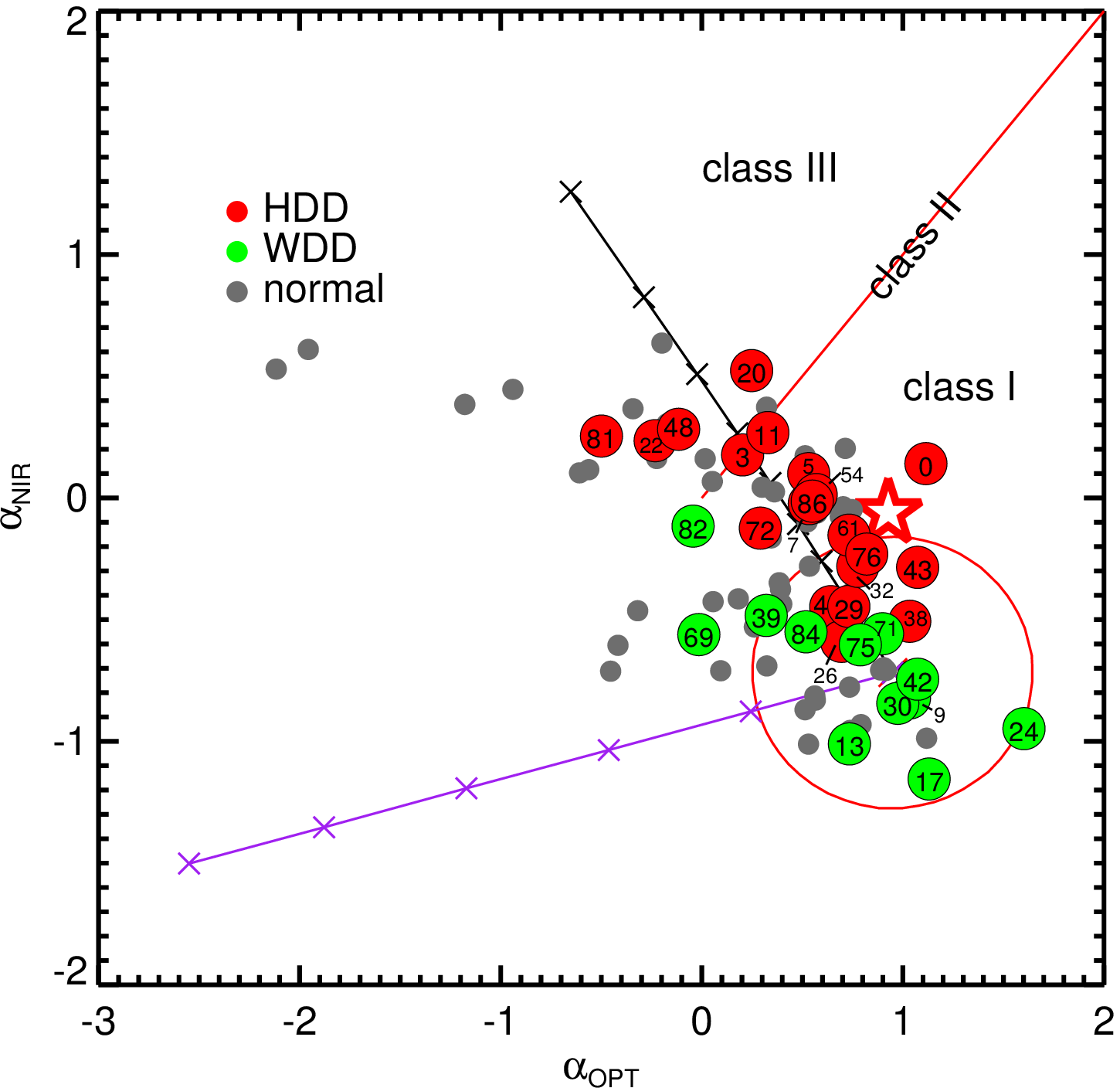} 
	\caption{ 
	  The \cite{Hao2010} selection of hot-dust-poor quasars applied to the
	  PG sample.  The red circle shows the \cite{Elvis1994} mean AGN SED
	  and the $1\sigma$ dispersion of the \cite{Elvis1994} samples. The
	  black line is a mixing curve of the AGN template and the galaxy
	  template (Spi4 in the SWIRE library, see Figure 1 of
	  \citealt{Hao2010}).  The purple line shows the reddening vector for
	  the \cite{Elvis1994} template. The straight solid line shows where
	  $\alpha_{\rm OPT}=\alpha_{\rm NIR}$.  The big red dots are selected
	  HDD PG quasars. We also show the position of the HDD AGN template on
	  this diagram as the red five-pointed star.
	}
	\label{fig:hao10_select}
\end{figure}

Different selection criteria are another important reason for the discrepancies
in the HDD fractions in the literature. \cite{Hao2010} introduced the
AGN-galaxy mixing diagram, which takes account of both host galaxy
contamination and quasar reddening, to search for dust-deficient quasars.
Following \cite{Hao2010,Hao2011}, we fit power laws ($\nu F_\nu \propto
\nu^\alpha$) at 0.3-1$~\mum$ and 1-3$~\mum$ to derive an optical slope
($\alpha_{\rm OPT}$) and an NIR slope ($\alpha_{\rm NIR}$) for each quasar in
the PG sample, and show their distribution in the AGN-galaxy mixing diagram as
Figure~\ref{fig:hao10_select}. For the HDD PG quasars, four out of the seven
ambiguous cases, PG 0043+039,PG 1022+519,  PG 1341+258, and PG 2209+184 locate
below the AGN-galaxy mixing curve. Meanwhile, six other HDD PG quasars have
deviations of $\alpha_{\rm OPT}$ and $\alpha_{\rm NIR}$ within the 1$\sigma$
dispersion of the \cite{Elvis1994} sample. HDD quasars like PG 1100+772, PG
1121+422, PG 1115+407, PG 1216+069, PG 1302$-$102, and PG 1626+554 are not
revealed by the stringent selection. Besides PG 1011$-$040, the remaining HDD PG
quasars are grouped into class I, suggesting the \cite{Hao2010} selection only
recovers $\sim50\%$ of the HDD quasars identified through SED decomposition.
Additionally, the near-IR SEDs of class II and class III HDP quasars selected
from this diagram are likely dominated by host galaxy emission, making the
identification of the real dust-deficient quasars ambiguous.

\cite{Mor2011} defined the dust-covering factor as the ratio between the
luminosities of dust emission at 2-35~$\mum$ and the AGN power-law component,
$C_{\rm HD} = L_{\rm dust, \text{2-35}\mum}/L_\text{AGN,power-law}$, to look for
quasars with low dust-covering. They also apply a luminosity cut at rest-frame
$3000~\AA$, $L_{0.3}> 10^{45}~{\rm erg~s^{-1}}$ to remove quasars with possible
strong host contamination. As demonstrated in their Figure 3, normal quasars
have $C_{\rm HD}$ peaked around 0.23, and low dust-covering quasars have
$C_{\rm HD}$ = 0.1-0.13. These values are consistent with our Elvis normal AGN
template and
HDD template. The 15-20\% fraction in \cite{Mor2011} is also similar to our
value for HDD quasars (15-23\%). However, since they did not study the IR SED
shapes, the \cite{Mor2011} low dust-covering quasar sample is not guaranteed to
be purely HDD-like quasars.

\cite{Jun2013} required the luminosity ratio $L_{2.3}/L_{0.51} < 0.32$ to
define HDD quasars, whereas the HDD AGN template we derived based on the PG
sample has $L_{2.3}/L_{0.51} = 0.63$. Only the extreme HDD PG 0049+171 has a
$L_{2.3}/L_{0.51}\sim0.34$. As a result, \cite{Jun2013} only looked at the
extreme HDD and rare quasars. These quasars may be extremely optically blue HDD
quasars (as is the case for PG 0049+171).

\subsubsection{the HDD Quasar Fraction}

\cite{Richards2009} showed that optical-only selection would miss 50\% of the
whole type 1 quasar sample (their Figure 9). As a rough estimation, the 20\%
HDD fraction in the PG sample should translate into about 10\% in a complete
sample to similar luminosity. This value is similar to HDP quasar fraction of
the XMM-COSMOS AGN sample and the SDSS-{\it Spitzer} sample for AGN with
$\log(L_{\rm AGN}/L_\odot) > 12$ ($z\sim$0.5-4, \citealt{Hao2010, Hao2011}),
whose host stellar contamination is negligible. 

In the \cite{Leipski2014} $z\gtrsim5$ sample (selected by blue colors in the
rest optical bands), the fraction of dust-deficient quasars is $11/69\sim16\%$,
which is identical to that of our PG sample (optically selected). As shown in
Figure~\ref{fig:hiz_dd}, the SEDs of $z\gtrsim5$ HDD quasars are
similar to our HDD AGN template, suggesting a similar incidence of HDD AGNs.
The luminosities of these quasars are very high ($L_{\rm
AGN}\sim10^{13}~L_\odot$, \citealt{Lyu2016}), so that the host galaxy stellar
contamination can be ignored. However, most of these $z\gtrsim5$ quasars are
selected from SDSS and it is well-known that such a quasar sample could have
very low completeness \citep[e.g.,][]{McGreer2013}. As a result, we should be
cautious in comparing the $z\gtrsim5$ sample to the intermediate-redshift or
low-redshift results directly.

\cite{Hao2010} argued that the fraction of HDP AGNs shows a jump from $z<2$ to
$2<z<3.5$. Nevertheless, at $z<2$, there is a large number of AGN with
$\log(L_{\rm AGN}/L_\odot) < 12$ (Figure 2 in \citealt{Elvis2012}, assuming
$L_{\rm AGN}\approx 50 \nu L_\nu \text{[2~keV]}$). As shown above, the
\cite{Hao2010} selection may miss a large number of HDD quasars with strong
near-IR host galaxy contamination.  As a result, we do not think the redshift
evolution of the HDP quasar fraction in the XMM-COSMOS sample \citep{Hao2010}
is convincing since the degeneracy between the AGN luminosity and redshift was
not considered. In agreement, in the subsequent paper by \cite{Hao2011}, no
detectable redshift evolution of the HDP quasar fraction was reported. For the
\cite{Richards2006} SDSS-{\rm Spitzer} sample studied in the second paper
\citep{Hao2011}, the vast majority of AGNs have $\log(L_{\rm AGN}/L_\odot) > 12$
and therefore are bright enough that the host galaxies do not confuse the
identification of hot-dust-poor quasars.

In summary, the HDD fraction of quasars from an optical sample is about
15-20\%, with no strong evidence for redshift evolution. A purely NIR-selected
quasar sample could miss a large number of HDD quasars due to their abnormal IR
SEDs. In a complete sample, either selected by optical-IR techniques or by
X-ray emission, the HDD fraction is around 10\%.

\subsection{Dust-deficient Quasars: What is the Cause?}
\label{sec:dd_cause}

The standard unified model of active nuclei \citep[e.g.,][]{Urry1995} suggests
that the accretion disk lies within the central hole of a circumnuclear torus
that is optically thick in the optical and near infrared; the torus is in turn
surrounded by cold interstellar clouds. Within the main body of the torus,
where much of the dust is shielded from direct illumination by the central
source, the temperature decreases with increasing radius and the resulting SEDs
are complex and depend on radiative transfer and viewing angle
\citep[e.g.,][]{Fritz2006}. With this classical picture, several possibilities
have been proposed in the literature to explain the dust-deficient behavior of
AGNs. 

For example, \cite{Haas2003} proposed that quasars with weak near-IR and mid-IR
emission might arise due to reduced optical-UV emission from the accretion
disk, reducing the energy to be absorbed by the hot dust.  However, we find
that the UV/optical SEDs of the HDD quasars are similar to those of normal
quasars, and their luminosities are not low (also see, e.g.,
\citealt{Hao2010,Hao2011, Jiang2010, Mor2012, Jun2013}), contrary to this
hypothesis.

\cite{Jiang2010} suggested that the $z\sim6$ extremely IR-weak quasars could be
the first-generation of quasars that live in a dust-free medium and hence do not
have the torus structure. However, as we have shown, similar quasars can also be
found in the nearby Universe (Section~\ref{sec:highz}) and there is no evidence
for strong redshift evolution of the fraction of HDD quasars
(Section~\ref{sec:hdd-z-evolution}).  Additionally, considering that quasars
are already metal-rich \citep{Nagao2006,Nagao2012, Jiang2007,Juarez2009} at
$z\sim$5-6, and very dusty quasars already exist at $z\sim7$
\citep{Barnett2015}, a significant population of dust-free quasars at $z\sim6$
would be surprising.

On the other hand, \cite{Kawakatu2011} argued that the weak near-IR dust
emission of quasars is associated with a super Eddington ratio and stated that
the IR-weak quasars in \cite{Jiang2010} have super Eddington ratios as a
support for this model. However, we note that accurate Eddington ratios are
hard to get for these quasars because of their lowest luminosities among
quasars with UV to far-IR SED constraints at $z\gtrsim5$ \citep{Lyu2016} and
the uncertainties in the calibration of the black hole mass estimators at high
redshifts \citep[e.g., see review by][]{Shen2013}. More importantly, our
analysis of the HDD quasars suggests the opposite: the deficiency of hot dust
emission is more easily seen in quasars with low accretion rates. 

The receding torus model \citep[e.g.,][]{Lawrence1991, Simpson2005, Assef2013}
has been frequently invoked to explain the decrease in the infrared-to-optical
luminosity ratios of quasars with increasing AGN luminosity
\citep[e.g.,][]{Maiolino2007, Roseboom2013, Mateos2016}. The classical picture
assumes the torus has an approximately constant scale height
\citep[e.g.,][]{Lawrence1991}. The size of the inner wall of the torus is
determined by the dust sublimation radius, $R_{\rm sub, 0} \propto L_{\rm AGN,
UV}^{0.5} T_{\rm sub}^{-2.8} a^{-0.5}$, where $T_{\rm sub}$ is the dust
sublimation temperature and $a$ is the dust grain size
\citep[e.g.,][]{Barvainis1987}. With increasing AGN luminosity $L_{\rm AGN,
UV}$, $R_{\rm sub}$ will grow, increasing the solid angle through which energy
from the central engine can escape. As shown in Section~\ref{sec:char}, the
fraction of WDD quasars grows with increasing AGN luminosity, seemingly
consistent with this prediction.

However, the relatively stronger silicate emission feature observed in the WDD
quasars (see Section~\ref{sec:char}) may be inconsistent with the assumption of
a constant scale height for the torus. The silicate emission arises from
warm dust relatively far from the central engine and of moderate optical depth
\citep[e.g.,][]{Fritz2006, Nenkova2008a}. If the torus scale height were
constant, with the receding of the torus we should observe similar silicate
emission strength due to the identical structure of the warm dust emission
region. In fact, we can explain the stronger silicate emission of
dust-deficient quasars if the scale height of the warm dust above the torus
mid-plane is reduced and the tori intercept less energy from their accretion
disks (reducing the mid-IR continuum, while the outermost zones responsible for
the silicate emission are relatively unchanged). Observations show the hydrogen
column density for high-luminosity AGNs is lower than that for low-luminosity
AGNs \citep[e.g.,][]{Ueda2003, Barger2005, La_Franca2005, Akylas2006}, also
favoring a smaller torus scale height with increasing AGN luminosity.

\begin{figure}[htp]
	\centering
	\includegraphics[width=1.0\hsize]{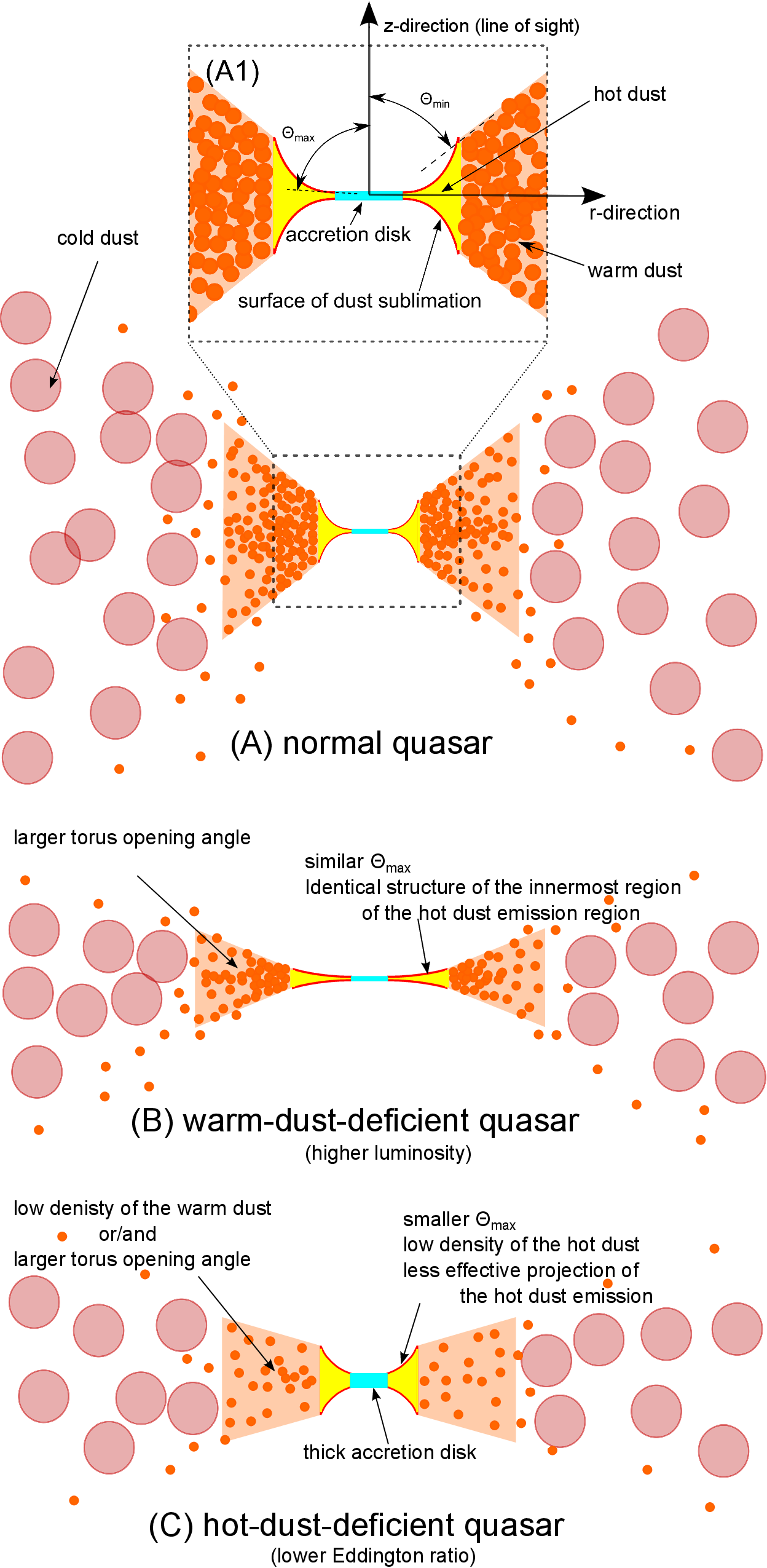} 
	\caption{ 
	    Schematic demonstration of the torus geometry of normal, WDD and
	    HDD quasars. The shape of each component is exaggerated to reflect
	    its distinct features. See the text for details.
	}
	\label{fig:torus_geo}
\end{figure}

Finally, we propose a schematic model for the torus geometry evolution to
explain the observed behavior of dust-deficient quasars through modifications
in the standard model. As pointed out by, e.g., \citealt{Stalevski2016}, if the
torus structure is not changed, the AGN luminosity alone does not modify the IR
SED shape of the torus. Consequently, we focus on the geometry of the torus,
not on the self-similar scale expansion of various structures with increasing
AGN luminosity. In cartoon A of Figure~\ref{fig:torus_geo}, we picture the key
components of our model. Heated by the UV-optical emission from the nucleus, the
dust that makes up the torus assumes a temperature gradient as a function of
distance to the accretion disk, resulting in a distribution of cold dust, warm
dust and hot dust.  Due to gravity, the dusty structures (e.g., clumpy clouds
or a smooth distribution) have a higher density closer to the accretion disk
and torus equatorial plane. In the innermost region of the torus, since the
emission from the black hole accretion disk is anisotropic
\citep[e.g.,][]{Netzer1987}, the dust grains sublimate and form a concave
structure for the hot dust emission region \citep{Kawaguchi2010}.  Following
\cite{Kawaguchi2010}, we also assume the innermost region of the hot dust
emission region is connected with the outermost region of the accretion disk.
In the zoom-in plot of the inner region of the torus (panel A1), we show the
concave part of the host dust emission is truncated at $\theta_{\rm min}$ and
$\theta_{\rm max}$, controlled by the thickness of the torus and the thickness
of the accretion disk, respectively.

For WDD quasars, an increase of AGN luminosity results in a smaller torus scale
height or larger opening angle, reducing the volume of the warm dust emission
region (cartoon B in Figure~\ref{fig:torus_geo}) and causing its output to
drop. This reduction may be enhanced according to the suggestion by
\cite{Honig2007}, that dust clouds at large torus scale heights would be
destroyed by their exposure to the increased AGN luminosity. However, the
structure of the innermost range of the hot dust emission region is identical
to that of normal quasars due to a constant $\theta_{\rm max}$.  Thus, there is
little change in hot dust emission.

The HDD quasars tend to have lower Eddington ratios. For the outer part of the
accretion disk, where gas pressure provides most of the vertical support, the
scale height $h\propto (L/L_{\rm Edd})^{-0.7}$ \citep[e.g.,][]{Krolik1999}.
Consequently, the hot dust emission region has a reduced $\theta_{\rm max}$ and
smaller dust density with increasing $h$ (cartoon C in
Figure~\ref{fig:torus_geo}). In addition, the projection of hot dust emission
in the polar direction is less effective due to the relatively high skewness of
the dust sublimation surface. As a result, we will not observe much hot dust
emission. We may also expect that the cloud density in the warm dust emission
region is decreased or that the opening angle of warm dust emission is
increased, which would explain the observed stronger silicate emission (see
Section~\ref{sec:char}) as well as the weak mid-IR emission.

In brief, we suggest that the observed deficiency of emission by warm and hot
dust in some quasars can result from modifications of the torus structure: (1)
WDD quasars can arise from an increase in the torus opening angle and the
resulting reduction in the volume of warm dust; and (2) HDD quasars can arise
from the increase of the accretion disk thickness for quasars with relatively
low Eddington ratios, and the accompanying adjustment in the geometry of the
regions dominating hot dust emission. 

\section{Summary}\label{sec:summary}

\cite{Elvis1994} proposed what has become the classic template for quasar SEDs,
shown since to be appropriate for the majority of luminous AGNs. However, a
minority of objects seem to depart from this standard in the infrared
\citep[e.g.,][]{Hao2010, Hao2011, Jiang2010}. We have studied the AGN infrared
SED variations in an archetypical optically selected sample composed of 87
$z<0.5$ PG quasars. The optical to far-IR SEDs of these quasars were
investigated with the aid of a three-component model including the AGN
emission, the host galaxy stellar emission, and the host galaxy infrared
emission powered by star formation. The host galaxy properties derived from the
SED decomposition are consistent with those determined through other methods:
the stellar contributions in the near-IR band based on the SED analysis are
similar to the values derived from image decompositions on the high-resolution
images obtained by HST or ground-based adaptive optics; and the infrared
luminosities of the host galaxies yield SFRs consistent with the strength of
the mid-IR 11.3$~\mum$ aromatic features.  Our SED model fits generally have
residuals within 0.3 dex of the observed SEDs at 0.2-500~$\mum$ and reproduce
the 0.5-70~$\mum$ region especially well. The principal results can be
summarized as follows.

\begin{enumerate}

{\item  The intrinsic infrared SEDs of the PG quasars cannot be represented by
	one single SED template. Normal quasars occupy $\sim60\%$ of the sample
	and their AGN SEDs can be described by the \cite{Elvis1994} template
	reasonably well. In comparison, the AGN template derived by
	\cite{Assef2010} appears to have too deep a minimum near 1.0~$\mum$, as
	indicated by the overestimate of host galaxy brightness using it in SED
        deconvolution.
    }

{\item  There is a substantial fraction ($\sim$30-40\%) of abnormal quasars
	with weak infrared emission in the PG sample, which have often been
	overlooked by previous studies. These IR-weak quasars can be classified
	into two populations with distinct SED properties.
    }

{\item The hot-dust-deficient (HDD) quasars are characterized by a deficiency
	of dust emission at $\lambda > 2~\mum$. The AGN SEDs of these quasars do
	not have the typical hot dust emission peaked at $\sim2~\mum$ and
	present very weak warm and cold dust emission. In the PG sample, they
	contribute $\sim$15-23\% of the total population. 
    }

{\item The warm-dust-deficient (WDD) quasars have a similar near-IR SED bump
       peaked at $\sim2.0~\mum$ as is the case for normal quasars but have a
       very quick drop in the mid-IR. The fraction of the WDD quasars in the
       PG sample is $\sim$ 14-17\%.
    }

{\item Compared with normal quasars, the HDD quasars are similar in terms of
	AGN luminosities and black hole masses, but they tend to have lower
	Eddington ratios with a K-S probability of $\sim$0.025 of being drawn from
	the same distribution. We also find that the HDD quasar fraction is not
	luminosity dependent, consistent with previous work by \cite{Hao2010,
	Hao2011} and \cite{Mor2012}. These HDD objects account for roughly 10\% of
	quasar samples selected on the basis of bolometric luminosity.
    }

{\item The WDD quasars do not differ from normal quasars in Eddington ratio,
       but their fraction increases with AGN luminosity. The decreased mid-IR
	to optical luminosity ratios of quasars with increasing AGN luminosity
	(as found by, e.g., \citealt{Maiolino2007, Treister2008, Mor2011,
	Calderone2012, Ma2013, Gu2013, Roseboom2013}) may be mostly contributed
	by the more frequent appearance of WDD quasars.
    }

{\item Compared with normal quasars, both WDD and HDD quasars tend 
       to have stronger silicate emission features at $\sim10~\mum$, which can
       be explained by a reduced scale height of the warm dust above the
       equatorial plane of the circumnuclear torus.
    }

{\item   The high-$z$ dust-free or dust-poor quasars found by, e.g.,
	\cite{Jiang2010}, \cite{Hao2010} share similar SEDs to the HDD quasars
	in the PG sample. WDD quasars are also seen at $z=0.5\sim6$.
	Although the near- to mid-infrared SEDs of high-$z$ quasars vary to
	some degree, they show no obvious difference from the archetypal PG
        sample.
    }

{\item Considering biases in the parent sample and the selection methods, there
       is no evidence for a strong cosmic evolution of the dust-deficient
       quasars. Instead, we suggest that observed dust-deficient behavior of
       quasars is caused by a change of the torus structure controlled by AGN
       luminosity and Eddington ratio.
    }

\end{enumerate}

\hspace{1cm}
\acknowledgments

We thank Richard Green and the anonymous referee for helpful suggestions and
Lei Xu for sharing the data of the LoCuSS {\it Spitzer}/MIPS AGN sample.
This work was supported by NASA grants NNX13AD82G and 1255094. This publication
has made use of data products from the {\it Wide-field Infrared Survey
Explorer}, which is a joint project of the University of California, Los
Angeles, and the Jet Propulsion Laboratory/California Institute of Technology,
funded by the National Aeronautics and Space Administration.  This publication
also makes use of data products from NEOWISE, which is a project of the Jet
Propulsion Laboratory/California Institute of Technology, funded by the
Planetary Science Division of the National Aeronautics and Space
Administration. This publication makes use of data products from the Two Micron
All Sky Survey, which is a joint project of the University of Massachusetts and
the Infrared Processing and Analysis Center/California Institute of Technology,
funded by the National Aeronautics and Space Administration and the National
Science Foundation.  This work is also based in part on data obtained as part
of the UKIRT Infrared Deep Sky Survey. We acknowledge the use of the NASA/IPAC
Extragalactic Database (NED) which is operated by the Jet Propulsion
Laboratory, California Institute of Technology, under contract with the
National Aeronautics and Space Administration. This work has also made use of
the VizieR catalog access tool, CDS, Strasbourg, France.

\bibliographystyle{apj.bst}

\end{document}